\documentclass[10pt,aps,amsmath,amssymb,pra,twocolumn,superscriptaddress]{revtex4-2}

\usepackage{graphicx}
\usepackage{dcolumn}
\usepackage{bm}
\usepackage[colorlinks=true,linkcolor=magenta,citecolor=blue]{hyperref}

\begin{document}
\title{Recent Advances in Laser Self-Injection Locking to High-\texorpdfstring{$Q$}{Q} Microresonators}

\author{Nikita M. Kondratiev}
\email{nikita.kondratyev@tii.ae}
\affiliation{Directed Energy Research Centre, Technology Innovation Institute, Abu Dhabi, United Arab Emirates}

\author{Valery E. Lobanov}
\author{Artem E. Shitikov}
\affiliation{Russian Quantum Center, Moscow, Russia}

\author{Ramzil R. Galiev}
\affiliation{Directed Energy Research Centre, Technology Innovation Institute, Abu Dhabi, United Arab Emirates}

\author{Dmitry A. Chermoshentsev}
\affiliation{Russian Quantum Center, Moscow, Russia}
\affiliation{Skolkovo Institute of Science and Technology, Skolkovo, Russia}
\affiliation{Moscow Institute of Physics and Technology, Dolgoprudny, Moscow Region, Russia}

\author{Nikita Yu. Dmitriev}
\affiliation{Russian Quantum Center, Moscow, Russia}

\author{Andrey N. Danilin}
\affiliation{Russian Quantum Center, Moscow, Russia}
\affiliation{Faculty of Physics, Lomonosov Moscow State University, Moscow, Russia}

\author{Evgeny A. Lonshakov}
\affiliation{Directed Energy Research Centre, Technology Innovation Institute, Abu Dhabi, United Arab Emirates}

\author{Kirill N. Min'kov}
\affiliation{Russian Quantum Center, Moscow, Russia}

\author{Daria M. Sokol}
\affiliation{Russian Quantum Center, Moscow, Russia}
\affiliation{Moscow Institute of Physics and Technology, Dolgoprudny, Moscow Region, Russia}

\author{Steevy J. Cordette}
\affiliation{Directed Energy Research Centre, Technology Innovation Institute, Abu Dhabi, United Arab Emirates}

\author{Yi-Han Luo}
\affiliation{Shenzhen Institute for Quantum Science and Engineering, Southern University of Science and Technology, Shenzhen 518055, China}
\affiliation{International Quantum Academy, Shenzhen 518048, China}

\author{Wei Liang}
\affiliation{Key Laboratory of Nanodevices and Applications, Suzhou Institute of Nano-Tech and Nano-Bionics Chinese Academy of Sciences, Suzhou, Jiangsu, China}

\author{Junqiu Liu}
\email{liujq@iqasz.cn}
\affiliation{Shenzhen Institute for Quantum Science and Engineering, Southern University of Science and Technology, Shenzhen 518055, China}
\affiliation{International Quantum Academy, Shenzhen 518048, China}
\affiliation{Hefei National Laboratory, University of Science and Technology of China, Hefei 230026, China}

\author{Igor A. Bilenko}
\affiliation{Russian Quantum Center, Moscow, Russia}
\affiliation{Faculty of Physics, Lomonosov Moscow State University, Moscow, Russia}

\date{\today}

\begin{abstract}
The stabilization and manipulation of laser frequency by means of an external cavity are nearly ubiquitously used in fundamental research and laser applications. 
While most of the laser light transmits through the cavity, in the presence of some back-scattered light from the cavity to the laser, the self-injection locking effect can take place, which locks the laser emission frequency to the cavity mode of similar frequency. 
The self-injection locking leads to dramatic reduction of laser linewidth and noise. 
Using this approach, a common semiconductor laser locked to an ultrahigh-$Q$ microresonator can obtain sub-hertz linewidth, on par with state-of-the-art fiber lasers. 
Therefore it paves the way to manufacture high-performance semiconductor lasers with reduced footprint and cost. 
Moreover, with high laser power, the optical nonlinearity of the microresonator drastically changes the laser dynamics, offering routes for simultaneous pulse and frequency comb generation in the same microresonator. 
Particularly, integrated photonics technology, enabling components fabricated via semiconductor CMOS process, has brought increasing and extending interest to laser manufacturing using this method.
In this article, we present a comprehensive tutorial on analytical and numerical methods of laser self-injection locking, as well a review of most recent theoretical and experimental achievements.
\end{abstract}

\maketitle

\section{Introduction}

Laser sources with narrow linewidth and low noise are of paramount importance for nearly all laser applications such as timing, communication, spectroscopy, metrology, navigation, as well as fundamental research. 
State-of-the-art chip-scale semiconductor laser diodes emit continuous-wave light of wavelength from ultraviolet to mid-infrared, sufficient optical power, and are produced with low cost and high volume. 
However, the Achilles' heel of them is the inevitable frequency fluctuations due to low cavity finesse. 
Several methods to frequency-stabilize diode lasers have been demonstrated. 
One solution to achieve simultaneously high laser power and narrow linewidth is to transfer the narrow frequency spectrum of a well-stabilized, but low-power master laser to a high-power broad-spectrum slave diode laser using optical injection \cite{Hadley:1986}. 
However, such a system is complicated to implement and very sensitive to ambient perturbation. 
The linewidth reduction can also be achieved by locking the laser diode to an external high-finesse reference cavity. 
Active locking, like Pound-Drever-Hall (PDH) technique \cite{Drever:1983, Telle:1989, Zhu:93}, is conventionally and widely used, requiring optical modulation and electronic feedback circuitry. 
The side-of-fringe stabilization \cite{Kourogi:1991} provides locking without optical modulation, but requires stable laser intensity and the reference level.

Passive stabilization of semiconductor lasers uses resonant optical feedback from an external optical element. 
One of the most effective approaches to stabilize laser frequency using an external cavity is based on the self-injection locking effect.
Self-injection locking is a profound phenomenon observed in oscillatory circuits. 
For many years this effect has been used in radio-physics, radio-engineering, and microwave electronics to improve the spectral purity of the devices \cite{Ohta1, Ohta2, Ota1, Chang1, Chang2, magnetron1, magnetron2, gyrotron1, gyrotron2, 6999932, 7119883}. 
It has also been widely applied for stabilizing laser sources, and has enabled various practical applications \cite{Dahmani:87,li:1988, Laurent1989} including high-resolution spectroscopy and high-precision metrology. 
Self-injection locking of a chip-scale semiconductor laser to an optical microresonator results in sub-kilohertz laser linewidth that is orders of magnitude smaller than the original linewidth of the free-running semiconductor lasers (typically megahertz to hundreds of megahertz) \cite{Liang:10, Liang2015}.

\begin{figure*}[t!]
\centering
\includegraphics[width=0.95\linewidth]{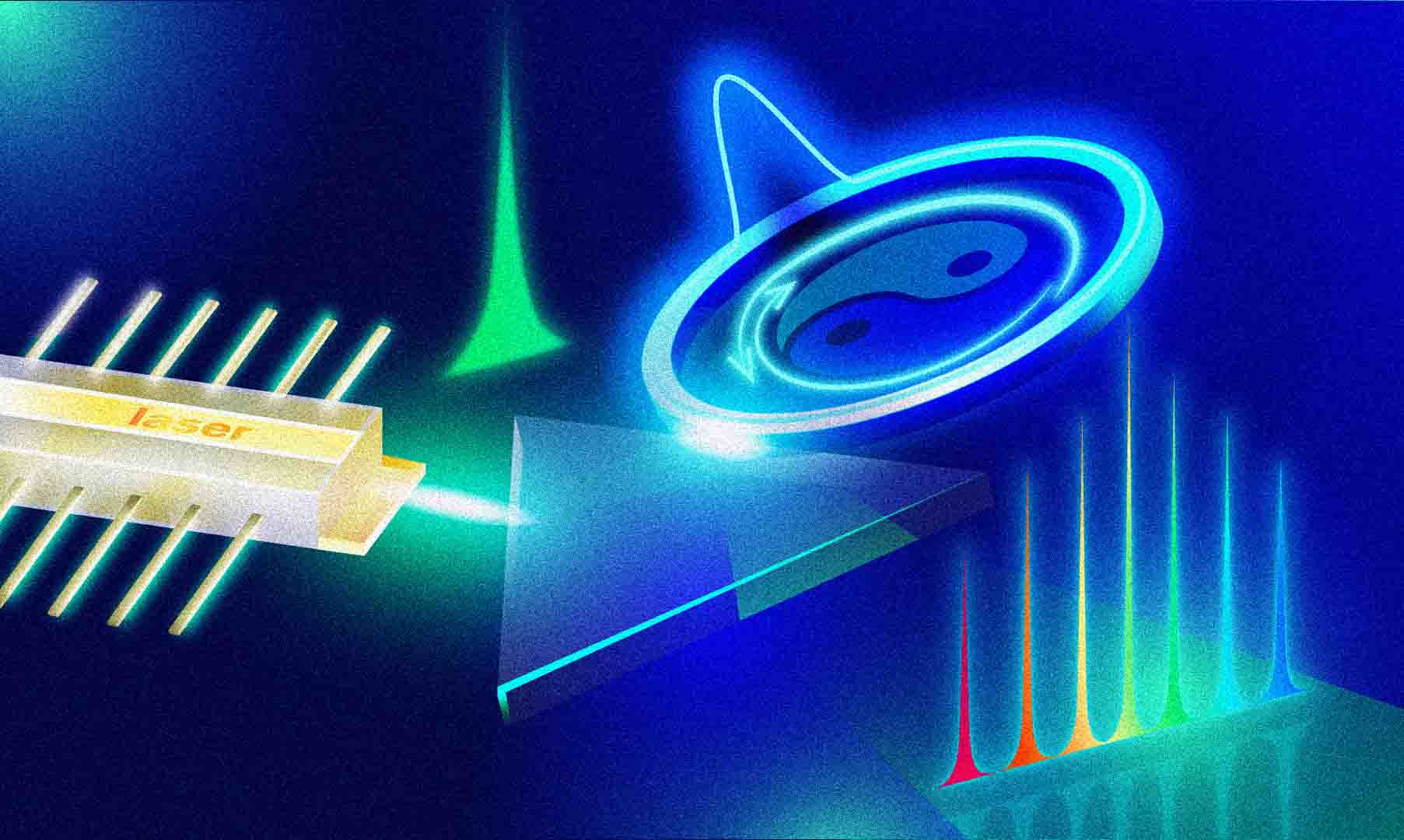}
\caption{
Artistic vision of self-injection locking of a diode laser to a microring resonator, which enables frequency comb generation seeded by its narrowed linewidth emission.
}
\label{SIL_illustration}
\end{figure*}

Self-injection locking of oscillators has been extensively studied for the last three decades. 
It was shown initially that adding a partially transparent mirror at the output of a Fabry-P\'erot (FP) laser can lead to laser noise reduction \cite{osti_6336783, Lang_Kobayashi, Belenov_1983, Patzak1983, Agrawal1984, Tkach1986}. 
However, this stabilization scheme has significant limitations due to the dynamic instability arising from the excessively strong optical feedback. 
Even the relative feedback power at the level of $10^{-4}$ can be sufficient to destabilize the system. 
The instability can be reduced if the feedback is frequency-selective. 
Although resonant reflectors are also used in this case (diffraction, Bragg or holographic gratings in the Littrow or Littman configuration \cite{Olesen1983, Saito1982}), the main effect comes from the resonators formed by one of the diode facets on one side, the external element used for stabilization on the other side, and the distance between them. 
In this case, the output edge of the diode is often covered with an anti-reflective coating. 
Such narrow-linewidth lasers are also called external cavity lasers. 
Regimes of the optical feedback in such a system have been heavily studied for simple mirror feedback \cite{Acket1984, Tkach1986, Petermann1995} and even used for distance measurement \cite{Selfmixing}.

Self-injection locking of a laser generation line to a high-quality-factor (high-$Q$) mode of an external resonator provides fast frequency-selective optical feedback, which leads to improved stabilization of the laser frequency \cite{Dahmani:87, Hollberg:1987, Li:1989, hemmerich90oc, Hemmerich:94}. 
This configuration is dynamically stable and can produce coherent light even when the relative feedback power exceeds tens of percent. 
It was initially demonstrated with tilted (or V-shaped) FP resonators \cite{Dahmani:87, Laurent1989,hemmerich90oc}. 
Then it was studied for other large resonators such as discrete mirror ring cavities \cite{Hemmerich:90ol} and monolithic total-internal-reflection resonators (TIRRs) \cite{Hemmerich:94}. 
It was shown that the locking results in the reduction of the phase and amplitude noises \cite{Dahmani:87,hjelme91jqe}, while simultaneously allows frequency tuning of the laser emission and facilitates efficient frequency doubling \cite{Hemmerich:94}. 
The laser linewidth can be narrowed by six orders of magnitude if a high-$Q$ microresonator is involved \cite{Liang2015, Zhao:11}. 
A theory of the self-injection locking was developed for larger optical cavities nearly three decades ago \cite{Li:1989,hjelme91jqe}, which indicated that a high $Q$ factor of the optical modes, low modal density, and a highly stable optical path are required to achieve prominent linewidth reduction. 
Unfortunately, the stabilization technique using large optical cavities has drawbacks due to the sensitivity of the cavities to the environment. 
It was shown recently that a high-$Q$ FP cavity can also be miniaturized to make compact self-injection-locked narrow-linewidth lasers \cite{Li:21}.

Whispering-gallery-mode (WGM) microresonators \cite{BRAGINSKY1989393, PhysRevA.70.051804, Matsko2006, Savchenkov:07,doi:10.1002/lpor.201000025, Lin:14, Henriet:15,strekalov:2016, GRUDININ200633, Lecaplain:16, Shitikov:18}, combining high $Q$ factors in a wide spectral range with small size, simple construction, and reduced environmental sensitivity, have proven to be suitable to implement self-injection locking. 
The first implementation of using a high-$Q$ optical microresonator for laser linewidth narrowing due to the optical feedback from the microresonator was reported in Ref. \cite{Vassiliev1998}. 
However, the term  ``self-injection locking" was not mentioned in this paper. 
The authors first encountered a parasitic effect that prevented the resonance curve from being obtained because the tunable laser was clanging to the resonance. 
They immediately realized that it is an effect similar to the frequency pulling of radio-frequency generators by additional high-$Q$ circuits and that the feedback is formed due to the resonant Rayleigh back-scattering in the resonator. 
Comprehensive analysis of the back-scattering and counter-propagating mode formation in WGM microresonators was performed in Ref. \cite{Gorodetsky:00}. 
The possibility of realizing robust and effective laser stabilization without an external electronic feedback chain was implemented in Ref. \cite{ilchenko1999high, oraevsky2001frequency}, where the existence of the optimal coupling and back-scratching coefficients was experimentally shown. 
When a laser is locked to a microresonator mode, the laser wavelength can be fine-tuned by changing the microresonator mode frequency, e.g. by mechanical compression or extraction or stretching the microresonator \cite{rezac2001locking, bilenko2002measurement}. 
Detailed theoretical analysis of thermodynamic and quantum limits of the resonance frequency stability of solid-state WGM microresonators was performed in Ref. \cite{matsko2007whispering}. 
An efficient method for numerical calculation of the thermo-refractive noise (TRN) was suggested in Ref. \cite{KONDRATIEV20182265}, and experimental characterization of TRN of integrated silicon nitride microresonators has been shown in Ref. \cite{Huang:19}. 
Substantial progress from the laboratory demonstrations to off-the-shelf devices started since Ref. \cite{Liang:10} demonstrated a narrow-linewidth DFB lasers self-injection-locked to a WGM microresonator in a package with 15 mm size,  3 mm thickness with instantaneous linewidth narrower than 200 Hz.

Recent studies have demonstrated the feasibility of using high-$Q$ optical WGM microresonators for passive stabilization of single-frequency \cite{Vassiliev2003, Liang:10, Liang2015, Dale:16, Xie:15, Savchenkov:19a, Savchenkov:2022} or even multi-frequency \cite{Pavlov:18, Donvalkar_2018, Galiev:18, Pavlov2018aip, Savchenkov:19, Savchenkov:2020, Yacoby:2021, Ji:22} semiconductor lasers to sub-kilohertz linewidth in different spectral ranges. 
Some of the lasers became commercial products \cite{OEwaves}. 
Very recent studies have shown photonic package of lasers on photonic integrated circuits (e.g. made of silica \cite{Lee:2012, Yang:2016, Wu:20} or silicon nitride \cite{Spencer:14, Xuan:16, Pfeiffer:16, Ye:19, Ji:2021}), where the WGM microresonators are in the form of high-$Q$ microrings \cite{Stern2018, Li:18, Gaeta2019, Raja2019, Shen:20, Raja:20, Kovach:20, Jin2021, Shim:2021, Lihachev:22, Siddharth:2022} or microspirals \cite{Guo2022}. 
Laser self-injection locking has also been demonstrated in fiber-ring resonators \cite{Korobko2017, Spirin:20}, but with worse stability due to small intermode distance and environmental sensitivity. 
Most of the configurations rely on the Rayleigh scattering in the microresonator. 
However, there are also schemes using add-drop ports or filters with symmetric couplers \cite{Li:18, Shao:21} or circulators \cite{Jiang:21_0, Jiang:21}.

Here we review some recent key advances in theoretical and experimental study of the physics and applications of laser self-injection locking. 
We note that most of the effects here are described on the example of WGM microresonators, but generally can also be implemented with any types of high-$Q$ optical resonators e.g. ring, FP and other types.

\section{Stabilization of lasers by means of self-injection locking}

Fast optical feedback of laser self-injection locking enables significant reduction in laser linewidth. 
First demonstrated with fused silica microspheres \cite{Vasiliev:96}, this method is now actively used to control the spectral characteristics, e.g. to narrow the linewidth and stabilize the frequency, of various laser sources \cite{Vassiliev1998, Vassiliev2003, Liang:15}, including fiber ring \cite{Sprenger:09}, and DFB lasers \cite{Liang:10}. 
Note that over the past decade, significant progress has been made in the applications of this technique. 
In 2010, it was reported that the linewidth of an external-cavity semiconductor laser was reduced by a factor of $10^4$ and an instantaneous linewidth of less than 200 Hz was achieved \cite{Liang:10}.
In 2015 the laser linewidth was further decreased by a factor of $10^7$ and reached a sub-Hertz level \cite{Liang2015}.

\subsection{Basics of self-injection locking}

\begin{figure*}[t!]
\centering
\includegraphics[width=.49\linewidth]{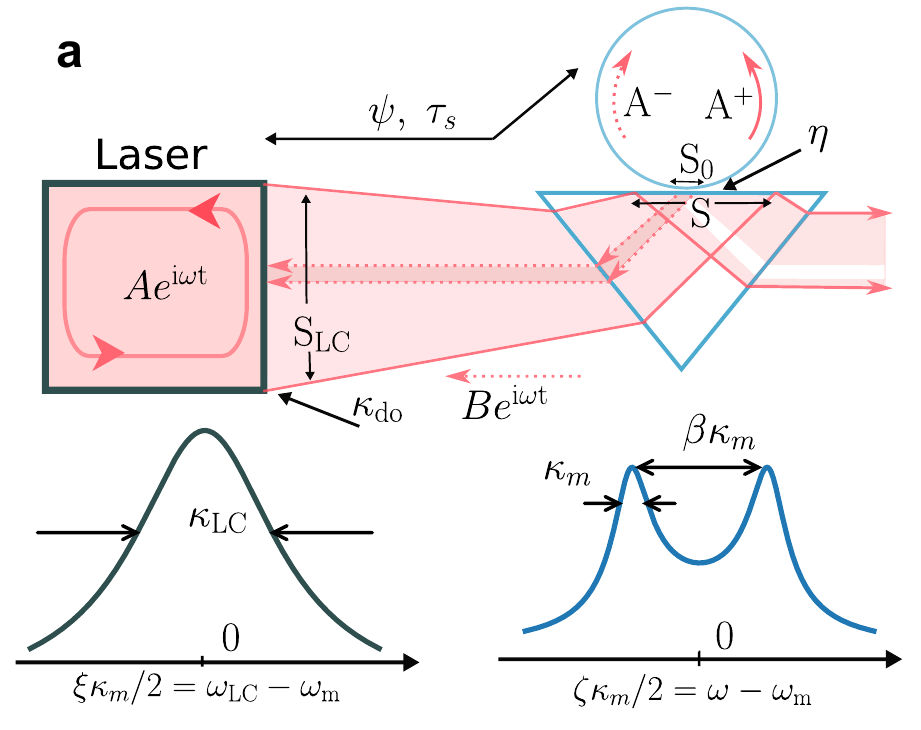}
\includegraphics[width=.5\linewidth]{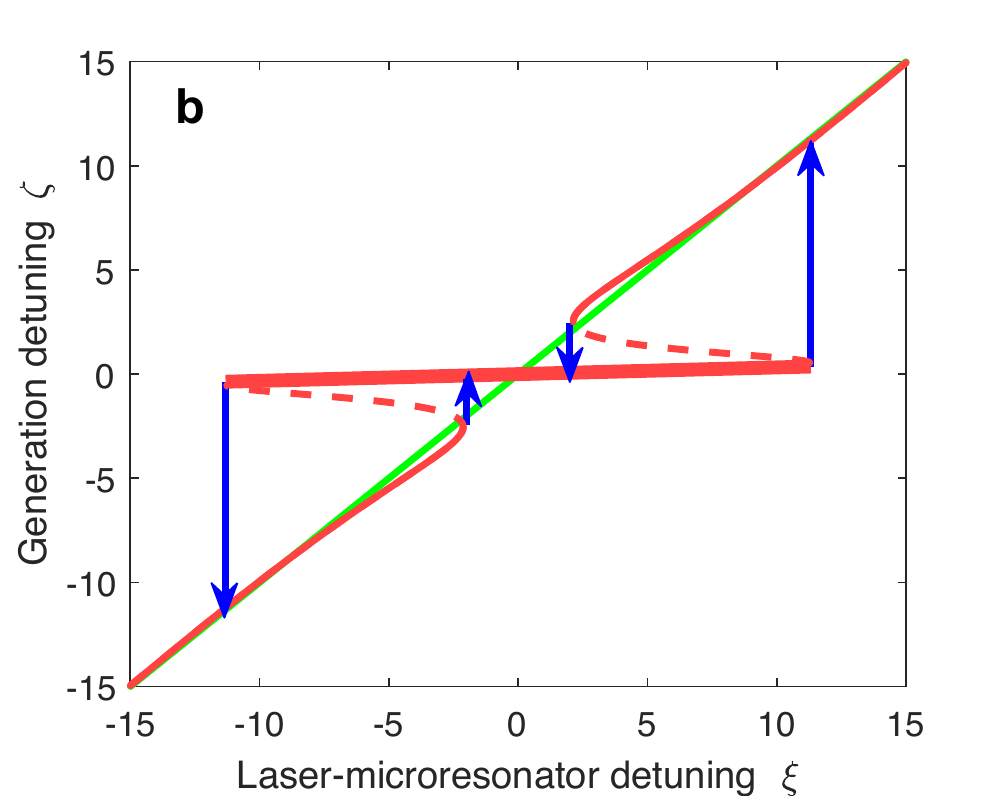}
\caption{
Schematics of laser self-injection locking.
\textbf{a}. Top panel: Schematics of laser self-injection locking to a WGM mode using a prism coupler. 
Bottom panel: Resonance profiles of the laser cavity (left) and the WGM reflected wave with normal mode splitting (right). 
$A$: the laser generation field amplitude
$\omega_{\rm LC}$ and $\kappa_{\rm LC}$: the laser cavity mode frequency and linewidth. 
$\kappa_{do}$: laser output mirror coupling rate.
$B$: back-reflected wave amplitude. 
$S_{\rm LC}$ and $S$: laser beam cross-section area at the laser aperture and on the prism surface.
$\tau_s$: round-trip time of feedback
$\psi$: locking phase.
$A^{+}$ and $A^{-}$: amplitudes of the forward and backward waves inside microresonator.
$\eta$: microresonator coupling coefficient. 
$\omega_m$ and $\kappa_m$: the microresonator mode frequency and linewidth.
$\omega$: laser generation frequency. 
\textbf{b}. An optimal self-injection locking curve with $\psi=0$, $\kappa_m\tau_s\ll1$, and $K_0=35$. 
The unstable branches are shown with dashed lines. 
The locking range is marked with a thick red line. 
The bi-stable transitions are marked with with blue arrows.
Panel \textbf{a} is taken from Ref. \cite{Galiev2020},  and Panel \textbf{b} from Ref. \cite{Kondratiev:17}.
}
\label{fig:scheme}
\end{figure*}

The schematics of laser self-injection locking is presented in Fig.~\ref{fig:scheme}, where a refocused laser beam is resonantly coupled to a high-$Q$ WGM microresonator via a prism coupler. 
Any other resonantly reflecting element can replace the WGM microresonator, but the general idea and qualitative results are the same. 
As shown in Fig.~\ref{fig:scheme}a, a part of the laser radiation is resonantly back-scattered (e.g. due to Rayleigh scattering in the WGM microresonators \cite{Gorodetsky:00}, or direct reflection in FP cavities) to the laser cavity, locking the laser radiation frequency to the frequency of the microresonator mode. 
Note that the bottom left panel of Fig.~\ref{fig:scheme}a shows the laser cavity resonance, not the free-running laser emission line, whose width $\delta\omega_{\rm free}$ is defined by the laser noises and limited by Schawlow-Townes relation. 

To describe the self-injection locking, one can start with the general phase and amplitude lasing criteria for a FP laser diode, with amplitude reflectivity coefficients of the output and end facets $R_o$ and $R_e$, as 
\begin{equation}
\label{panda}
\begin{cases}
\omega_{\rm LC} \tau_{\rm LC} + \arg(R_eR_o) + \alpha_g g \tau_{\rm LC} = 2\pi N,\\
g \tau_{\rm LC} + \ln|R_eR_o|  = 0,
\end{cases}
\end{equation}
where $\omega_{\rm LC}$ is the laser generation frequency of the free-running diode,
$\tau_{\rm LC}$ is the light round-trip time in the diode laser cavity,
$g$ is the diode medium gain,
$\alpha_g$ is the Henry factor, 
and $N$ is an integer number that can be attributed to the mode number of the system. 
If a reflector with amplitude reflectivity $\Gamma$ is introduced to induce self-injection, we can unite it with the output facet, composing an effective reflector. 
This effective reflector can now be considered as a FP cavity with the length corresponding to the diode-reflector round-trip time $\tau$, and with effective reflectivity
\begin{equation}
\label{eq:effective_reflection}
R_{\rm eff}(\omega) = \frac{ R_o-\Gamma e^{i\omega \tau_s}}{1 - \Gamma R_o e^{i\omega \tau_s}}.
\end{equation}
Similarly, the dependence of the self-injected laser generation frequency $\omega$ can be derived by inserting $R_{\rm eff}$ into Eq. \eqref{panda} instead of $R_o$. 
Solving both systems and using $2\pi N \approx \omega_{\rm LC} \tau_{\rm LC} \gg 1$ (see Ref. \cite{Kondratiev:Strong}), we obtain the relation between the free-running laser frequency and the locked laser frequency as following:
\begin{equation}
\omega_{\rm LC} \approx \omega + \frac{1}{\tau_{\rm LC}}\arg \left (\frac{R_{\rm eff}}{R_o} \right) - \frac{\alpha_g}{\tau_{\rm LC}} \ln\left | \frac{R_{\rm eff}}{R_o} \right |.
\label{general_tuning_curve}
\end{equation}

This relation is usually called the ``tuning curve''. 
It shows the dependence of the system generation frequency $\omega$ on the free-running laser generation frequency $\omega_{\rm LC}$ (i.e. without feedback). 
For free-running laser, the tuning curve is a 1:1 line. 
For weak self-injection locking, the tuning curve is approaching to the 1:1 line.
For strong self-injection locking, the tuning curve has nearly-horizontal parts -- locking ranges.
A common form of the tuning curve is shown in Fig.~\ref{fig:scheme}b (there frequencies are taken relative to the reflector resonance and normalized over its resonance width).
When the laser frequency is tuned (e.g. by changing the laser current) far from the resonance frequency of the reflector (i.e. the microresonator), the laser generation frequency follows the 1:1 line (see the green line in Fig. \ref{fig:scheme}b). 
When the laser frequency approaches the reflector resonance -- to the multi-stable part of the red curve in Fig.~\ref{fig:scheme}b -- it can jump to the stable central part of the curve (thick red line in Fig.~\ref{fig:scheme}b). 
In this regime, varying the laser cavity frequency (e.g. by changing the laser current or due to noises and fluctuations) result in negligible change in the system generation frequency, i.e. the laser is locked to the microresonator. 
Finally, the locking is lost if the laser cavity is tuned far from the microresonator resonance frequency (see outer blue arrows in Fig.~\ref{fig:scheme}b).

The inverse slope of the tuning curve in the locked region $K=\partial\omega_{\rm LC}/\partial\omega$ is called the ``stabilization coefficient'', and its square represents the locked laser linewidth narrowing factor \cite{Laurent1989, Kondratiev:17, Kazarinov1987}. 
This derivative should be averaged near the laser cavity detuning over its initial linewidth $K^{-1}=\int_{\omega_{\rm LC}-\delta\omega_{\rm free}}^{\omega_{\rm LC}+\delta\omega_{\rm free}}|\frac{\partial\omega}{\partial\omega_{\rm LC}}|\frac{d\omega_{\rm LC}}{2\delta\omega_{\rm free}}$ to avoid singularities, as the linewidth can be viewed as frequency fluctuation.
Note that the jump between the metastable branches can also happen before the turning point \cite{Kondratiev:20, Shitikov2022arx} in case of a higher microresonator $Q$ factor. 
However, such spontaneous locking usually happens to the branch with the highest stabilization coefficient.

\begin{figure*}[t!]
\includegraphics[width=0.495\linewidth]{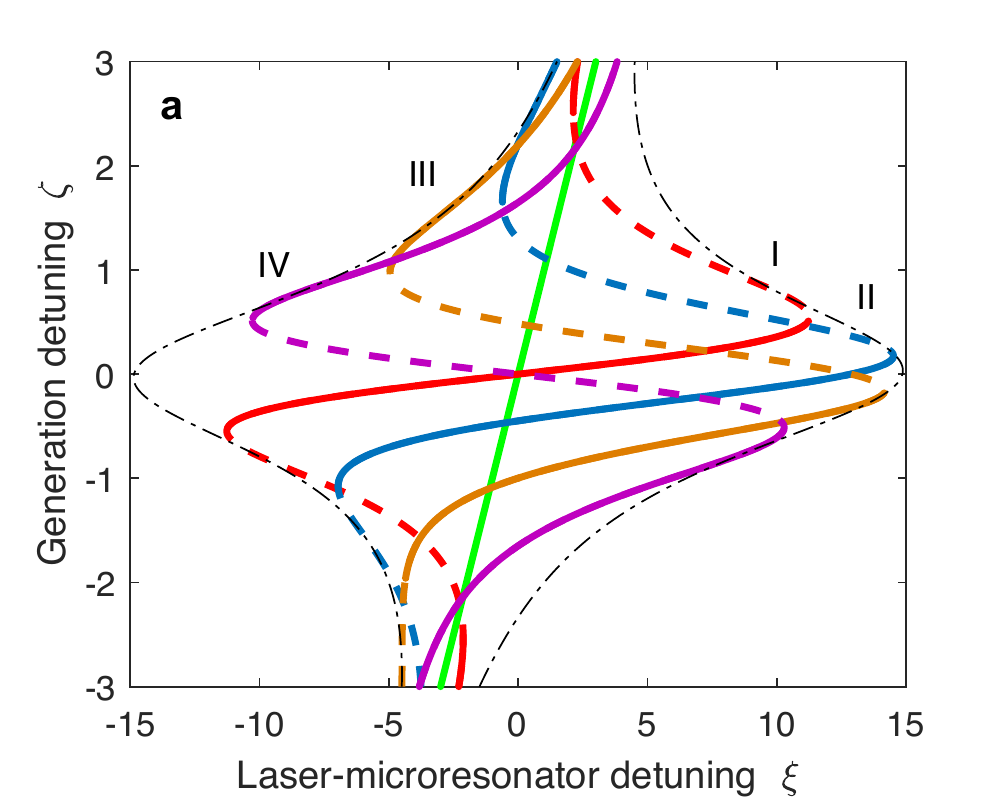}
\includegraphics[width=0.495\linewidth]{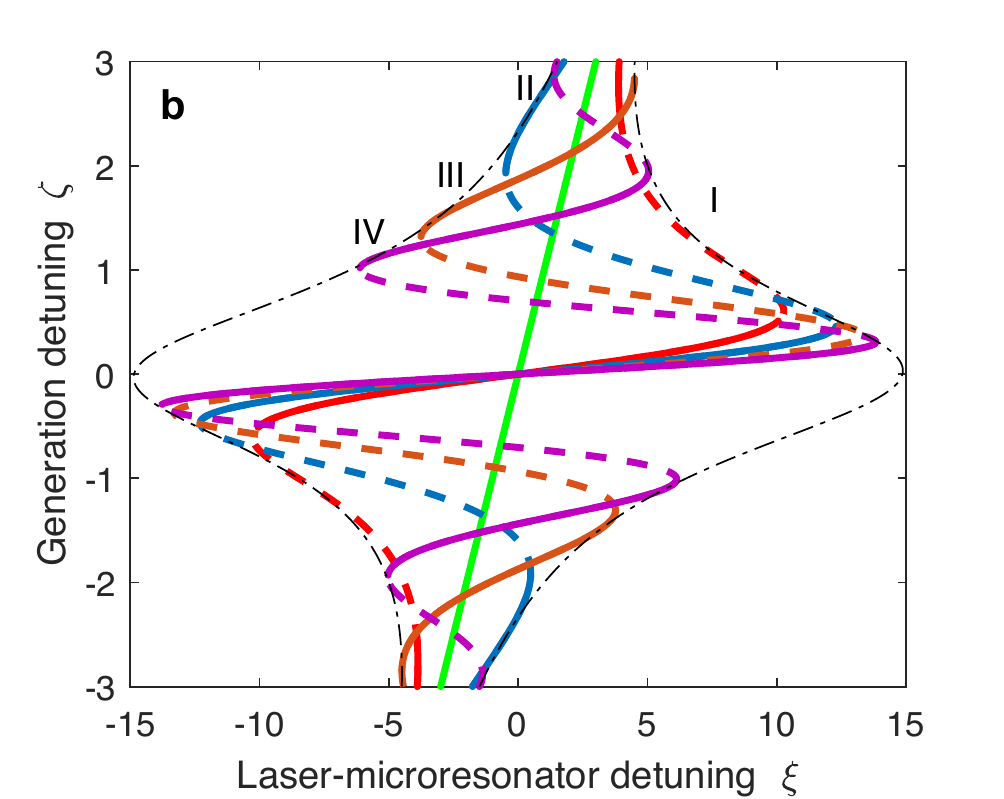}
\caption{
Illustration of the phase and normalized delay as functions of laser-microresonator detuning.
\textbf{a}. Tuning curves Eq. \eqref{master} for different initial phases $\psi$. 
Points I-IV correspond to phases $\psi=[0, \pi/3, 2\pi/3, \pi]$ with $\kappa\tau_s\ll1$. 
The envelope for the family of curves with different $\psi$ is shown with the black dash-dotted line. 
In comparison, the solid green line shows a tuning curve for a free-running laser. 
\textbf{b}. Tuning curves I-IV corresponding $\psi=0$ and long delay, so that $\kappa_m\tau_s/2 =[0,1,2,3]$. 
This figure is taken from Ref.  \cite{Kondratiev:17}.
}
\label{Pulling_delay}
\end{figure*}

For complex reflectors, Eq. \eqref{general_tuning_curve} is difficult to analyze. 
A simple and common approach was considered in Ref. \cite{Laurent1989} for tilted FP cavities and in Ref. \cite{Kondratiev:17} for WGM microresonators. 
The equivalence with Eq. \eqref{general_tuning_curve} in a specific range of feedback and front facet reflectivity was shown in Ref. \cite{Kondratiev:Strong}.
The system can be described by the nonlinear rate equation:
\begin{equation} 
\dot A+\left [\frac{\kappa_{\rm LC}}{2} -\frac{g}{2}(1+i\alpha_g) -i(\omega-\omega_{\rm LC})\right ] A=\kappa_{do}B,
\label{LaserC}
\end{equation}
where $\omega_{\rm LC}$ and $\kappa_{\rm LC}$ are the laser cavity eigenfrequency and loss rate, 
$\kappa_{do}=(1-R_o^2)/(\tau_{\rm LC}R_o)$ is the coupling rate of laser output mirror divided by front facet reflectivity, 
$g=g(|A|^2)$ is the laser gain, 
$\omega$ is the actual laser generation frequency, 
$A$ is the laser field slowly-varying complex amplitude 
and $B$ is the complex amplitude of the field reflected from the microresonator.
In the case of small feedback, $B$ can be described by the following equation
\begin{widetext}
\begin{equation}
B(t)=\Gamma(\zeta)A(t-\tau_s)e^{i\omega\tau_s}\approx\sqrt{\Theta}\frac{2i\eta\beta}{(1-i\zeta)^2+\beta^2}A(t-\tau_s) e^{i\omega\tau_s},
\label{B(t)}
\end{equation}
\end{widetext}
where $\Gamma(\zeta)$ is the microresonator amplitude reflection coefficient. 
It depends on the detuning of the laser oscillation frequency $\omega$ from the nearest microresonator eigenfrequency $\zeta=2(\omega-\omega_m)/\kappa_m$ ($\omega_m$ and $\kappa_m$ are the microresonator mode frequency and the loaded linewidth or loss rate). 
In Eq. \eqref{B(t)} we explicitly substituted the reflectivity of a WGM microresonator \cite{Gorodetsky:00} where
$\eta$ is the dimensionless pump coupling coefficient,
$\beta$ is the normalized mode-splitting coefficient (Rayleigh scattering), 
and $\Theta$ is the power mode coupling factor, proportional to the ratio of the laser aperture area $S_{\rm LC}$ to the final beam area $S$. 
The results for FP cavities are qualitatively the same \cite{Laurent1989}.
As discussed earlier, Eq. \eqref{B(t)} does not account for the power reflecting back and forth inside the front facet-microresonator facet, but it works fine for small feedback.

\begin{figure*}[t!]
\centering
\includegraphics[width=0.55\linewidth]{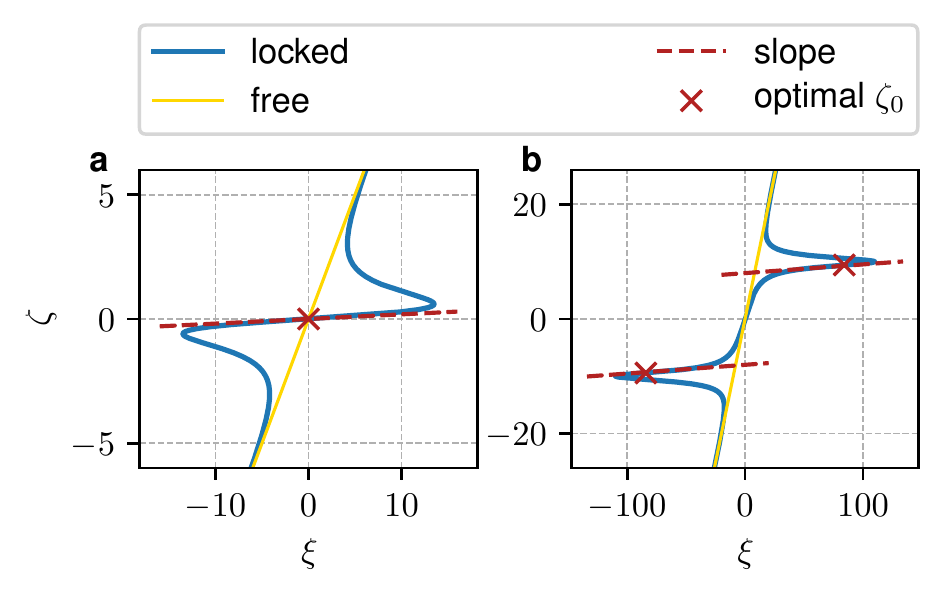}
\includegraphics[width=0.44\linewidth]{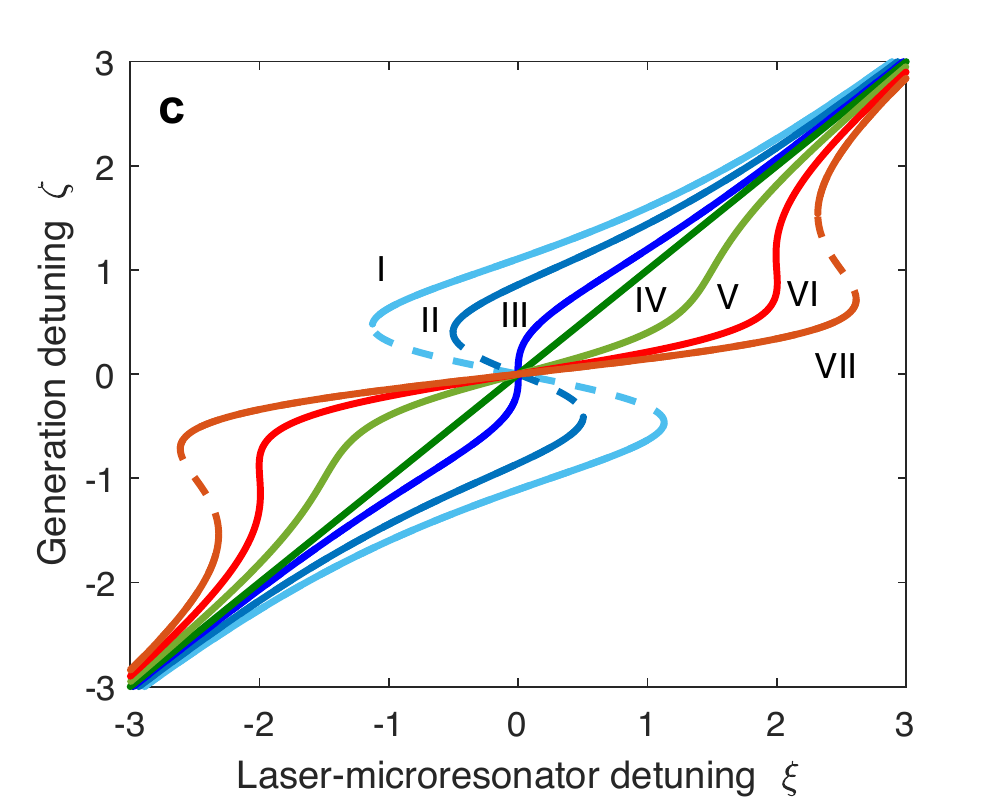}
\caption{
Analytical study of tuning curves with self-injection locking.
\textbf{a} and \textbf{b} show the tuning curves for $\beta=0.1$ and $\beta=10$ respectively, with $K_0/\beta=400$ and $\kappa_m\tau_s=0.011$. 
The red dashed lines show the slope of locking bands, and the red crosses mark the optimal points $\zeta=\zeta_0$ Eq. \eqref{zeta0}.
\textbf{c} shows the tuning curves for different $K$ values. 
Curves I-III correspond to $\psi=\pi$ and $K=[5,3,1]$, respectively. 
Curves IV-VII correspond to $\psi=0$ and $K=[0,2,4,6]$, respectively.
All quantities are plotted in dimensionless units. 
Panels \textbf{a,  b} are taken from Ref. \cite{Galiev2020},  and Panel \textbf{c} from Ref. \cite{Kondratiev:17}.
}
\label{fig:tune}
\end{figure*}

It is convenient to use the normalized tuning curve to analyze the self-injection locking effect. 
The curve shows the dependence of the effective frequency detuning $\zeta$ on the detuning of the laser cavity frequency $\omega_{\rm LC}$ from the microresonator eigenfrequency $\xi=2(\omega_{\rm LC}-\omega_m)/\kappa_m$. 
Equation \eqref{LaserC} should be split into the amplitude and phase parts, with the former discarded, and then the tuning curve can be determined. 
We will return to it in some sense in Section \ref{chap:MultifreqSec}. 
The tuning curve can be described by the following \cite{Kondratiev:17}:
\begin{align}
\label{master}
\xi=\zeta+\frac{K_0}{2}&\frac{2\zeta\cos\bar\psi+(1+\beta^2-\zeta^2)\sin\bar\psi}{(1+\beta^2-\zeta^2)^2+4\zeta^2},\\
\label{masretphase}
&\bar\psi=\psi+\frac{\kappa_m \tau_s}{2}\zeta,
\end{align}
where $\psi=\omega_m \tau_s - \arctan \alpha_g - 3/2\pi$ is the locking phase \cite{Kondratiev:17}, determined by the round-trip time $\tau_s$ from the laser to the microresonator, the microresonator resonant frequency $\omega_m$, and the Henry factor $\alpha_g$.
$K_0=\frac{8\eta\beta}{\kappa_m}\kappa_{do}\sqrt{\Theta}\sqrt{1+\alpha_g^2}$ is the zero-point stabilization coefficient, and the Henry factor-related laser cavity frequency shift is included into $\omega_{\rm LC}$.
The coupling coefficients also can be expressed in terms of more common coupling rates $\eta = \kappa_c/(\kappa_0 + \kappa_c)$ and $\beta = 2\gamma/(\kappa_0 + \kappa_c)$, where $\kappa_c$ and $2\gamma$ are the pump and forward-backward wave coupling rates and $\kappa_0$ is the intrinsic microresonator loss rate ($\kappa_m=\kappa_c+\kappa_0$). 
It was shown in Ref. \cite{Kondratiev:Strong} that Eq. \eqref{master} can be obtained from Eq. \eqref{general_tuning_curve} assuming $R_o|\Gamma|\ll1$ and qualitatively valid in even broader region.

Both parts on the right-hand side of Eq. \eqref{masretphase} depend on the feedback round-trip time $\tau_s$. 
In the following, we consider that $\psi$ is ``independent'' on $\tau_s$ as the self-injection locking process is periodic in the locking phase, and thus its absolute value is trivial. 
The scales of $\kappa_m\tau_s$ and $\omega_m\tau_s$ (which is a part of $\psi$) are quite different for high-$Q$ microresonators, thus these values need to be treated separately. 
More formally, the parameter $\psi$ can also be independently tuned with locking mode frequency. 
Figure \ref{Pulling_delay} illustrate the influence of the phase $\psi$ and normalized delay $\kappa_m\tau_s/2$. 
Essentially, Eq. \eqref{master} is sinusoidal along the 1:1 line, which can be attributed to an external cavity \cite{Kondratiev:Strong} filtered with the resonant envelope. 
So the change in the phase moves the sinusoid along the 1:1 line, while the normalized delay changes its frequency. 
The latter also can bring undesired fringes inside the resonant envelope, making the system more multistable.

\begin{figure*}[t!]
\centering
\includegraphics[width=1\linewidth]{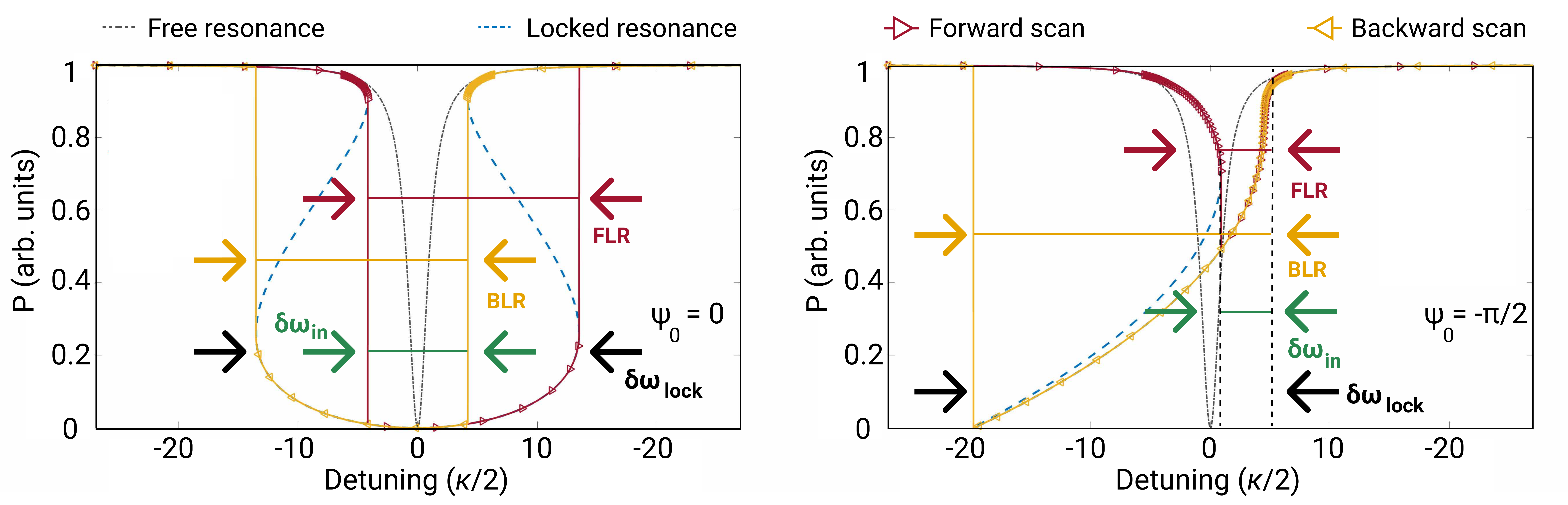}
\caption{
Calculated normalized transmitted power dependence on the detuning of the laser frequency from the WGM frequency for the different locking phases. 
The sum of the forward and backward locking ranges (FLR and BLR) can be measured experimentally. 
It is a good approximation of the total locking range $\delta\omega_{\rm lock}$ in the case of a high $Q$ factor when the inner overlap band $\delta\omega_{\rm in}$ is negligible. 
This figure is taken from Ref. \cite{Shitikov2020}.
}
\label{fig:Detuning}
\end{figure*}

The value $K_0$ is a universal constant for self-injection locking \cite{Kondratiev:17}. 
First, it defines the stabilization coefficient in the case of optimal phase and zero laser cavity detuning $\left.\frac{d \xi}{d \zeta}\right|_{\psi=0,\zeta=0} = K_0+1$. 
It leads to a realistic estimation of maximal line narrowing $\delta\omega\approx\delta\omega_{\rm free}/K_0^2$.
Second, it defines the locking range $\delta\omega_{\rm lock}/\kappa_m = 3\sqrt{3}K_0/16\approx0.32 K_0$ for small $\beta$. Note that if the locking range is greater than the double finesse (ratio of the microresonator intermode distance to its linewidth), it can overlap with the locking ranges of neighboring modes.
Third, the tuning curve's criterion for having a pronounced locking region is $K_0>4$. 
In this sense, the zero-point stabilization coefficient $K_0$ is analogous to the feedback parameter $C$ used in the theory of simple mirror feedback \cite{Petermann1995, Selfmixing}, where self-injection locking is achieved with the frequency-independent reflector forming an additional FP cavity. 
However, in the resonant feedback setup, the self-injection locking coefficient does not depend on the laser-to-reflector distance, but depends on the parameters of the reflector instead. 
Though the system has qualitatively similar regimes as the simple one \cite{Tkach1986}, their ranges and thresholds are different \cite{Laurent1989, Kondratiev:17, Galiev2020}. 
The value of $K_0>4$ is required for the pronounced locking with sharp transition, naturally becoming a locking criterion. 
Figure \ref{fig:tune}c shows the tuning curves for small $K_0$ values. 
It also includes several curves with different $\pi$ phase values. 
In the case of high-$Q$ microresonators $K_0$ can be no less than several hundred.

The illustrative tuning curves for high and low values of the mode-splitting coefficient $\beta$ are presented in Fig.~\ref{fig:tune}(a, b). 
Note that the tuning curve experiences splitting similar to the resonance splitting at the increased forward-backward wave coupling \cite{Gorodetsky:00}. 
The splitting impacts the self-injection locking process, and the stabilization can be worse at larger splitting values.

Experimentally, it is more convenient to control the transmission resonance of the system, when the light from the output of the coupler is directly registered with a photo detector and oscilloscope while tuning the frequency. 
The diagrams are also called ``light-current (LI) curve'' because the frequency is usually tuned by changing the injection current of laser diodes.
The theoretical LI curve for a WGM microresonator can be obtained from the transmission resonance curve, given by \cite{Gorodetsky:00}:
\begin{equation}
\label{Transmission}
B_t\approx B_{\rm in}\left[1-2\eta\frac{(1-i\zeta)}{(1-i\zeta)^2+\beta^2}\right].
\end{equation}
Note that if the detuning is controlled by the injection current $I$, the power $B_{\rm in}\propto I\propto -\zeta$ and the curve is slightly tilted.
Calculated dependencies of the transmitted power on the laser detuning from the WGM frequency are shown in Fig. \ref{fig:Detuning}. 
A free resonance is observed in the absence of back-scattering (black dotted line in Fig. \ref{fig:Detuning}), and a locked resonance appears when the back-scattering causes self-injection locking (blue dashed line in Fig.~\ref{fig:Detuning}). 
In the self-injection locking regime, one can determine the locking range: the bandwidth on the LI curve where self-injection locking suppresses the frequency change. 
Sharp edges bound the locking range at the locking phases $\psi\in[-\pi/2;\pi/2]$. 
For locking phases $\psi\neq n\pi$, the shape of the LI curves during forward and backward frequency scans are different, which can be observed experimentally. 
We need both scans to capture the full locking band correctly. 
The theoretical predictions are shown in Fig. \ref{fig:Detuning} with solid lines with triangle markers. 
The sum of the forward locking range (FLR) measured with forward scan, and backward locking range (BLR) measured with backward scan, is connected with the locking range $\delta\omega_{\rm lock}$ as ${\rm FLR}+{\rm BLR}=\delta\omega_{\rm lock}+\delta\omega_{\rm in}$ (see Fig. \ref{fig:Detuning}). 
The FLR-BLR overlap or the inner band $\delta\omega_{\rm in}$ is significant only for low intrinsic quality factor $Q_{\rm int}$ or for overcoupled microresonator \cite{Shitikov2020}.

\subsection{Multi-frequency laser locking}
\label{chap:MultifreqSec}

Self-injection locking is efficient in laser linewidth reduction. 
However, in the case of a diode laser self-injection-locked to a high-$Q$ WGM microresonator, a broad multi-frequency emission spectrum can collapse to a single line with a linewidth of kilohertz level or even below. 
Due to mode competition, the total initial power redistributes in favour of the locked mode (or, in some cases, in favour of several locked modes), providing a single-frequency (or few-frequency) regime with energy concentration reaching 96\% \cite{Galiev:18}. 
For example, a multi-frequency laser diode with 1535 nm central wavelength, 100 mW output power, a spectrum initially consisting of 50 lines and linewidth of the order of megahertz is locked to a high-$Q$ magnesium fluoride (MgF$_2$) microresonator.
One can obtain a single-frequency laser with a power of 50 mW and a linewidth of less than 1 kHz. 
This also brings out the so-called Bogatov effect \cite{Bogatov,BogatovIEEE} -- non-uniform energy distribution inside the suppressed modes.

The amplitude relations of the modes should be considered to model self-injection locking of a multi-frequency laser. 
The standard multi-mode laser model can be represented as a system of differential rate equations \cite {Yamada1989,Lang_Kobayashi}:
\begin{align} 
\dot N= & \frac{I}{e} - \frac{N}{\tau_s} -\sum_{l}G_l^{(1)} S_l,\label{N_free}\\
\dot S_l = &\left(G_l(N) - G_{\rm th}\right) S_l + NF_l, \label{S_free}	
\end{align}%
where $I$ is the diode current, $e$ is the electron charge, $N$ is the number of excited electrons, $ \tau_s$ is the lifetime of the excited electron, $S_l$ is the number of photons, $G_l$ is the stimulated emission coefficient in the laser mode $l$ and $ G_{\rm th} $ is the threshold gain (or in other words the total losses), and $F_l$ is the spontaneous emission contour. 
Note that Eq. \eqref{N_free} and \eqref{S_free} are real. 
They can be considered as some form of the amplitude part of Eq. \eqref{LaserC} that was discarded while deriving the tuning curve.
The magnitude of the threshold gain is determined by the design features of a particular laser, and in the simplest case for a laser cavity consisting of two mirrors with reflection coefficients $R_0$ and $R_e$, one can obtain the following expression:
\begin{equation}
\label{thr_gain}
G_{\rm th} =   \frac{1}{\tau_{\rm LC}}L_{\rm LC}\alpha_{\rm loss} + \frac{1}{\tau_{\rm LC}}  \ln \frac{1}{R_o R_e} ,
\end{equation}
where $ L_{\rm LC} $ is the length of the diode, $ \tau_{\rm LC} $ is the diode round-trip time and $\alpha_{\rm loss}$ is material loss factor.
The gain in each mode depends on a combination of such effects as stimulated emission of photons, spectral hole burning due to neighboring modes, and asymmetric mode interaction. 
For the gain factor $G_l$, the following expression can be written as \cite {intraband_yamada_1981,Yamada1989}:
\begin{equation}
G_l(N) = G^{(1)}_l(N) - G^{(3)}_l S_l  - \sum_{k \neq l} (G^{(3)}_{l(k)} + G^{\rm Bogatov}_{l(k)}) S_k,
\end{equation}
where $G^{(3)}_l$ is the self-saturation coefficient (spectral hole burning), $G^{(3)}_{l(k)}$ is the coefficient of symmetric cross-saturation (spectral hole burning due to neighboring modes), $G^{\rm Bogatov}_{l(k)}$ is the asymmetric mode interaction coefficient (Bogatov effect) \cite{bogatov_effect_HIROSHI}.

The coefficient $G_l^{(1)}$ is determined by both the number of excited electrons and the dispersion of the linear gain:
\begin{equation}
\label{linearGain}
G_l^{(1)} = g_l (N - N_g - D_g (\lambda_l - \lambda_{\rm peak})^2),
\end{equation}
where $ g_l $ is the differential gain, $ N_g $ is the number of excited electrons when the laser diode becomes optically transparent, $D_g$ is the linear gain dispersion coefficient, $ \lambda_l $ is the wavelength of the mode $l$, and $ \lambda _ {\rm peak} $ is the central wavelength of the laser.

The effect of asymmetric mode interaction was first described by Bogatov in Ref. \cite {Bogatov, BogatovIEEE}, where a model of stimulated scattering of laser light on the dynamic electron density inhomogeneities was introduced as the theoretical explanation of this effect. 
The model proposed by Bogatov describes the change in the permittivity $ \delta {\epsilon} $, caused by the dynamic inhomogeneity of the electron density due to the stimulated emission of the excited electrons under the influence of mode interference.
The expression obtained by Bogatov for the variation of the dielectric constant can be rewritten in terms of the gain of a laser active region. 
Thus the expression to describe the coefficient of asymmetric gain (Bogatov coefficient) is:
\begin{equation}
\label{BogatovGain}
 G^{\rm Bogatov}_{l(k)} = \frac{3}{4} g_l^2 ( N - N_g)
\frac{\frac{1}{\tau_s} + \frac{3}{2} g_l S + 
\alpha_g \Omega_{l(k)} }
{(\frac{1}{\tau_s} + \frac{3}{2}g_l S)^2 + 
\Omega_{l(k)}^2 },
\end{equation}
where $ \Omega_ {l (k)} = \omega_l - \omega_k $ are laser modes offsets, $ S = \sum S_l $ is the total number of photons, and $ \alpha_g $ is the linewidth enhancement factor.

In the self-injection-locked regime, the feedback term for electric field amplitude $E_l$ is introduced. 
Taking into account that the photon number $ \dot S_l \propto 2 \dot E_l E_l $, the expression for the $ \delta S _ {\rm feedback} $ contribution to the dynamics of the mode intensity in Eq. \eqref {S_free} can be obtained as following:
\begin{widetext}
\begin{equation}
\label{Sfeedback}
\delta S_{\rm feedback} = 2\kappa_{dl} \sqrt{S_l (t - \tau) S_l (t) }\cos(\psi_l + \phi_l(t) - \phi_l(t - \tau)),
\end{equation}
\end{widetext}
where $\kappa_{dl} \approx \frac{1 - R_o^2}{R_o\tau_{\rm LC}}\Gamma(\omega_l)$ is the total feedback rate, 
$ \phi_l (t) $ is the phase of the mode, 
$ \psi_l = \omega_l \tau + \arg (\Gamma (\omega)) $ with $ \tau $ being the round-trip time from laser to the reflector and back.

\subsubsection{Spectrum Collapse}

For the case when the high-$Q$ microresonator acts as an external mirror, it is sufficient to replace the reflection coefficient of the mirror $\Gamma(\omega_l)$ in the expression for the total feedback rate in Eq. \eqref{Sfeedback} with an expression for the frequency-selective reflection coefficient of the WGM microresonator \cite{Gorodetsky:00}.
Each laser mode is assumed to interact efficiently with only one mode of the microresonator. 
This assumption is evidently justified for the case when the free spectral range (FSR) of the laser is larger than the resonator mode spacing.
By tuning the laser frequency, one can achieve the regime when a certain mode of the laser $ \omega_ {l = p} $ becomes close enough to some mode of the optical microresonator $\omega_m$. 
In this case, the feedback to this laser mode from the microresonator increases dramatically, and the laser mode locking to the high-$Q$ mode of the microresonator occurs.
In stationary regime we can assume that $ \Gamma (\widetilde {\omega} _ {l \neq p}) \ll 1 $, so the feedback expression is simplified $\delta S_{\rm feedback} = \delta_{lp}2 \kappa_{dl} S_l\cos(\psi_l)$,
where $ \delta_ {lp} $ is the Kronecker symbol, meaning that the feedback is added only to the mode closest to the frequency of the WGM.

\begin{figure*}[t!]
\centering
\includegraphics[width=1.0\linewidth]{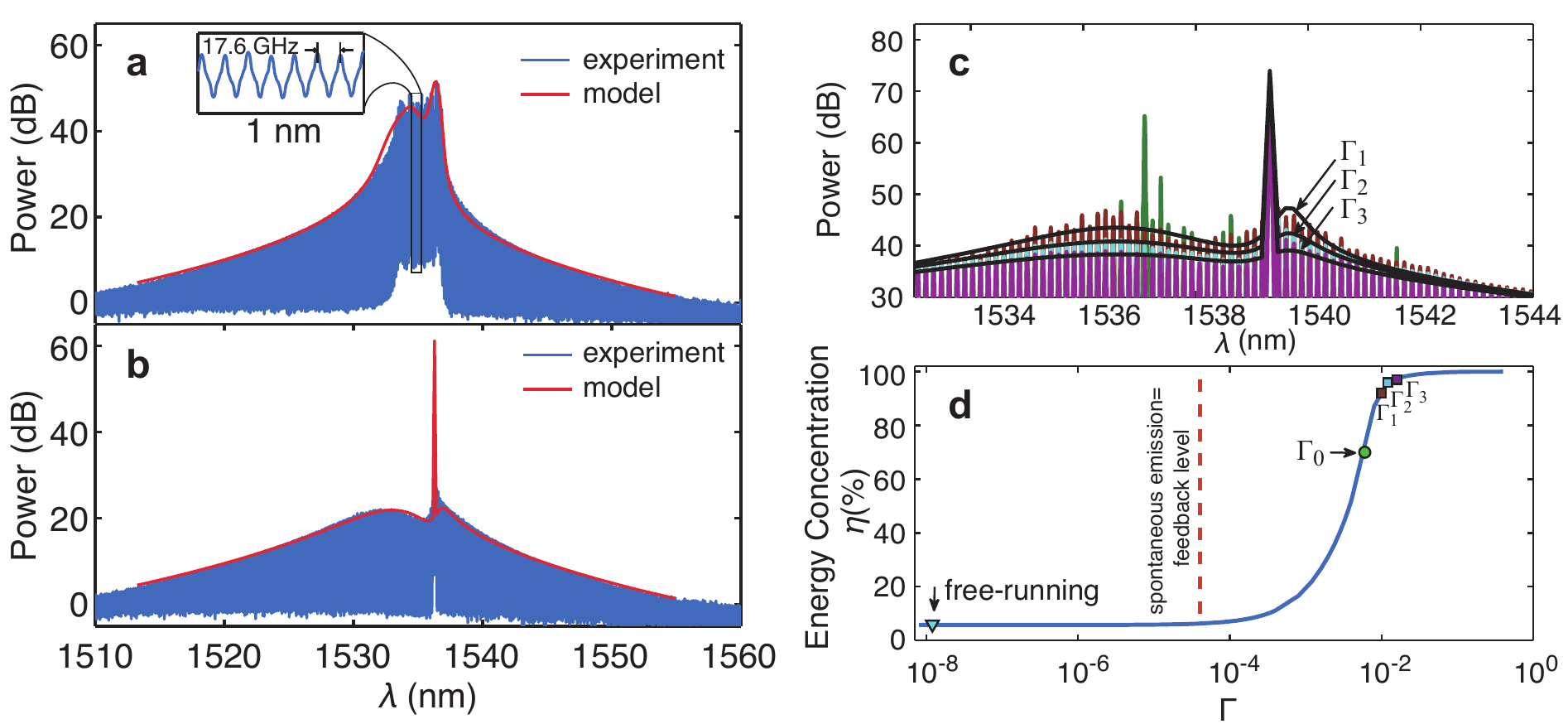}
\caption{
Experimental and numerical study of laser spectra with and without self-injection locking.
\textbf{a}. Experimental (blue line) and numerically calculated (red line) emission spectrum of the free-running multi-frequency diode laser [Eq. \eqref{N_free}, Eq. \eqref{S_free}].
\textbf{b}. Experimental (blue line) and numerically calculated (red line) emission spectrum of the self-injection-locked multi-frequency diode laser.
\textbf{c}. Experimentally obtained spectra of the self-injection locked laser at different feedback levels (coloured solid lines) with numerically calculated envelopes (black lines) for $\Gamma_1 = 1\times 10^{-2}$, $\Gamma_2= 1.2 \times 10^{-2}$, $\Gamma_3 =1.5\times 10^{-2}$, respectively. 
The green spectrum is not approximated well.
\textbf{d}. Numerically calculated dependence of the single-mode energy concentration $\eta$ on the feedback level (blue line), and the experimentally obtained energy concentration points (squares). 
The circle corresponds to the green spectrum in Panel \textbf{c}, and the triangle corresponds to the free-running laser. 
This figure is taken from Ref. \cite{Galiev:18}.
}
\label{fig:spectrum}
\end{figure*}

The emission spectrum envelope in the model of a free-running multi-frequency laser [Eq. \eqref{N_free}-Eq. \eqref{S_free}] is mainly defined by the linear gain dispersion Eq. \eqref{linearGain} and spontaneous emission $F_l$. 
In the model of a laser with optical feedback, the frequency-selective feedback coefficient introduced in Eq. \eqref{Sfeedback}, in addition to the dispersion of the linear gain, also plays an important role. 
If one laser mode $p$ is self-injection-locked to the microresonator mode, the feedback coefficient of this mode can compensate for the dispersion term of the linear gain. 
The total gain of this mode then exceeds the gain of the central mode (zero-dispersion mode) with wavelength $\lambda_{\rm peak}$. 
This enhances the power of the mode $p$ to that comparable to the central mode. 
Further feedback enhancement can lead to the strong feedback -- ``complete'' suppression of other modes $S_l \ll S_p$.
In this case, the mode $p$ uses all the excited electrons produced by the injection current, which simplifies the electron/photon dynamics [Eq. \eqref {N_free}], allowing the summation to be omitted. 
Consequently, this process effectively transfers the energy of the other laser modes into the locked mode. 
It is the strong feedback when a multi-frequency laser becomes a single-frequency one effectively.
Figure~\ref{fig:spectrum}(a,b) show the comparison of the analytical solution of the model with [panel b] and without [panel a] the feedback term together with the experimental measurement \cite{Galiev:18}. 
The non-locked spectrum (Fig.~\ref{fig:spectrum}a) became matching automatically after the parameters of modelling were slightly adjusted for the Bogatov spectrum (see Fig.~\ref{fig:spectrum}b) to match.

To obtain a strong feedback condition, we derive $N$ from $G_p^{(1)}$ from the stationary form of Eq. \eqref{S_free} for the locked mode and substitute into $G_l^{(1)}$ in stationary form of Eq. \eqref{S_free} for other modes. 
In this way, the relation between $S_p$ and $S_l$ is found.
For the consistency of our initial assumption and solution, it is necessary that the condition $S_l \ll S_p$ should follow from this solution. 
In this way, we get the following criterion for the strong feedback:
\begin{equation}
2 \kappa_{dp} S_p\gg N F_p.
\label{threshold_feedback}
\end{equation}
The physical meaning of this statement is that for efficient spectrum conversion, the strong feedback should be greater than the spontaneous emission rate.

A series of measurements of the self-injection-locked multi-frequency laser emission spectrum at different feedback levels have been carried out. 
Figure \ref{fig:spectrum}c shows several experimentally obtained states of the self-injection locking regime when the optical feedback level was controlled via the gap between the microresonator and the prism. 
Performing numerical modelling with different feedback levels, the theoretical curves shown in Fig.~\ref{fig:spectrum}c was obtained. 
A good correspondence between theory and experiment is observed.
In the strong coupling regime (see the purple curve in Fig. \ref{fig:spectrum}c) a single narrow line and the maximum suppression of other spectrum lines are observed. 
When the gap increases, the suppression decreases. 
The green, red and blue lines in Fig.~\ref{fig:spectrum}c show that the intensity of the suppressed modes begins to grow at lower coupling efficiency. 
At a certain threshold level of the feedback, other lines start to appear in the optical spectrum (see the green curve in Fig.~\ref{fig:spectrum}c), above which the locking is destroyed. 
The feedback is weak near the threshold level. 
In other words, the power of the backward wave is not enough for stable self-injection locking.

\begin{figure*}[t!]
\centering
\includegraphics[width=1\linewidth]{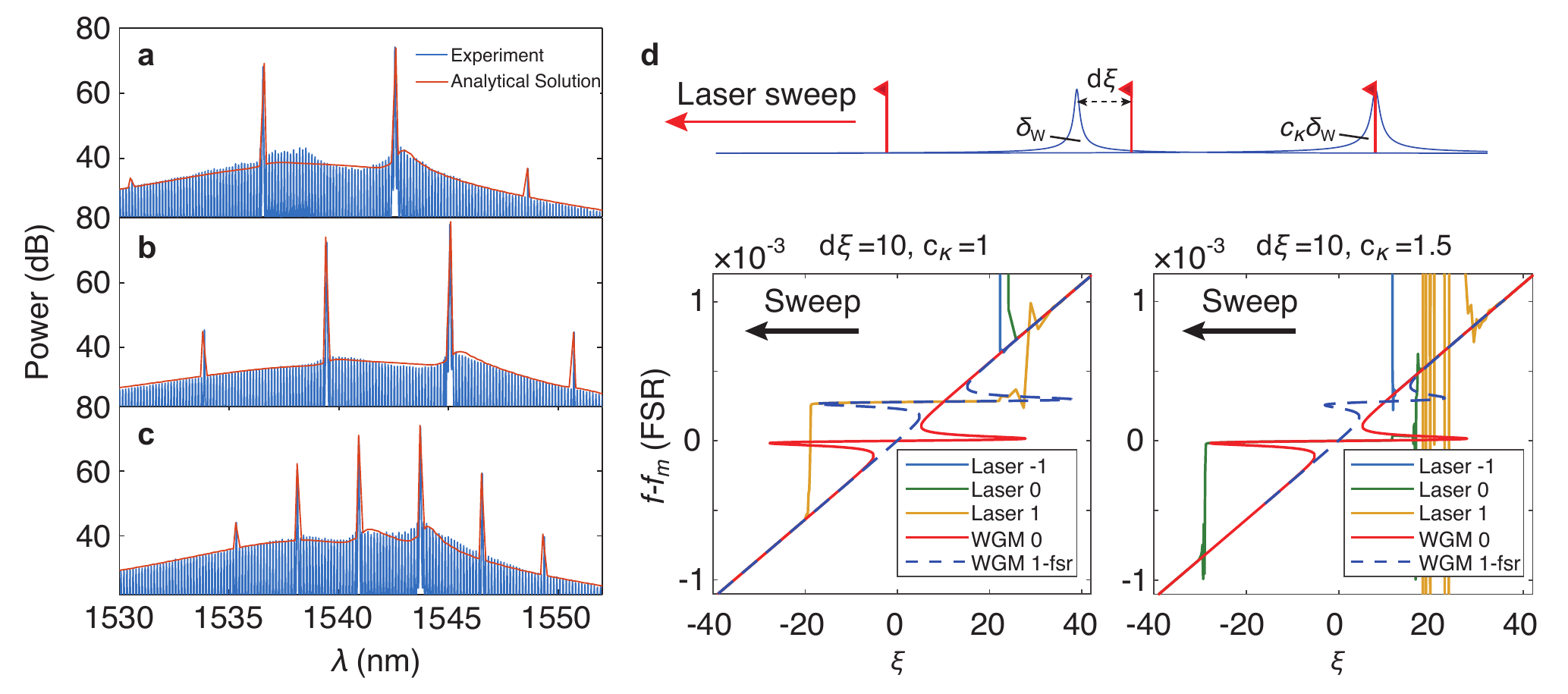}
\caption{
Optical spectra of the multi-frequency emission of the self-injection-locked laser. 
\textbf{a}. two-frequency regime; 
\textbf{b}. four-frequency regime; 
\textbf{c}. six-frequency regime. 
Additional feedback is added to corresponding modes. 
Blue data is the experimental data, orange curve is from the analytic model.
\textbf{d}. Upper panel: scheme of the laser(red) and WGM (blue) mode frequencies in model. 
$\mathrm{d}\xi$ -- difference between laser and microresonator intermode distances, 
$c_\kappa$ -- first to second microresonator mode width ratio.
Lower panel: theoretical (red and blue-dashed) and modelled (blue, green, yellow) tuning curves for the same WGM mode widths (left) and different mode widths (right). 
Panels \textbf{a -- c} are taken from Ref. \cite{Galiev:18}, and Panel \textbf{d} from Ref. \cite{Kondratiev:21fio}.
}
\label{fig:spectrum2}
\end{figure*}

To estimate the efficiency of the spectrum collapse parameter  $\eta = S_p/\sum S_l$ describing the concentration of the energy in the locked mode was introduced and calculated using the developed model. 
The curve in Fig.~\ref{fig:spectrum}d shows the numerical estimation of $\eta$ for different feedback level $\Gamma$ together with the experimental dots. 
Note that after entering the strong feedback regime (according to Eq. \eqref{threshold_feedback}) the concentration quickly grows and reaches the value of about 96\%, which corresponds to the transition to a single-frequency generation. 
The energy concentration near the threshold level can only be considered an estimated value. 
All numerical results obtained from the developed model are in good agreement with the experimental data.  

The measured power feedback level $|\Gamma(\omega_p)|^2$ (about $\sim 10^{- 4}$) was sufficient for single-frequency lasing. 
Similar measurements when the laser diode was stabilized by other WGM modes give estimates of the optical feedback level of about $10^{-4}$ to $10^{-3}$. 
The same feedback level has been demonstrated with a DFB laser locked to high-$Q$ microresonator \cite{Vasiliev:96}. 
The obtained results suggest that a higher level of single-mode energy concentration and line narrowing can be achieved by developing a technique for increasing the feedback. 
It should be noted that an arbitrary feedback increase will decrease the output power of the stabilized laser. 

\subsubsection{Multi-frequency locking}
\label{chap:Multifreq}

In addition to the collapse of the diode laser's multi-frequency spectrum into one narrow line, simultaneous self-injection locking of a few laser modes to different microresonator modes can be achieved \cite{Galiev:18}. 
This effect results in effective discrimination of these locked modes and transformation of the initial spectrum into a spectrum with just a few locked narrow lines. 
The locking occurs on modes spaced by an integer number of the microresonator FSR interval $\Delta f_{\rm WGR}$ and laser FSR interval $\Delta f_{\rm d}$, i.e. $\Delta f_{\rm mult}=M\Delta f_{\rm WGR}=N\Delta f_{\rm d}$ (which sometimes is called Vernier effect). 
In this case, the mode competition near each diode frequency acts the same way as the case of single-frequency self-injection locking: the spectrum in the vicinity of the resonant frequency is suppressed, and energy is redistributed in favour of the locked spectral line. 
The spectral width of the locked mode also decreases significantly. 
This situation is depicted in Fig.~\ref{fig:spectrum2}, which shows a two-frequency regime (panel a), a four-frequency regime (panel b) and a six-frequency regime (panel c). 
Note that if different mode families in the microresonator have slightly different FSRs \cite{Demchenko:13,Lucas_18}, different spacings between locked modes can be observed [Fig.~\ref{fig:spectrum2} (b--c)]. 
Good correspondence with the experiment is achieved if the feedback is added for the corresponding modes in the numerical model Eq. \eqref{Sfeedback}.

\begin{figure*}[t!]
\centering
\includegraphics[width=0.95\linewidth]{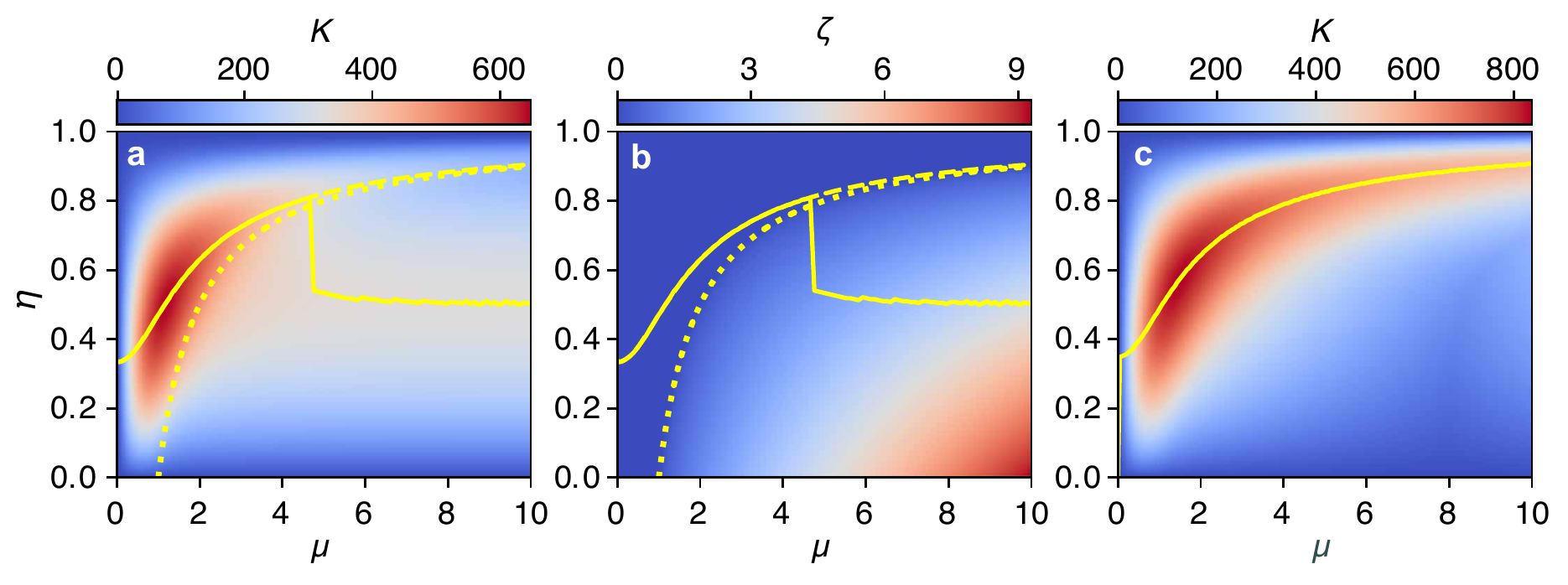}
\caption{
Numerical study of parameters for optimal self-injection locking.
The stabilization coefficient $K$ (panel \textbf{a}) and the optimal detuning $\zeta$ (panel \textbf{b}) for $\psi = 0$, $\kappa_0\tau_s=0$, $\kappa_{\rm do}/\kappa_0=1000$. 
Panel \textbf{c} shows the stabilization coefficient for $\kappa_0\tau_s=0.4$ and the same other parameters. 
The solid line is the numerical maximum of $K$ with respect to $\eta$. 
The dashed line is the second optimum branch for the zero-phase case.
The dotted line corresponds to $\beta=1$. 
All quantities are plotted in dimensionless units. 
This figure is taken from Ref. \cite{Galiev2020}.
}
\label{fig:inphase}
\end{figure*}

In Ref. \cite{Kondratiev:21fio} numerical modelling for three laser modes and two microresonator modes was performed. 
The three laser modes' frequencies were sweeping through the microresonator modes with different mode spacings and widths (see Fig.~\ref{fig:spectrum2}d upper panel). 
The results showed that the self-injection locking occurs to the mode that is the first to match with the laser mode (see Fig.~\ref{fig:spectrum2}d, lower left -- orange curve locks) or to the narrowest mode (Fig.~\ref{fig:spectrum2}d, lower right -- green curve locks). 
Then all laser modes begin to oscillate on the locking WGM frequency stably without switching. 
At the same time, if several WGM microresonators have similar parameters (linewidth and detuning from the closest laser mode), self-injection locking can happen to both WGM modes, and the laser modes oscillate on both frequencies simultaneously. 
It was demonstrated that for the multi-mode locking, the allowed linewidth discrepancy is $\pm0.005\kappa$, and the allowed difference between laser-WGM modes distances is $\mathrm{d}\xi<1$. 
This somehow explains why the multi-locking occurs only within a single transverse mode family: this is the only way for the modes to have close enough losses and regular spacing.

\subsection{Optimization of the self-injection locking regime} 

\begin{figure*}[t!]
\centering
\includegraphics[width=0.95\linewidth]{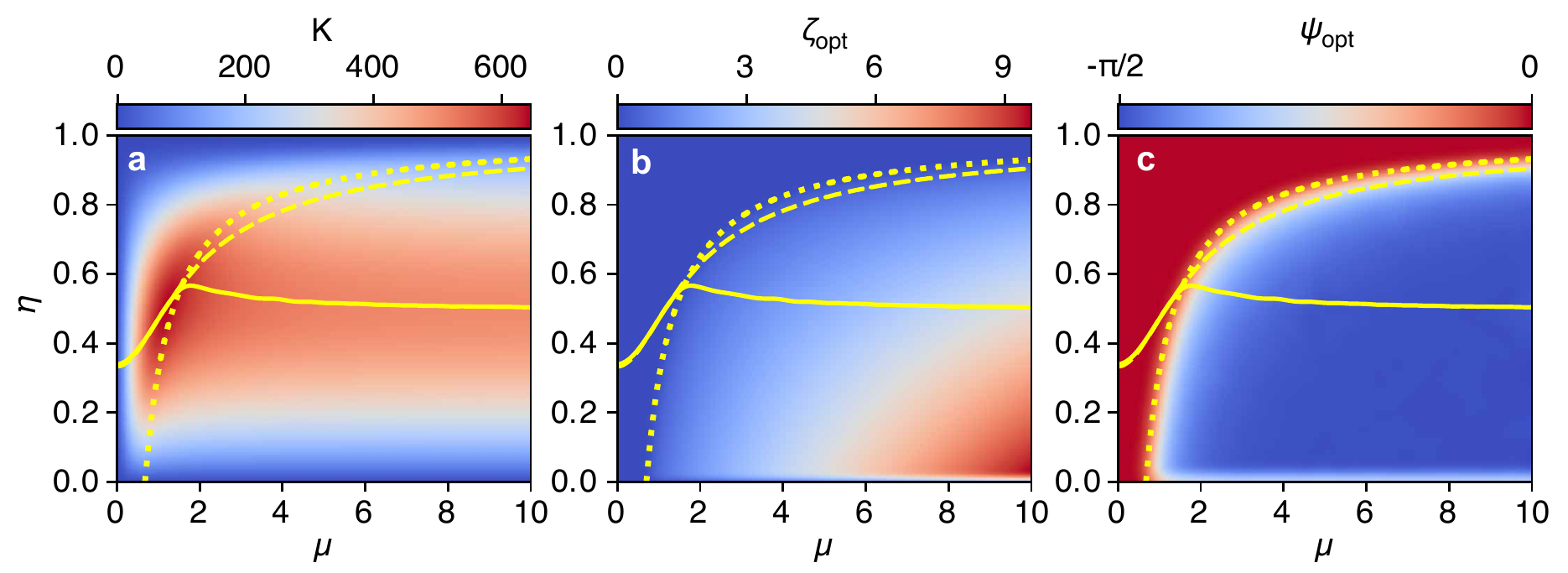}
\caption{
Numerical study of parameters for optimal self-injection locking.
The stabilization coefficient $K$ (panel \textbf{a}), the optimal detuning $\zeta_{\rm opt}$ (panel \textbf{b}) and optimal $\psi_{\rm opt}$ (panel \textbf{c}) for $\kappa_0\tau_s=0$, $\kappa_{\rm do}/\kappa_0=1000$. 
The solid line is the numerical maximum with respect to $\eta$.
The dashed line is the second optimum branch for the zero-phase case.
The dotted line corresponds to $\beta=\beta_{\rm cr} \approx0.68$. 
All quantities are plotted in dimensionless units. 
This figure is taken from Ref. \cite{Galiev2020}.
}
\label{fig:full}
\end{figure*}

One can see that the five main parameters define the laser performance of in the self-injection locking regime, which are: 
1. The coupling strength of the forward and backward waves in the cavity $\beta$; 
2. The locking phase $\psi$ determined by the optical path between the laser and the microresonator, and the frequency of the microresonator locking mode; 
3. The optical round-trip time $\tau_s$ between the laser and the microresonator; 
4. The laser cavity-microresonator frequency detuning $\xi$; 
and 5. The pump coupling efficiency $\eta$. 
In what follows, we consider effective detuning $\zeta$ instead of the normalized frequency difference between the laser cavity mode and the WGM, $\xi$, since $\xi \ll 1$ in the case of tight locking.

The laser linewidth is reduced proportionally to the square of the stabilization coefficient \cite{Laurent1989, Spano1984} determined by the slope of the tuning curve $K(\eta,\beta,\zeta,\psi)  = \partial \xi/\partial \zeta$. 
The free-running and locked laser linewidths are related as 
\begin{equation}
\delta\omega_{\rm locked}=\frac{\delta\omega_{\rm free}}{K^2}\approx \delta\omega_{\rm free}\frac{Q_d^2}{Q_m^2}\frac{1}{64\eta^2\beta^2(1+\alpha_g^2)}.
\end{equation}
The last simple formula for the linewidth reduction was obtained in Ref. \cite{Kondratiev:17} under small back-scattering $\beta\ll1$, zero locking phase $\psi=0$, resonant tuning $\zeta=0$ and critical coupling $\eta=0.5$. 
This formula has been tested in several works \cite{photonics5040043,Savchenkov:2022}. 
In Ref. \cite{Galiev2020} a five-parameter ($\psi, \ \zeta,\ \eta,\ \beta, \ \kappa_0 \tau_s$) optimization study of the stabilization coefficient was performed. 
In contrast with the common knowledge, the increase of the back-scattering, described by the parameter $\beta$, does not monotonously enhance the stabilization coefficient but leads to its eventual saturation (see Fig.~\ref{fig:inphase}). 
The optimal selection of the system parameters reduces the laser linewidth by several orders of magnitude. 
Here we consider parameters $\psi = 0$ and $\kappa_0\tau_s \ll 1$ as the most common and illustrative. 
In this case, the resonance curve of the locking mode (for a laser diode, it coincides with the light-current curve) can be observed while the laser frequency is broadly scanned in and out of the locking range and has a nearly rectangular shape. 

To optimize the laser performance, we look for the effective detuning $\zeta_0$ that maximizes the stabilization coefficient $K$. 
The expression $\left.\partial K(\zeta,\psi=0,\eta,\beta)/\partial\zeta\right|_{\zeta_0}=0$ results in a $\zeta$-multiplied bi-cubic characteristic equation, dependent on $\beta$ only. 
Solving it, we obtain for different $\beta$ values
\begin{equation}
\label{zeta0}
\zeta_0=
\begin{cases}
0, &\beta\leq1\\
\sqrt{\frac{3}{5}(\beta^2-1)}, &\beta\in(1;1.48)\\
\beta - \frac{1}{\sqrt{3}}+... &\beta>1.48
\end{cases}
\end{equation}
This expression has a simple physical meaning. 
The linear interaction of the counter-propagating waves leads to the resonance splitting \cite{Gorodetsky:00}. 
The splitting value is approximately equal to $\beta\kappa_m$ (see Fig.~\ref{fig:scheme}a) \cite{Galiev:21}. 
The locking band also splits into two for the large $\beta$ [see Fig.~\ref{fig:tune}b, the crosses mark the points $\zeta=\pm(\beta-1/\sqrt{3})$, and the tips of the peaks are close to $\zeta=\pm\beta$].  

The slope has a maximum inside its validity range if $\beta\leq1$. 
For the case $ \kappa_0\tau_s \ll 1$ the maximum of the stabilization coefficient is given by \cite{Galiev2020}
\begin{eqnarray}
\label{global_expansion}
\beta_{\rm max}\approx \frac{1}{\sqrt{3}}  +\frac{2\sqrt{3}}{27}\kappa_0\tau_s \nonumber, \\ 
\eta_{\rm max}\approx 1/2 +\frac{1}{6}\kappa_0\tau_s, \\ 
K_{\rm max}\approx \frac{3\sqrt{3}}{8}\frac{\tilde\kappa_{do}}{\kappa_0}+1 +\frac{\tilde\kappa_{do}}{\kappa_0}\frac{\sqrt{3}}{4}\kappa_0\tau_s,\nonumber 
\end{eqnarray}
where $\tilde\kappa_{do} = \kappa_{do}\sqrt{\Theta}\sqrt{1+\alpha_g^2}$.
This expression shows that the optimal stabilization coefficient increases with the laser-microresonator distance. 
However, one should not increase the distance uncontrollably as the increase is responsible for decreasing the laser signal quality and producing redundant metastable fringes on the tuning curve \cite{Kondratiev:17}. 
The criterion of the stable operation was approximated as $\kappa_m\tau_s<9.4(8\eta(1-\eta)\beta\tilde\kappa_{do}/\kappa_0)^{-0.36}$.

It is helpful to introduce another parameter $\mu = 2\gamma/\kappa_0$ (so $\beta = \mu (1 - \eta)$) that is a constant for a given resonator. 
Using this notation, it is possible to perform a complete parametric optimization of the stabilization coefficient. 
Figure \ref{fig:inphase}a shows the results of the numerical optimization for zero phases $\psi=0$ and optimal frequency detuning Eq. \eqref{zeta0}. 
It can be seen that the critical coupling is optimal for the short laser-microresonator distance  ($\kappa_0\tau_s<0.1$) and for the large back-scattering $\beta\geq1$. 

The dependence of the optimal pump coupling coefficient $\eta$ on the normalized forward-backward coupling rate $\mu$ is shown by the solid line. 
While increasing $\mu$, we should increase the load to keep $\beta<1$, preventing the resonance splitting ($\beta=1$ is shown in Fig. \ref{fig:inphase} with the dotted line). 
It is also clearly seen in the map of the optimal detuning (see Fig. \ref{fig:inphase}b). 
At some point (at $\mu \approx 5$ for the considered parameters), the detuning increase is no longer advantageous, and the critical coupling becomes optimal. 

The stabilization coefficient can be optimized for the locking phase $\psi$ \cite{Galiev2020}.
For small $\beta$ ($\beta\leq \beta_{\rm cr}\approx0.68$), and, thus, small $\mu$, we get the same $\zeta_{\rm opt}=0$ (also see Fig.~\ref{fig:full}b) as for the zero-phase case. 
The critical value $\beta_{\rm cr}$ increases with the round-trip time and pump coupling coefficient ($\kappa_m\tau_s$) but always stays less than unity \cite{Galiev2020}. 
It was shown also in that $\zeta_{\rm opt}=0$ provides that $\psi_{\rm opt}=0$ (see  Fig.~\ref{fig:full}c). 
The map of the stabilization coefficient under optimal detuning and locking phase conditions is shown in Fig.~\ref{fig:full}a for different combinations of $\eta$ and $\mu$. 
The maps of optimal detuning and optimal phase are shown in Figs.~\ref{fig:full}(b, c). 
Zero phase ($\psi=0$) is an exact optimum for $\beta<0.68$ (see Fig. \ref{fig:full}c), which is connected to the optimal condition $\zeta=0$ (see Fig. \ref{fig:full}b). 
Since the optimal value of $\beta$ found earlier for the zero-phase case $\beta_{\rm max}=3^{-1/2}$ is less than $0.68$, the maximum stabilization coefficient value for the zero locking phase is a global maximum.

\begin{figure}[t!]
\centering
\includegraphics[width=\linewidth]{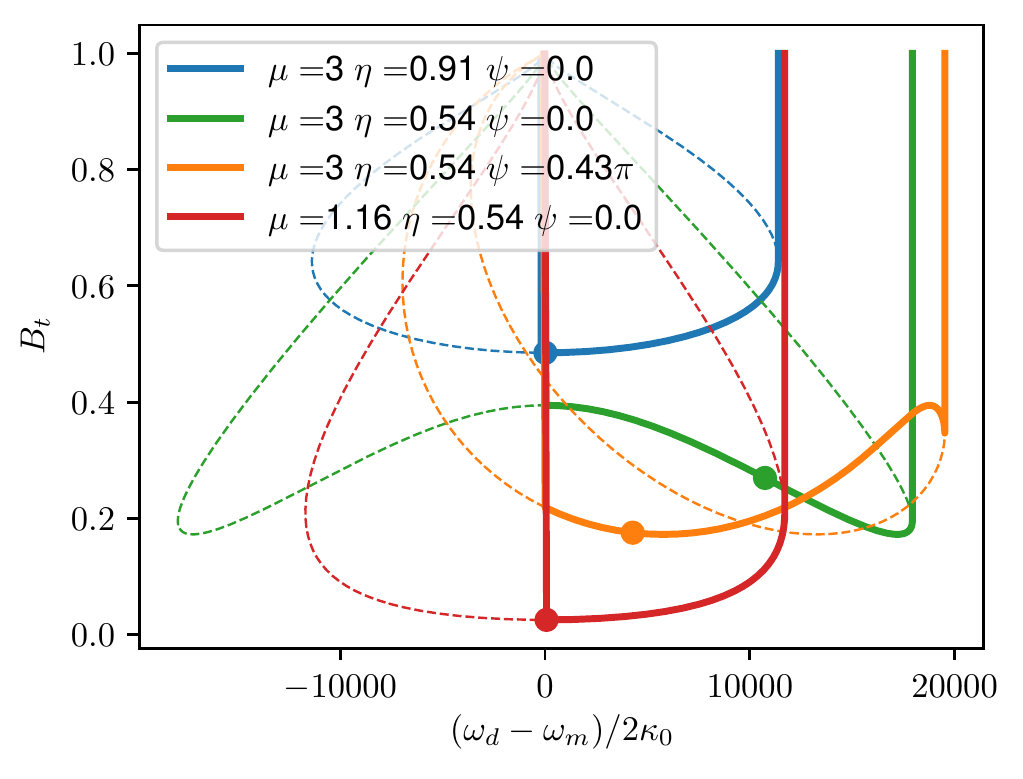}
\caption{
Numerically obtained the transmission resonance curves Eq. \eqref{Transmission} for the parameters taken from Ref. \cite{Liang2015}.
The parameter $\zeta(\xi)$ was evaluated using Eq. \eqref{master} (the dashed lines), and the corresponding LI curves were evaluated for the increasing frequency (the solid lines). 
The points on the transmission curves mark the optimal detuning values. 
All quantities are plotted in dimensionless units. 
This figure is taken from Ref. \cite{Galiev2020}.
}
\label{exampleLI}
\end{figure}

According to our model, optimizing the self-injection locking can result in a significant reduction of the laser linewidth compared with the best experimental results. 
For example, a diode laser linewidth reduction from 2 MHz in the free-running regime to below 100 Hz in the locked regime was demonstrated in Ref. \cite{Liang2015}. 
The linewidth reduction in the case of the optimal parameters $\eta_{\rm opt}(\mu), \ \psi_{\rm opt}(\mu)$ and $\zeta_{\rm opt}(\mu)$ for $\mu = 3$ can be improved by $15$ times, which is at least an order of magnitude better than the result obtained for the non-optimal coupling. 
Furthermore, if the mode of the resonator is optimally selected ($\mu = 1.16$), the linewidth reduction can be improved by $94$ times. 

It was shown the non-monotonic saturation of the stabilization coefficient $K$ with respect to the back-scattering (see Fig. \ref{fig:full}a). 
The maximum value of the stabilization coefficient is reached at {$\beta\in[3^{-1/2};1]$}, which determines the optimal ``semi-split'' mode. 
The saturation happens due to the doublet back-scattering resonance's formation because of the microresonator's counterpropagating modes. 
Microresonator modes with high back-scattering rates require the laser frequency to be tuned to the inner slope of the doublet back-scattering resonance to achieve the highest stabilization coefficient (see Fig.~\ref{fig:full}b).

In the experiment, one can choose and control the corresponding parameters by analysing light-current (LI) or transmission-resonance curves (see Fig. \ref{exampleLI}).
If the detuning and phase in the experiment are close to the optimal $\zeta = 0$ and $\psi = 0$, it leads to the typical close-to-rectangular LI curve (see the blue curve in Fig. \ref{exampleLI}). 
The optimal detuning is marked with a point. 
For a sufficient $Q$ factor of the microresonator, the locking region entrance point is close to zero detuning, which also helps to tune to the optimum. 
Decreasing the pump coupling to the optimal value, we set it close to the critical coupling. 
Thus transmittance will be reduced. 
If $\beta$ is greater than the critical value $\beta_{\rm cr}\approx0.68$, the optimal phase becomes nonzero. 
The correct phase also can be controlled using the transmittance resonance form ([see the difference between the green and orange curves in Fig. \ref{exampleLI}). 
If the laser-microresonator distance is unchangeable in a particular setup, the phase can be tuned by switching the operational mode. 
This step, however, can also modify the scattering coefficient $\mu$. 
The LI curve analysis also suggests the optimal mode, which should have a particular shape (see Fig.~\ref{exampleLI}, red curve).

In general, the following optimization recommendations based on the developed theoretical model can be made:
\begin{enumerate}
    \item If we can estimate $\mu$, we select a mode with the optimal $\mu$.
    \item We set up the critical coupling regime, which is indicated by the nearly maximal depth of the dip, at which the LI-curve width is also maximal.
    \item We adjust the phase so that the LI-curve acquires the correct shape -- the first angle (counting in the frequency scanning direction) should be sharp, and the second -- with rounding in the scanning direction.
    \item If we do not know the $\mu$ and have not yet selected a mode and/or cannot change the laser position, then we can look for the resonance with the correct shape (see above).
\end{enumerate}
Any particular experimental realization of the laser requires an adjustment of the optimization algorithm in accordance with the theoretical model described above.

\begin{figure*}[t!]
\centering
\includegraphics[width=0.80\linewidth]{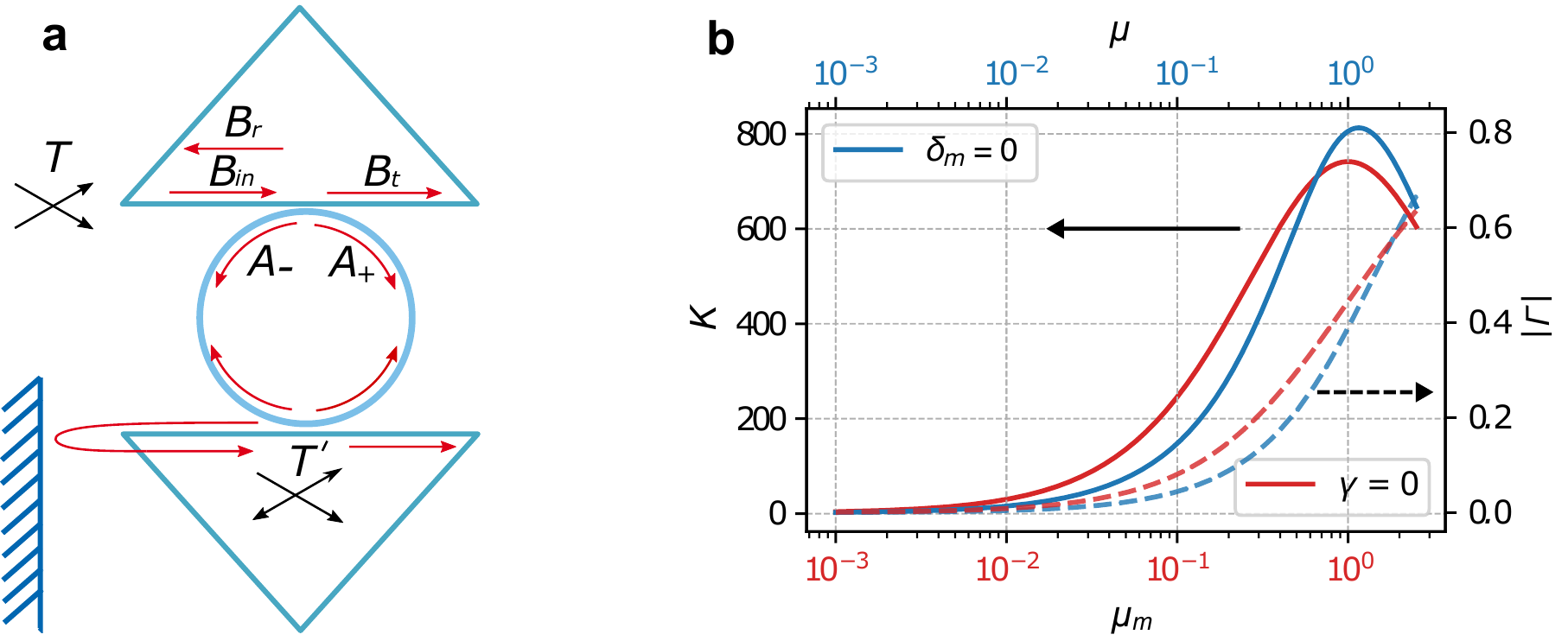}
\caption{
Schematic of laser self-injection locking with variable optical feedback.
\textbf{a}. The self-injection locking scheme with additional mirror and coupling prism. 
The amplitudes at the input port coupling point: 
$B_{\rm in}$ is the pump amplitude; 
$B_{t}$ and $B_{r}$ are amplitudes transmitted through and reflected from the prism coupler.
$A_{+}$ and $A_{-}$ are the amplitudes of counter-circulating modes of the microresonator. 
\textbf{b}. Solid lines -- left y-axis: Comparison of the stabilization coefficient for the case of absent Rayleigh scattering (red line) and the case of absent mirror coupling (blue line). 
Dashed lines -- right y-axis: Comparison of the $|\Gamma|$ for the case of absent Rayleigh scattering (red dashed line) and for the case of absent mirror coupling (the blue dashed line). 
This figure is taken from Ref. \cite{Galiev:21}.
}
\label{fig:mirror_sil_figure}
\end{figure*}

In some cases, optimal parameters are hard to achieve. 
Recently, it has been proposed the scheme of the self-injection locking of a laser via a high-$Q$ WGM microresonator, in which the drop-port-coupled mirror adjusts the optical feedback \cite{Galiev:21} (see Fig.~\ref{fig:mirror_sil_figure}). 
The adjustment enables tuning the stabilization coefficient and optimizing it for any level of Rayleigh scattering. 
In this way, the self-injection locking scheme with a mirror solves the problem of the non-ideal Rayleigh back-scattering rate, which is highly suppressed in high-$Q$ crystalline WGM microresonators. 
It has been noticed that the optimal regime of the proposed scheme is far from critical coupling (unlike the classic self-injection locking scheme), which results in less radiation losses. 

Based on the quasi-geometrical approach, the resonant optical feedback for the mirror-assisted scheme (see Fig. \ref{fig:mirror_sil_figure}a) was derived \cite{Galiev:21}: 
\begin{equation}
B_{\rm r} = - \frac{2i\eta(\beta + 2 \beta_m)}{(1 + \beta_m -i\zeta)^2+\beta(\beta + 2 \beta_m)} B_{\rm in},
\label{eq:mirror_sil_gamma}
\end{equation}
where $\beta_m = \kappa_{\rm mirror}/\kappa_m$ is the ratio between the mirror-coupling rate and rate of total losses in the microresonator. 
It was found out that for the stabilization coefficient optimization problem, the ratio between mirror-coupling rate and rate of the internal losses in the microresonator ($\mu_m = \kappa_{\rm mirror}/\kappa_0$) is more convenient \cite{Galiev:21}. 
We note that such scheme can also be implemented on-chip \cite{Shim:2021} by means of a Sagnac reflector.

The maximal values of the stabilization coefficient for the self-injection locking scheme with drop-port coupled mirror and for the classic scheme with the optimal Rayleigh scattering reported in Ref. \cite{Galiev2020} are approximately the same (see Fig.~\ref{fig:mirror_sil_figure}b). 
However, for the classic scheme, the maximal level of laser stabilization needs precise Rayleigh scattering rate tuning, which is not a trivial task compared to the drop-port mirror coupling rate tuning. 

\subsection{Broadening usable spectral range}

Though first experiments and applications where focused on the telecommunication ($1.55 \mu $m) and visible band, there are no principal limitations for implementation of self-injection locking at other wavelength.
A number of main directions can be distinguished in the area. 
A gallium nitride (GaN) semiconductor FP laser diode operating at a wavelength of 446.5 nm with a linewidth of less than 1 MHz has been demonstrated \cite{Donvalkar_2018} as well as the sub-100-kHz UV laser at 370 nm \cite{Savchenkov:19}. 
It is worth noting that there were crystalline MgF$_2$ microresonators with $Q$ factors exceeding $10^9$. 
Also, a hybrid integrated laser composed of a GaN-based laser diode and a Si$_3$N$_4$ photonic chip-based microresonator operating at record low wavelengths of 410 nm in the near-ultraviolet wavelength region was reported \cite{Siddharth:2022}. 
It is suitable for addressing atomic transitions of atoms and ions used in atomic clocks, quantum computing, or for underwater LiDAR. 
By self-injection locking of the FP diode laser to a high-$Q$ ($0.4\times10^6$) photonic integrated microresonator, the optical phase noise at 461 nm was reduced by a factor greater than 100, limited by the device $Q$ factor and back-reflection. 
A chip-scale visible lasers platform was created by using tightly-confined, micrometer-scale Si$_3$N$_4$ resonators and commercial FP laser diodes \cite{Corato-Zanarella:2021}. 
A tunable and narrow-linewidth lasers in the deep blue (450 nm), blue (488 nm), green (520 nm), red (660 nm) and near-IR (785 nm) with coarse wavelength tuning up to 12 nm, fine tuning up to 33.9 GHz, linewidth down to sub-kilohertz, side-mode suppression ratio $>35$ dB, and fiber-coupled power up to 10 mW was achieved. 

\begin{figure*}[t!]
\centering
\includegraphics[width=0.9\linewidth]{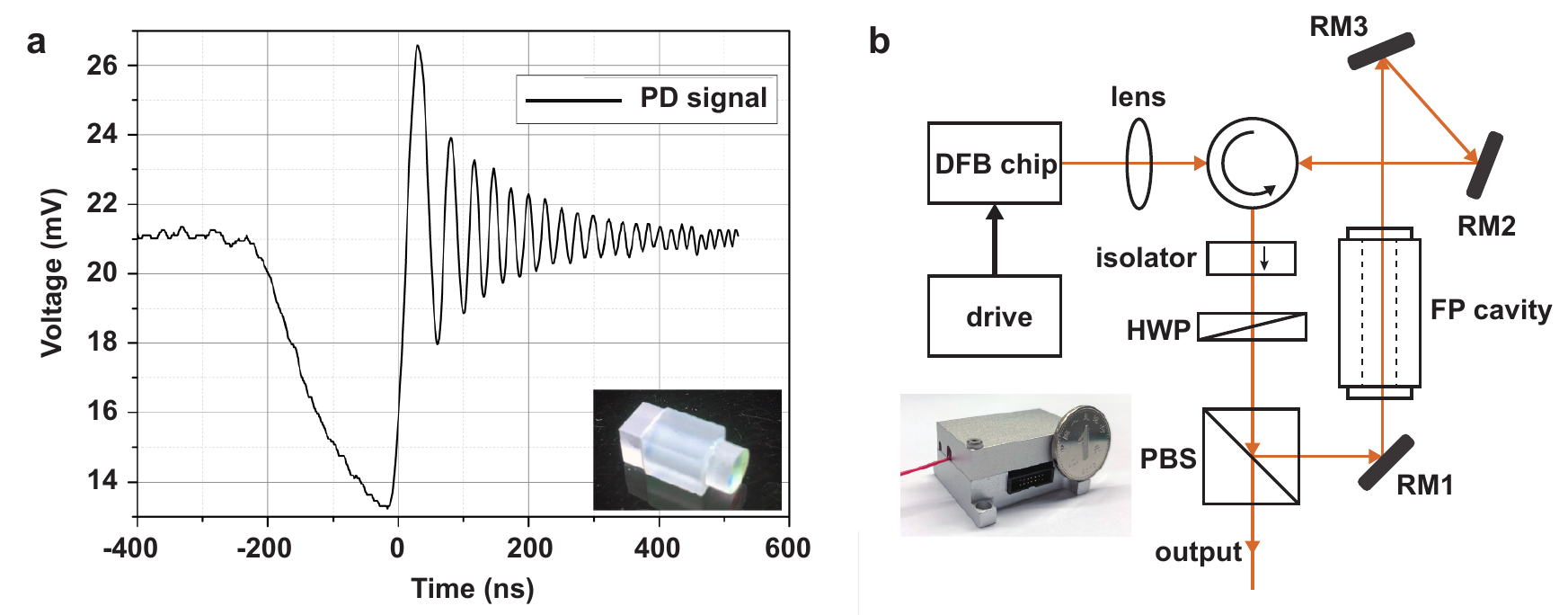}
\caption{
Experiments of laser self-injection locking to Fabry-P\'erot cavities.
\textbf{a}. Ring-down measurement of the $Q$ factor of a miniature FP cavity. 
\textbf{b}. Schematic diagram of the narrow-linewidth laser employing the miniature FP cavity. 
This figure is taken from Ref. \cite{Li:21}.  
}
\label{fig:SIL_FP}
\end{figure*}

The possibility of stabilizing lasers at longer wavelengths is also of interest. 
In particular, the stabilization of a GaSb-based diode laser with distributed feedback at a wavelength of 2.05 $\rm \mu m$ by a WGM microresonator was studied \cite{Dale:16}. 
The measured frequency noise of the stabilized laser are below 100 $\rm Hz/Hz^{1/2}$ in the range from 10 Hz to 1 Hz. 
The instantaneous linewidth decreased by four orders of magnitude compared to the free-running laser and amounted to 15 Hz at a measurement time of 0.1 ms. 
The integral linewidth was 100 Hz. 
Even better results were obtained in later work \cite{Savchenkov:19a}: the frequency noise of the laser was below 50 $\rm Hz/Hz^{1/2}$ at 10 Hz, reaching 0.4 $\rm Hz/Hz^{1/2}$ at 400 kHz. 
The instantaneous linewidth of the laser improved by almost four orders of magnitude compared to the free-running laser and amounted to 50 Hz at a measurement time of 0.1 ms. 
The Allan deviation of the laser frequency was about $10^{-9}$ from 1 to 1000 s. 
In addition, the possibility of stabilizing a quantum cascade laser at a wavelength of 4.3 $\rm \mu m$ was studied. 
The linewidth decreased to 10 kHz for integration times from 1 ms to 1 s \cite{Cumis16}. 
The $Q$ factor of the CaF$_2$ microresonator at this wavelength was about $2.2\times 10^7$. 
The possibility of frequency tuning in the 1 GHz band by controlling the cavity temperature was also demonstrated. 
Ww note that the $Q$ factor of crystalline microresonators made of many standard materials, including MgF$_2$ and CaF$_2$, decreases in the mid-IR due to multi-phonon absorption \cite{Lecaplain:16}, which limits their application for laser frequency stabilization in this range. 
One of the possible ways is to use a microresonator made from crystalline silicon, which has comparable $Q$ \cite{Shitikov:18}, and self-injection locking to crystalline microresonator made from silicon was demonstrated at 2.64 $\mu$m \cite{Shitikov2020}. 
Also, a tunable, single-mode, mid-IR laser at 3.4 $\mu$m using a tunable high-$Q$ silicon microring cavity and a multi-mode interband cascade laser was developed \cite{Shim:2021}. 
Single-frequency lasing with 0.4 mW output power via self-injection locking and a wide tuning range of 54 nm with 3 dB output power variation was achieved. 
An upper-bound effective linewidth of 9.1 MHz was estimated and a side mode suppression ratio of 25 dB from the locked laser was measured.

\subsection{Additional thermal stabilization}

The thermo-refractive coefficient as well as thermal expansion of the microresonator imposes the major limitations on the self-injection-locked laser performance, causing both thermal drift and inevitable thermodynamic fluctuations \cite{lim2017chasing,Savchenkov:2022}.
One way to increase thermal stabilization efficiency is the development of methods to study the microresonators to thermal bath connection. 
For example, WGM microresonators are usually characterized by a relative temperature sensitivity frequency of the order of $10^{-5}/^\circ$C. 
The ambient temperature should be stabilized at the level of the $\mu$K so that the linewidth of the cavity-based laser is less than 10 kHz for 1 s. 
The problem can be solved by using a thermally self-compensating resonator. 
Thermal compensation is carried out by using a specially developed design of the composite resonator. 
For microresonators made of MgF$_2$, a sandwich structure with layers of Zerodur was developed, which led to a decrease in sensitivity to fluctuations by a factor of 7 \cite{lim2017chasing}. 
For CaF$_2$, the use of a composite structure with layers of Zerodur provided a threefold decrease in sensitivity to fluctuations. 
However, for CaF$_2$, it turned out to be more promising to create a composite structure with ceramic layers with a negative coefficient of thermal expansion \cite{savchenkov2018calcium}. 
This approach made it possible to reduce the sensitivity to thermal fluctuations by more than 100 times, and to achieve a frequency stability level of $10^{-12}$ at normal atmospheric pressure with an integration time of 1 s \cite{photonics5040043}. 
To further improve the laser stability, the integration of lasers in evacuated thermally stabilized packages, the introduction of active stabilization of the optical path, and the use of high-$Q$ thermally compensated cavities are promising. 
For a thermally compensated MgF$_2$ microresonator in a rigid evacuated shell, we obtained a linewidth of less than 25 Hz and a relative frequency stability of $1.67 \times 10^{-13}\, (5.0 \times 10^{-12})$ for an integration time of 0.1 (1.0) s for 191 THz optical frequency \cite{lim2017chasing}. 
Active methods of thermal stabilization are also being developed. 
In particular, using cross-polarized two-mode temperature stabilization for a birefringent high-$Q$ WGM microresonator, the long-term stability was improved by a factor of 51 at an integration time of 1000 s \cite{lim2019probing}. 
A cavity temperature instability level of 10 $\mu$K has been achieved even up to an integration time of 1000 s, allowing this compact optical cavity module to serve as a high-performance frequency reference in potential metrology, synchronization, and frequency transmission applications.

\subsection{Self-injection locking to Fabry-P\'erot cavities} 

Performance of narrow-linewidth laser based on self-injection locking to WGM microresonators and on-chip microring resonators are ultimately limited by the absorption, various nonlinear effects and thermal expansion of the hosting material \cite{Ilchenko1992ThermalNE,Liang:15a,strekalov:2016}. 
For example, although crystalline WGM microresonators has reached maximum $Q$ factor of $10^{11}$, they typically have to be loaded to $Q <10^9$ to avoid Kerr nonlinear effects \cite{Liang2015}. 
On the other hand, Fabry-P\'erot (FP) cavities has been the best optical cavities ever made and frequently used as the optical reference in the most demanding applications including optical clock, frequency combs, and precision measurements \cite{ludlow2007compact}. 
In terms of $Q$ factor and frequency stability, hollow FP cavities made of super-mirror and ultra-low-expansion supporting material are still far ahead of WGM microresonators and on-chip microring resonators. 
Compared with WGM microresonators or on-chip microring resonators, hollow FP cavity have the following advantages. 
Firstly, the only part of the cavity that bears the very high intensity of the build-up optical field inside the cavity is the coating of the mirror, with typical field penetration depths of a few micrometer. 
Combined with the fact that the modal volume is orders of magnitude larger, hollow FP cavity possesses much lower thermal effect, nonlinear effects and induced frequency drift. 
Hence it is possible to use FP cavities of very high finesse and $Q$ factor in self-injection locking without inducing nonlinear effects. 
Secondly, the body of an FP cavity can be built with zero-expansion material such as ULE or Zerodur, which significantly reduces its long-term frequency drift due to temperature fluctuation. 

\begin{figure*}[t!]
\centering
\includegraphics[width=0.9\textwidth]{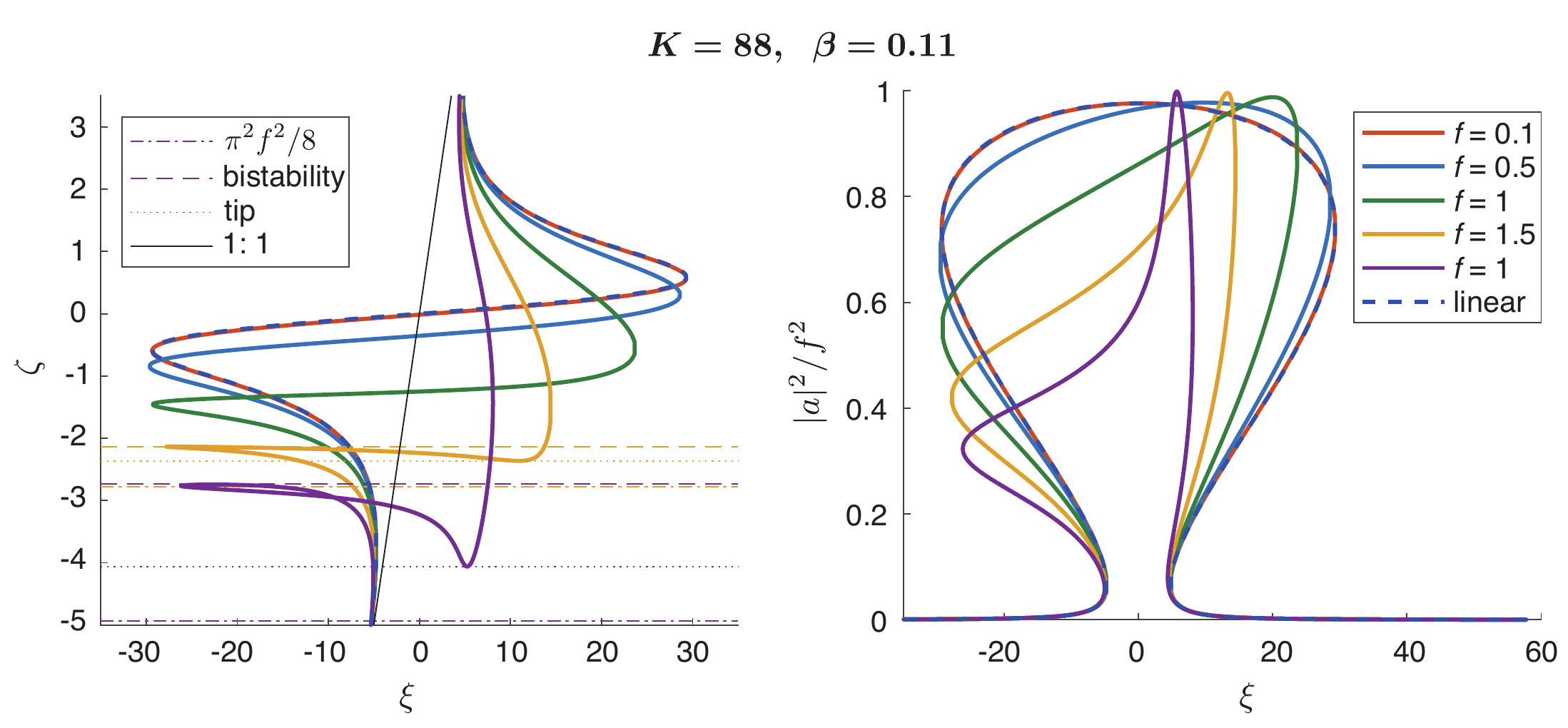}
\caption{
Typical theoretical solutions for tuning (left) and resonance (right) curves are shown for different pump values $f$ (see right panel legend). 
In the left panel, dashed horizontal lines show the bi-stability start detuning,
dotted lines shows the bi-stability end detuning, 
and dash-dotted lines shows the soliton existence border. 
This figure is taken from Ref.  \cite{Kondratiev2020aip}.
}
\label{fig:nltune}
\end{figure*}

An FP cavity is also the first type of external high-$Q$ cavity used to demonstrate linewidth narrowing of diode lasers utilizing self-injection locking \cite{Dahmani:87,li:1988,Laurent1989}. 
Almost ten years ago, self-injection locking of a single laser and two lasers to a bulky FP cavity with high finesse have been demonstrated \cite{Zhao:11,zhao2012100,Zhao2012} for TIRR, linear and V-shaped configurations to produce very narrow linewidth. 
However it has never matured into a viable product due to the large FP cavity and optical bench used in the experiments. 
To take this technology out of lab and make commercial products similar to the size of other competing narrow-linewidth laser technologies, a miniaturized FP cavity of high-$Q$ factor needs to be used. 
Some recent research efforts have been put into developing miniature FP cavities and using them as reference in PDH lock scheme \cite{Newman:19,Guo2022}, and hertz-linewidth has been achieved. 
Despite the excellent results, the cavity needs to be stabilized in a vacuum and acoustic isolated chamber and the complexity makes the system only available in lab. 
In addition, there are also reports of using 10-mm-length confocal FP (CFP) cavity with self-injection locking to achieve sub-kilohertz linewidth \cite{lewoczko2015ultra,christopher2017narrow}. 
However finesse of CFP cavities can not reach very high level and the package size is still many times of a standard butterfly package. 
In a recent effort, a miniature FP cavity of sub-milliliter volume and $10^8 \sim 10^9$ $Q$ factor was developed and utilized in self-injection locking to make compact narrow-linewidth lasers \cite{Li:21}. 
Figure \ref{fig:SIL_FP}a shows a ring-down measurement of the linewidth of the FP cavity. 
Figure \ref{fig:SIL_FP}b shows the schematic diagram of a narrow-linewidth laser locked to the FP cavity. 
The earlier result \cite{Li:21} utilizing $\mu$-FP cavity of $10^8$ $Q$ has already shown its potential to beat the performance of a fiber laser with a very compact size. 
By improving $Q$ factor to $7.7\times10^8$  \cite{Liang2022compact}, we also demonstrated that the frequency noise is much better than leading commercially available narrow linewidth laser products such as NKT E15 and OE4040-XLN, it's also ahead of the recently reported heterogeneous integrated narrow linewidth laser utilizing high $Q$ microring resonator \cite{Guo2022}, if only self injection lock technique is employed. Our work marks a major step toward a new category of compact narrow linewidth lasers of
superior performance utilizing ultra-high $Q$ miniature FP cavity.

It is also worth to mention that, other than hollow FP cavities, solid FP cavities based on low-loss fibers have also been used in self-injection locking to achieve narrow linewidth laser \cite{favre1985spectral,hao2021narrow}. 
This platform has also demonstrated a Kerr frequency comb with external fiber laser pumping \cite{jia2020photonic}. 
In a recent work, a compact Kerr frequency comb engine based on self-injection locking of an 80 mW DFB laser to fiber FP resonators of various FSRs from $1\sim10$ GHz has been demonstrated \cite{Jia:22}.

\section{Influence of the microresonator nonlinearity} 

Because of the high $Q$ factor and small mode volume of optical microresonators, Kerr nonlinearity and thermal effect should be considered in the case of sufficient intracavity power. 
Above a certain pump power the tuning curve will be distorted. 
The locking coefficient and bandwidth will change, and the generation frequency will be shifted. 
Furthermore, if the modulation instability threshold is exceeded, comb states can be generated. 
The self-injection locking was found to be beneficial for soliton generation \cite{Voloshin2021, Shen:20, Kondratiev:20}, when the pump is stabilized at the desired regime. 
Moreover, self-injection locking allows for solitonic pulse generation even in microresonators with normal group velocity dispersion (GVD) \cite{Kondratiev:20, Jin2021, Lihachev:22a}, which otherwise requires specialized techniques \cite{Lobanov2015, Lobanov2019, Xue2015}.
On the other hand, the manifestation of thermal effects, such as thermo-optic and thermal expansion, are inevitable in high-$Q$ optical microresonators \cite{Ilchenko1992ThermalNE, Fomin:05} at high power. 
Microresonator thermal effects are often considered parasitic, especially in the context of nonlinear optical processes, where thermally induced drifts, fluctuations, and instabilities \cite{Fomin:05, Carmon:04, Diallo:15, PhysRevA.103.013512} can strongly impact the generation of optical frequency combs and solitonic structures \cite{Herr2014, Bao:17, Lobanov:21}.

\subsection{Kerr nonlinearity}
\label{chap:nlpull}

It is reasonable to begin with the analysis of the Kerr effect. 
Consider the microresonator coupled mode equations \cite{Herr2014} with back-scattering \cite{Kondratiev:20_BW} with forward and backward (clockwise and counter-clockwise propagating) mode amplitudes $a_\mu$ and $b_\mu$, which are analogous to the linear self-injection locking model \cite{Kondratiev:17} with additional nonlinear terms:
\begin{widetext}
\begin{align}
\dot{a}_\mu=&(-1+i\zeta_\mu) a_\mu+i\beta b_{\mu}+i\hspace{-1em}\sum_{\mu'=\nu+\eta-\mu}\hspace{-1em} a_{\nu}a_{\eta}a^{*}_{\mu'} 
+2i\alpha_xa_{\mu}\sum_\eta|b_{\eta}|^2 +f \delta_{\mu 0},\nonumber\\ 
\label{CMES}
\dot{b}_\mu=&(-1+i\zeta_\mu) b_\mu+i\beta a_\mu+i\hspace{-1em}\sum_{\mu'=\nu+\eta-\mu}\hspace{-1em} b_{\nu}b_{\eta}b^{*}_{\mu'}++2i\alpha_xb_{\mu}\sum_\eta|a_{\eta}|^2,
\end{align}
\end{widetext}
where $f$ is normalized pump amplitude ($f=1$ means the modulation instability threshold), 
$\beta$ is the normalized coupling rate between forward and backward modes (mode splitting in the unit of mode linewidth), 
$\alpha_x$ is a cross-modulation coefficient derived from mode overlap integrals \cite{Kondratiev:20_BW}, 
$\zeta_\mu=2(\omega_\text{eff}-\omega_\mu+\mu D_1)/\kappa$ is the normalized detuning between the laser emission frequency $\omega_\text{eff}$ and the $\mu$-th cold microresonator resonance $\omega_\mu$ on the FSR-grid (with $\mu=0$ being the pumped mode and $D_1/2\pi$ being the microresonator FSR. 
Note that in Ref. \cite{Voloshin2021}, as in other previous experimental works, the detuning $\zeta$ is defined with the opposite sign to be co-directional with the diode pump current and the wavelength. 
Here one conventionally sticks to the definition where the detuning is co-directional with frequency. 
For numerical estimations, $\alpha_x=1$ can be used for the modes with the same polarization. 
Equation \eqref{CMES} provides a nonlinear resonance curve and the soliton solution \cite{Herr2014, Kondratiev:20_BW}. 
For analyzing self-injection locking, we combine Eq. \eqref{CMES} with the standard laser rate equations Eq. \eqref{LaserC}, which is similar to the Lang-Kobayashi equations \cite{Lang_Kobayashi} but with resonant feedback \cite{Kondratiev:17}. 
The pumped mode corresponding to $\mu=0$ is of main interest. 
We search for the stationary solution:
\begin{align}
\label{stationary}
&\left(-1+i\zeta\right)a+i\beta b+i a(|a|^2+2\alpha_x|b|^2)+f=0,\nonumber\\
&\left(-1+i\zeta\right)b+i\beta a+i b(|b|^2+2\alpha_x|a|^2)=0,
\end{align}
where $a=a_0$, $b=b_0$ and $\zeta=\zeta_0$ for simplicity. 
These equations define the complex reflection coefficient of the WGM microresonator used for self-injection locking theory.
To solve Eq. \eqref{stationary} for the reflection coefficient and make the resemblance to the linear case \cite{Gorodetsky:00, Kondratiev:17}, the {nonlinear detuning shift} $\delta\zeta_{\rm nl}$ and {nonlinear coupling shift} $\delta\beta_{\rm nl}$ are introduced.
Then, we further transform $\bar \zeta=\zeta+\delta\zeta_{\rm nl}$, ${\bar\beta}^2=\delta\beta_{\rm nl}^2+\beta^2$ to achieve the reflection coefficient in the same form as in the linear self-injection locking model Eq. \eqref{B(t)}. 
After redefinition $\bar \xi=\xi+\delta\zeta_{\rm nl}$, the nonlinear tuning curve in the new coordinates $\bar \xi$-$\bar \zeta$ becomes the same as Eq. \eqref{master}:
\begin{equation}
\label{masterNL}
\bar \xi=\bar \zeta+\frac{K_0}{2}\frac{2\bar \zeta\cos\bar\psi-(1+{\bar\beta}^2-{\bar \zeta}^2)\sin\bar\psi}{(1+{\bar\beta}^2-{\bar \zeta}^2)^2+4{\bar \zeta}^2},
\end{equation}
Note that the laser cavity resonant frequency $\omega_{LC}$ as $\xi$ is also assumed to include the Henry factor in its definition.
The $\kappa\tau_s/2$ is usually considered small, i.e. $\kappa\tau_s/2\ll1$, so the locking phase $\bar\psi\approx\psi=\omega_0\tau_s-\arctan \alpha_g-3\pi/2$ depends on both the resonance frequency $\omega_0$ and the round-trip time $\tau_s$ from the laser output facet to the microresonator and back.
The nonlinear detuning and coupling can be expressed as
\begin{align}
\label{NLzeta}
\delta\zeta_{\rm nl}&=\frac{2\alpha_x+1}{2}f^2\frac{1+(\bar\zeta+\delta\beta_{\rm nl})^2+\beta^2}{(1+{\bar\beta}^2-{\bar \zeta}^2)^2+4{\bar \zeta}^2},\\
\label{NLbeta}
\delta\beta_{\rm nl}&=\frac{2\alpha_x-1}{2}f^2\frac{1+(\bar\zeta+\delta\beta_{\rm nl})^2-\beta^2}{(1+{\bar\beta}^2-{\bar \zeta}^2)^2+4{\bar \zeta}^2}.
\end{align}
Equations \eqref{masterNL}-\eqref{NLbeta} can be solved numerically and plotted in $\zeta = \bar \zeta - \delta\zeta_{\rm nl}$, $\xi = \bar \xi - \delta\zeta_{\rm nl}$ coordinates.
One can observe that the calculated tuning curve in the nonlinear case, where Kerr nonlinearity is present, differs drastically from the tuning curve predicted by the linear model.
Also, it can be seen from Eq. \eqref{NLzeta} that the nonlinear detuning shift is positive and allows for larger negative detuning $\zeta$ (proportional to the pump power).
The detuning in the locked state can be estimated by assuming $\bar\zeta=0$ in Eqs. \eqref{masterNL}-\eqref{NLbeta}. 
For low $\beta\ll1$ and $\bar\psi=0$  \cite{dmitriev2022hybrid}:
\begin{align}
\nonumber\delta\zeta_0=&-\frac{2\alpha_x+1}{2\alpha_x-1}\frac{2}{\sqrt{3}}\\
\label{zeta_locked}&~~~\times{\,\rm sinh\!\!}\left(\frac{1}{3}{\,\rm arcsinh\!\!}\left(\frac{3\sqrt{3}}{4}(2\alpha_x-1)f^2 \right)\right)
\end{align}
It is a reasonable estimation if the pump is moderate, where the tuning and resonance curves are symmetric, without self-intersections, and good stabilization can be achieved (see intersection of black 1:1 line with the tuning curves in the left panel of Fig.~\ref{fig:nltune}). 
It can be shown that this detuning is always inside the bi-stability region.
The proposed nonlinear self-injection locking model is valid for anomalous and normal GVD.

The theoretical solutions for tuning and resonance curves are shown for different pump values $f$ in Fig.~\ref{fig:nltune}. 
It can be seen that the locking band is decreased with the pump power, but the effective detuning occurs in the comb region. 
One can also note the points with infinite derivative $\partial\omega_{\rm LC}/\partial\omega$ suggesting formally for infinite stabilization coefficient. 
However, as the linewidth can be viewed as the fluctuation of frequency, this derivative should be averaged near the laser cavity detuning over its initial linewidth to get the real line narrowing.

Complex transient dynamics can happen in the blue-detuned region, where the instability of the lower branch of the unperturbed curve appears earlier than the bistability (second stable branch) of the nonlinear curve (compare violet and blue dashed curves in Fig. \ref{fig:nltune}, right panel). A spontaneous transition from the linear model curve (blue dashed line in Fig. \ref{fig:nltune}, right panel) to the locked state happens \cite{Shitikov2022arx,Kondratiev:Thermal}. However, the intracavity power increases in the locked state and in nonlinear regime there is no stable stationary solution at this laser detuning (see orange or purple curves in Fig. \ref{fig:nltune}, right panel). So the frequency unlocks and power tries to go back causing dynamical oscillations \cite{Kondratiev:Thermal}. Note that this effect comes from any nonlinearity, e.g. from both thermal and Kerr one and is not connected with the nonlinear generation (single-mode regime was checked in \cite{Kondratiev:Thermal}). If the scan is stopped in this oscillatory regime the detuning continues to perform periodic evolution.

\subsection{Thermal nonlinearity}

\begin{figure*}[t!]
\centering
\includegraphics[width=0.9\linewidth]{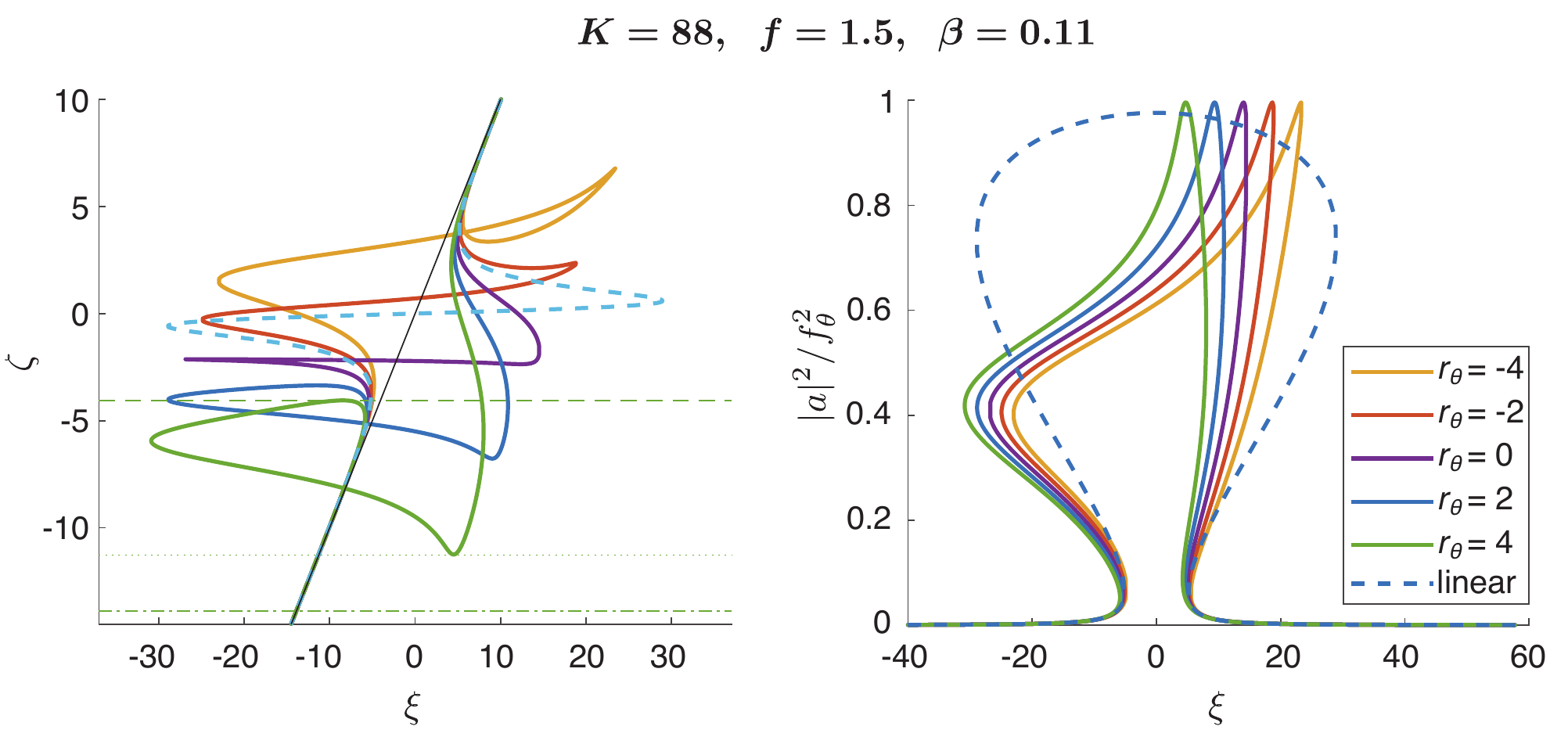}
\caption{
Tuning curve (left) and resonance curve (right) for different $r_\theta$. 
The cross-modulation coefficient is $\alpha_x=1$. 
Green dashed, dotted and dash-dotted lines in the left panel define the bi-stability begin, end and soliton region end, respectively, over $\zeta$ for $r_\theta=4$. 
This figure is taken from Ref. \cite{Kondratiev:Thermal}.
}
\label{results}
\end{figure*}

The above theory can also be modified in the case of thermal nonlinearity. 
The thermal equation can be added to the system Eq. \eqref{CMES} as following \cite{Lobanov:16}:
\begin{widetext}
\begin{align}
\label{Model_T_a}
\frac{da_\mu}{d\tau}=&-(1-i\zeta_\mu-i\theta) a_\mu+i\beta b_{\mu}+i \widehat{\widetilde a\widetilde a\widetilde a^*}_\mu+i2\alpha_xa_{\mu}P_b +f \delta_{\mu 0},\\  
\label{Model_T_b}
\frac{db_\mu}{d\tau}=&-(1-i\zeta_\mu-i\theta) b_\mu+i\beta^* a_\mu+i \widehat{\widetilde b\widetilde b\widetilde b^*}_\mu+i2\alpha_xb_{\mu}P_a,\\
\label{thermal_eq}
\frac{d\theta}{d\tau}~=&~\frac{\kappa_\theta}{\kappa}\left(r_\theta(P_a+P_b)-\theta\right). 
\end{align}
\end{widetext}
Here $\widehat{\widetilde a\widetilde a\widetilde a^*}_\mu$ and $\widehat{\widetilde b\widetilde b\widetilde b^*}_\mu$ stand for the Kerr nonlinear summations, and $P_{a}$ and $P_b$ stand for the average forward and backward wave powers compared with system Eq. \eqref{CMES}. 
The thermal variable $\theta$ is the temperature averaged with the optical field mode power and the thermo-refractive coefficient over the cavity volume, thus representing the thermal frequency shift.
The thermal parameters $\kappa_\theta$ and $r_\theta$ are inverse thermal relaxation time and thermal-to-Kerr-nonlinearity coefficient ratio \cite{Guo2017}. 
Note that negative thermo-refractive coefficient corresponds to negative $r_\theta$ and, consequently, negative $\theta$.
From the above equations, we can see that the comb and soliton parameters are now governed by $\zeta_{\rm eff}=\zeta+\theta$ rather than $\zeta$. 

In stationary regime, Eq. \eqref{thermal_eq} yields $\theta=r_\theta(P_a+P_b)$ and the system can be solved in a way similar to the above case, introducing the nonlinear detuning shift $\delta\zeta_{\rm nl}$ and the nonlinear coupling shift $\delta\beta_{\rm nl}$. 
In this case, the $2\alpha_x$ in Eq. \eqref{NLzeta} will be modified with $2r_\theta$ \cite{Kondratiev:Thermal}.
It can also be shown that after renormalization of the fields $a,b=a,b/\sqrt{1+r_\theta}$ in modified equations, the system is reduced to Eq. \eqref{CMES} with the effective pump and cross-modulation coefficient 
\begin{eqnarray}
f_\theta=\sqrt{1+r_\theta}f,\\
\alpha_\theta=\frac{\alpha_x+r_\theta/2}{1+r_\theta}.
\end{eqnarray}
This result, however, leads to formal complexities for the cases of $r_\theta\leq-1$, but the solution remains correct. We observe that thermal effects further deform the tuning and resonance curves of the self-injection locking. 
Figure \ref{results} shows the curves for different thermal-to-Kerr-nonlinearity coefficient ratio values. 
We can see that the locking band slope increases with $r_\theta$, meaning less stabilization efficiency. 
At the same time, the allowed generation detunings range grows. 
Another important note is that for the positive thermo-refraction, the points of $\partial \zeta/\partial \xi=0$ (infinite stabilization coefficient), which are also natural boundaries, holding the locking range inside, exactly correspond to the bi-stability criterion of the nonlinear resonance over generation detuning $\zeta$ with effective normalized pump amplitude $f_\theta$ (see green dashed and dash-dotted lines in the left panel of Fig.~\ref{results}). 

This stationary model has some issues related to the time scales.
As the Kerr nonlinearity is much faster than the thermal one, the stationary regime of the equations will occur at different times. So first, the system will come to the stationary regime of the non-thermalized model ($r_\theta=0$) and then will evolve to the presented one. 
The above consideration suggests that from the point of view of the comb dynamics, this will result in the effective incoming pump line broadening and additional detuning increase. 
However, this additional detuning does not drive the system from the soliton existence range.
Another issue is that while the Kerr nonlinearity constitutes the complex triple sum of the modal amplitude products, allowing to soliton formation \cite{Kondratiev:20_BW}, the thermal part depends on the total power. 
This difference makes the frequency comb parameters depend on the $f$ and $\zeta_{\rm eff}=\zeta+\theta$ rather than $f_\theta$ and $\zeta$. 
It can be shown that $\partial \zeta_{\rm eff}/\partial \xi=0$ points of tuning curves for the effective detuning $\zeta_{\rm eff}$ coincide with the bi-stability criterion for nonlinear resonance with pump $f$, similar to $\zeta$ and $f_\theta$. 
It directly shows that if the working point is in the self-injection locking state, it is automatically inside the bi-stability range, corresponding to the soliton existence domain. 
Note also that the locked state corresponds to small effective comb detuning.

\subsection{Frequency comb generation}
\label{chap:Combs}

\begin{figure*}[t!]
\center
\includegraphics[width=1\textwidth]{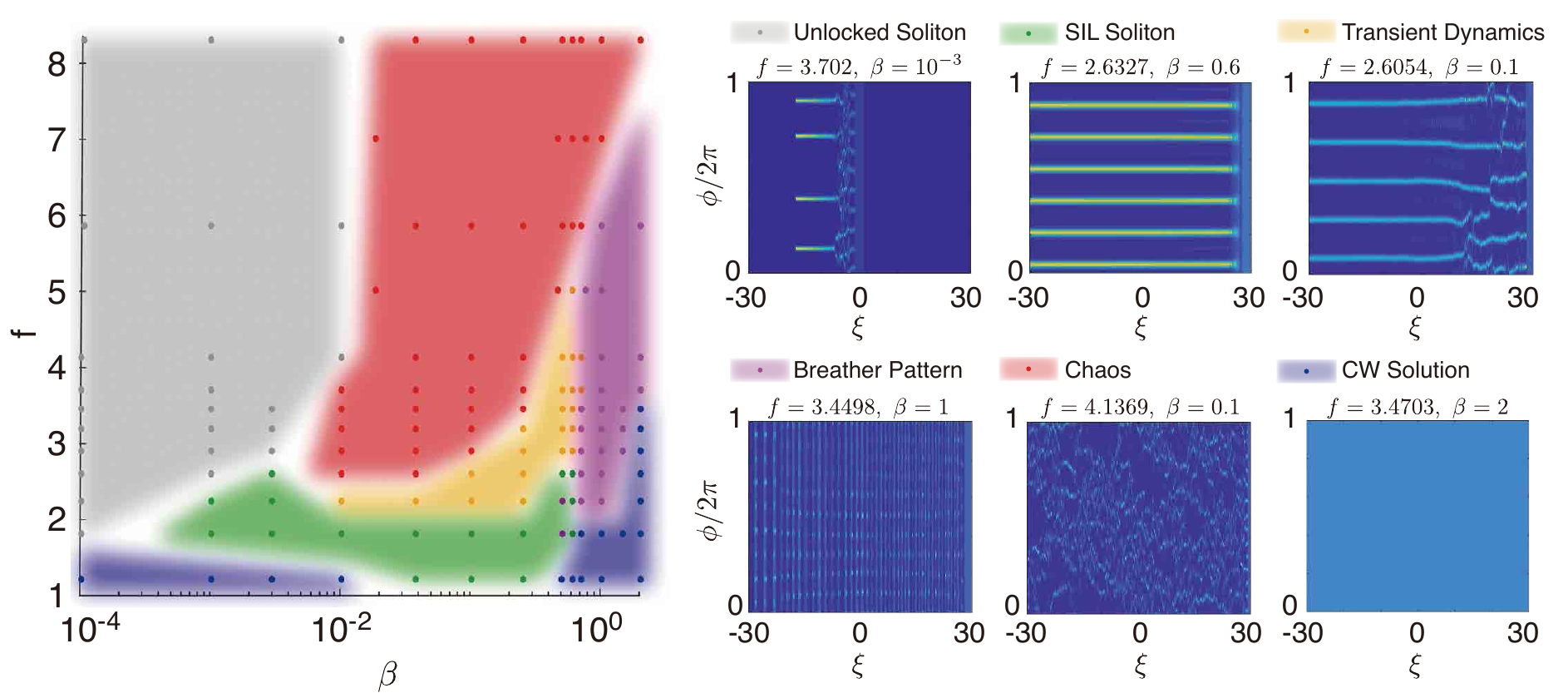}
\caption{
The diagram of the regimes arising upon pump frequency sweep with anomalous GVD for different combinations of the pump amplitude $f$ and back-scattering coefficient $\beta$.
The dots show the calculated parameters.
The shaded regions mark the extrapolation of the regime type. 
The white areas are either transient regions, or too far from the plotted points, to be certain. 
The legend is shown on the right and consists of panels with examples of characteristic regimes. 
This figure is taken from Ref. \cite{Kondratiev:20}.
}
\label{fig:SolitonDiagram}
\end{figure*}

We consider the coupled mode equations with back-scattering for the forward and backward mode amplitudes $a_\mu^+$, $a_\mu^-$ (having frequencies $\omega_\mu=\omega_0+D_1\mu+D_2\mu^2+D_3\mu^3+...$) with laser rate equations for the normalized carrier density $\tilde N_l$ 
and laser field $a_l$ \cite{Kondratiev:20}: 
\begin{widetext}
\begin{align}
\label{Model_carr_n}
\frac{d\tilde N_l}{d\tau} =&~r_{\rm g/\kappa}f_l^2(\tilde\kappa_l-\tilde N_l  |a_l|^2)+\tilde\kappa_N(\tilde\kappa_l - \tilde N_l),\\
\label{Model_las_n}
\frac{da_l}{d\tau}=&\left[-i\xi_{l}+(1+i\alpha_g)\tilde N_l-\tilde \kappa_l\right]a_l-e^{i\Omega_l t_{s}}\sum_\mu \frac{\tilde K_l}{f_l}a^-_\mu e^{i\omega_{l\mu} (t-t_{s})},\\
\label{Model_forw_n}
\frac{da^+_\mu}{d\tau}=&\left(-1+i\zeta_{\mu}\right)a^+_\mu +i\beta_\mu a^-_\mu +iS^+_\mu -e^{i\omega_\mu^{(1)} t_{s}}f_l a_le^{-i\omega_{l\mu} (t-t_{s})},\\
\label{Model_back_n}
\frac{da^-_\mu }{d\tau}=&\left(-1+i\zeta_{\mu}\right)a^-_\mu +i\beta_\mu a^+_\mu +iS^-_\mu,
\end{align}
\end{widetext}
where $\omega_\mu^{(1)}=\omega_0+D_1\mu$ is the microresonator mode first-order estimated oscillation frequency (FSR-grid),
$\omega_{l\mu}=\Omega_l-\omega_\mu^{(1)}$ is the laser cavity to FSR-grid mismatch, $\Omega_l$ is the laser cavity frequency,
$t_{s}=\tau_s/2$ is the one-way-trip-time from the laser to the microresonator defining the locking phase,
$\kappa/2\pi$ is the WGM resonance linewidth ($Q=\omega_0/\kappa$ is the loaded $Q$ factor),
$\tau=\kappa t/2$ is loss-normalized time,
$r_{\rm g/\kappa}$ 
is a combination of laser gain to microresonator nonlinearity ratio and laser to microresonator coupling coefficient \cite{Kondratiev:20},
$f_l$ is the normalized pump amplitude,
$\tilde\kappa_l$ and $\tilde\kappa_N$ are the normalized laser optical and carrier loss rates,
$\alpha_g$ is the laser Henry factor,
$\xi_{l}$ is the normalized laser cavity detuning from its initial value $\Omega_l$,
$\zeta_{\mu}=2(\omega_\mu^{(1)}-\omega_\mu)/\kappa$ is the microresonator effective detuning,
$\beta_\mu$ is the normalized forward-backward mode coupling (back-scattering coefficient) for the $\mu$-th mode (equal to the mode splitting in units of $\kappa_0$).
The $\tilde\kappa_W$ is the normalized laser-to-microresonator back coupling rate, and $\tilde K_l$ is the resonator to laser coupling coefficient that can be defined by the stabilization coefficient $K_l$ \cite{Kondratiev:17}:
\begin{equation}
\tilde K_l=\frac{K_l}{2\beta\sqrt{1+\alpha_g^2}}.
\end{equation}
\begin{figure*}[t!]
\center
\includegraphics[width=1\textwidth]{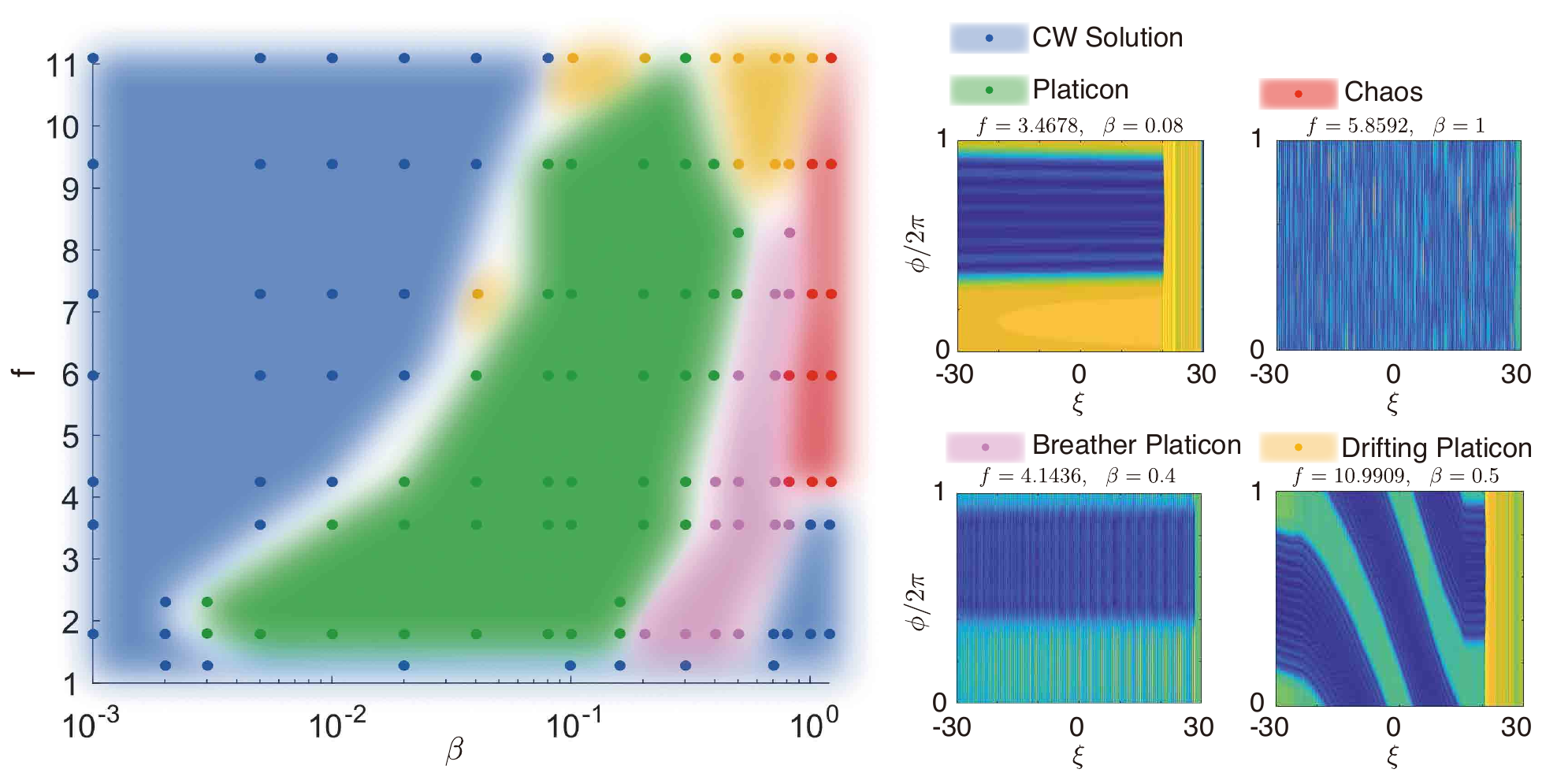}
\caption{
The diagram of the regimes arising upon pump frequency sweep with normal GVD for different combinations of the pump amplitude $f$ and back-scattering coefficient $\beta$.
The dots show the calculated parameters. 
The shaded regions mark the extrapolation of the regime type. 
The white areas are either transient regions, or too far from the calculated points to be certain. 
The legend is shown on the right and consists of panels with examples of characteristic regimes. 
This figure taken from Ref. \cite{Kondratiev:20}.
}
\label{fig:PlaticonDiagram}
\end{figure*}
The terms $S_\mu^\pm$ in \eqref{Model_forw_n}-\eqref{Model_back_n} represent the nonlinear sums, including the self- and cross-phase modulation \cite{Kondratiev:20, Kondratiev:20_BW}.
Equation \eqref{Model_carr_n} describes the carrier concentration dynamics, and Eq. \eqref{Model_las_n} describes the field amplitude in the laser. The laser field is normalized to the stationary solution of non-fedback equation, so that in stationary regime $a_l$ will be close to 1.
The last term of Eq. \eqref{Model_las_n} is the sum of the fields coming from the WGM microresonator (backward wave). 
In practice, the sum in Eq. \eqref{Model_las_n} is calculated so that $\left|\Omega_l-\omega_\mu^{(1)}-{\rm Im}\left[\frac{da_\mu/dt}{a_\mu}\right]\right|<5\kappa_l$, where $\kappa_l/2\pi$ is the laser cavity linewidth, to avoid modeling of redundant fast oscillating terms.
The second pair Eqs. \eqref{Model_forw_n}-\eqref{Model_back_n} describe the WGM field in the high-finesse limit \cite{Kondratiev:20_BW} [note the $\delta$-symbol in the pumping term of Eq. \eqref{Model_forw_n}]. 
The terms with $\beta$ stand for the forward-backward mode coupling. 
The backward wave Eq. \eqref{Model_back_n} is excited only through this term. 
The forward wave has two pumps: the backward wave ($i\beta_\mu a^-_\mu$ term) and the laser (the last term). 
In Ref. \cite{Kondratiev:20}, the Lugiato-Lefever equation (LLE) was obtained. 
The main feature here is that the feedback amplitude was written as if gathered from the point rotating around the microresonator. 
It is because the symmetry of the microresonator is broken with the introduction of the coupling element, providing the origin of the azimuthal angle at the touching point. 
It is where the field going to the laser originates in the laboratory frame, and the rotation goes from the fact that LLE is usually written in the rotating frame.
Equations \eqref{Model_carr_n}-\eqref{Model_back_n} can be easily expanded to the case of several lasers just treating the index $l$ as numerating the lasers and adding a summation over it in Eq. \eqref{Model_forw_n}.

The generation of the dissipative Kerr solitons (at anomalous GVD) and platicons (at normal GVD) was demonstrated numerically for the self-injection-locked pump \cite{Kondratiev:20}. 
Different regimes of different combinations of the locking phase (laser-microresonator round-trip time), back-scattering coefficient, and pump power exist (see Fig.~\ref{fig:SolitonDiagram} and Fig.~\ref{fig:PlaticonDiagram}). 
Generation of both types of the considered solitonic pulses was shown to be possible in a certain range of the locking phase and become less stable at high pump powers. 
Generation of the dissipative Kerr solitons was not very sensitive to the normalized back-scattering $\beta$, while for the platicon generation, this is a key parameter. 
The threshold value of the back-scattering coefficient was found to grow with the pump power. 
Some nontrivial dynamics, such as drift and breathing dynamics of the self-injection-locked platicons, were revealed. 
Self-injection locked solitons were demonstrated in Ref. \cite{Pavlov:18, Raja2019, Shen:20, Voloshin2021, Xiang:21} and platicons in Ref. \cite{Lihachev:22a,Jin2021}.

\subsection{Multilaser locking} 
\label{chap:Multilaser}

\begin{figure*}[t!]
\centering
\includegraphics[width=0.7\linewidth]{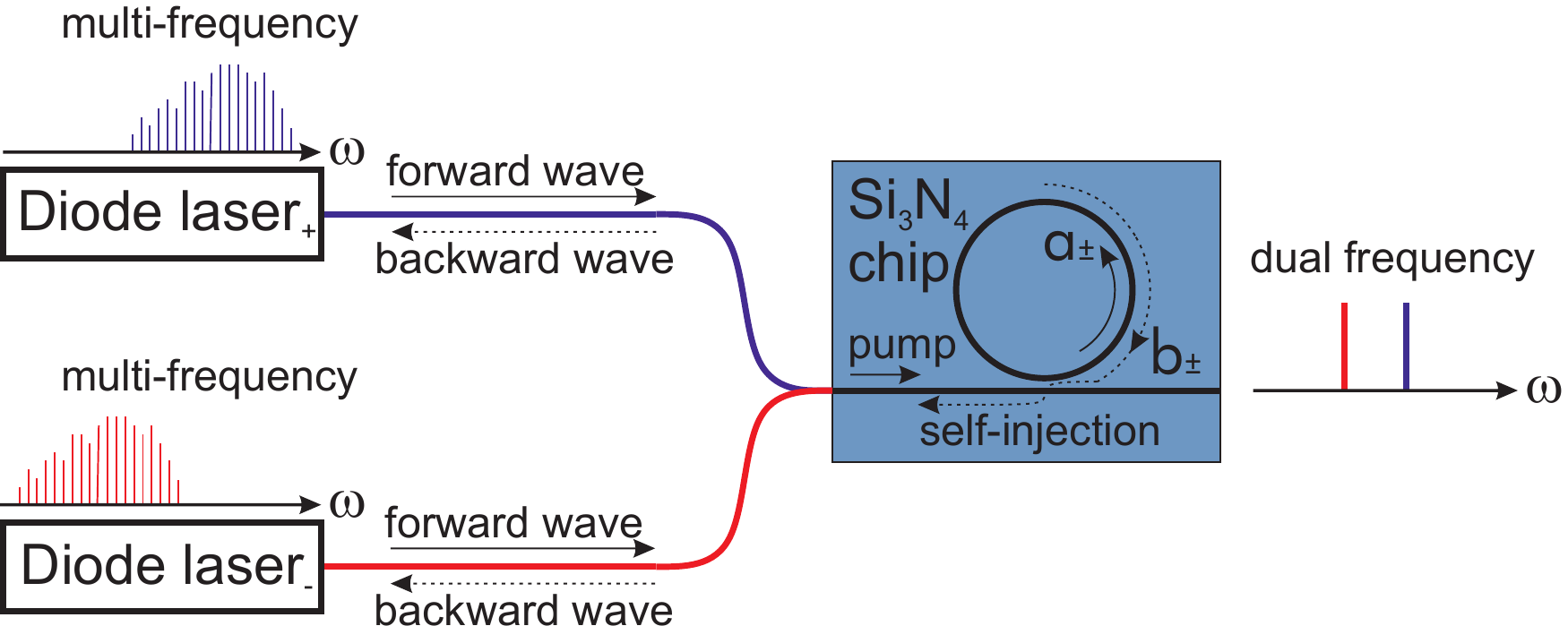}
\caption{
Dual-self-injection-locking concept. 
Two laser diodes are coupled to an integrated high-$Q$ microresonator. 
Both lasers are simultaneously self-injection-locked to different frequency modes of the microresonator, resulting in a stable and narrow-linewidth bichromatic output. 
This figure is taken from Ref. \cite{Chermoshentsev2022}.
}
\label{fig:2SILconcept}
\end{figure*}

Though most studies have been performed to lock a single laser to one or more intrinsic modes of a high-$Q$ microresonator, it cannot provide a multi-frequency pump with controllable detuning in each mode line. 
A multi-frequency pump can be realized by coupling two or more diode lasers to different resonances of a single microresonator with nonlinearity (see Fig. \ref{fig:2SILconcept}). 
This phenomenon is called ``multilaser self-injection locking'' or ``dual-self-injection-locking" (in the case of two lasers). 
Driving a single microresonator with two pump lasers features advantages in applications such as high-performance and compact photonic RF generators on chip \cite{Liu2020, Khan2010}, and frequency combs \cite{Hu2017, Wang2016, Wen2019, Wen2019a}. 
Moreover, based on dual-pump, compact optical parametric oscillator (OPO) can be developed \cite{Okawachi2015, Okawachi2016, Vaidya2020, Arrazola2021}. 
The OPO is useful for various quantum applications such as quadrature squeezing \cite{Zhao2020}, random number generation \cite{Okawachi2016}, Gaussian Boson sampling \cite{Arrazola2021}, and fully-optical implementation of the Ising model \cite{Okawachi2020}. 
However, in all above cases, two stable narrow-linewidth lasers are required that can be successfully replaced with laser diodes in self-injection locking. 

\begin{figure*}[t!]
\centering
\includegraphics[width=1\linewidth]{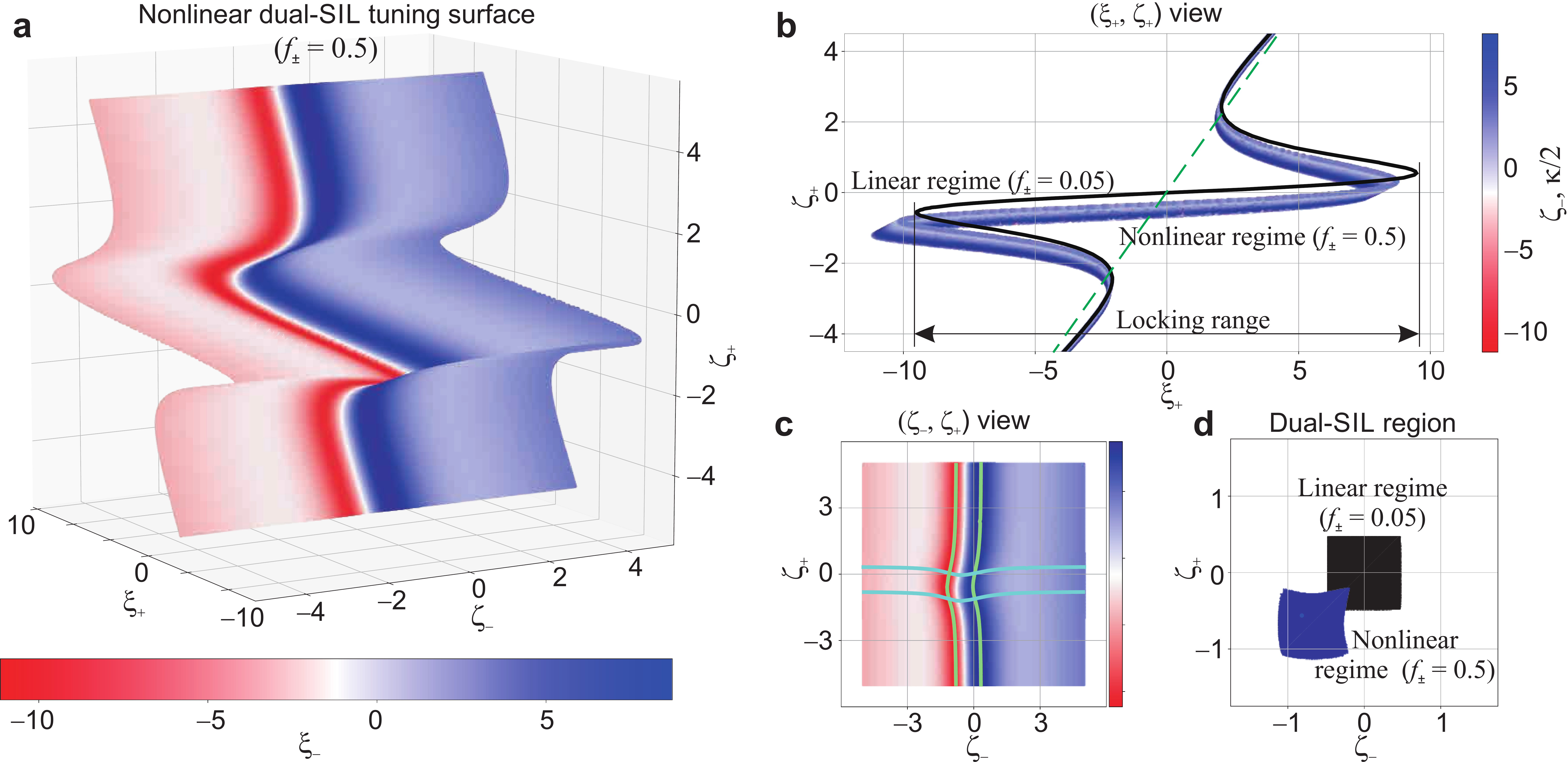}
\caption{
Tuning surface of the dual-laser self-injection locking in different views and a locking region diagram. 
The surface is calculated for normalized pump amplitudes $f_+=f_-=0.5$, where moderate nonlinear effects are present. 
Color in panels \textbf{a--c} represents the ``$-$" laser intracavity detuning $\xi_{-}$. 
The black curve in panel \textbf{b} and black square in panel \textbf{d} correspond $f_\pm=0.05$, where nonlinear effects are insignificant. 
The aqua and light-green curves correspond to the ``$+$" and ``$-$" laser individual locking regions. 
Dashed green line corresponds to $\xi_{+} = \zeta_{+}$. 
This figure is taken from Ref. \cite{Chermoshentsev2022}.
}
\label{fig:stationary_nonlinear}
\end{figure*}

Dual-self-injection-locking is more complicated than the single-laser case because of the nonlinear interactions between lasers inside the microresonator. 
The reason for such interaction is that the resonance shift caused by one pump applies to all microresonator modes. 
It shifts the other laser resonance, changing the frequency of back-scattered waves and self-injection locking dynamics.
The theoretical description of multilaser self-injection locking can be started with Eq. \eqref{Model_carr_n}-\eqref{Model_back_n}, that can be easily expanded to the case of several lasers just treating the index $l$ as numerating the lasers and adding a summation over it in Eq. \eqref{Model_forw_n}. 
A semi-analytical study was performed in Ref. \cite{Chermoshentsev2022} for two lasers locked to two separate modes. 
The solution was developed analogous to the study of a single-laser case (see Ref. \cite{Voloshin2021} and Section \ref{chap:nlpull}), but the number of equations and detunings doubled. 
Exactly Eq. \eqref{masterNL} was obtained for the two lasers detunings $\bar\xi_\pm$ and two-generation detunings $\bar\zeta_\pm$. 
The equations for the nonlinear shifts $\delta \zeta_{\pm}$ and $\delta \beta_\pm$ were found to be coupled (see the Supplemental Materials of Ref.~\cite{Chermoshentsev2022}):
\begin{widetext}
\begin{equation}\label{zeta_eq}
\begin{gathered}
(\alpha_x+1)\delta\zeta_{\pm}-\frac{2\alpha_x+1}{2}\delta\zeta_{\mp}=\frac{4\alpha_x+3}{4}f_{\mp}^2\frac{1+(\bar\zeta_{\mp}+\delta\beta_{\mp})^2+\beta_{\mp}^2}{(1+\bar\beta_{\mp}^2-\bar\zeta_{\mp}^2)^2+4\bar\zeta_{\mp}^2},\\
(\alpha_x-1)\delta\beta_{\pm}-\frac{2\alpha_x-1}{2}\delta\beta_{\mp}=\frac{3-4\alpha_x}{4}f_{\mp}^2\frac{1+(\bar\zeta_{\mp}+\delta\beta_{\mp})^2-\beta_{\mp}^2}{(1+\bar\beta_{\mp}^2-\bar\zeta_{\mp}^2)^2+4\bar\zeta_{\mp}^2}.
\end{gathered}
\end{equation}
\end{widetext}
Equations \eqref{zeta_eq} were solved numerically, and the values of nonlinear shifts were determined. 
As a result, a 4D dual-self-injection-locking tuning surface can be obtained as shown in Fig. \ref{fig:stationary_nonlinear}.

The  $(\xi_+, \zeta_+)$ view in Fig.~\ref{fig:stationary_nonlinear}b is shown to simplify the interpretation of this surface. 
The black curve is obtained for $f_+ = f_- = 0.05$ and approaches the weak-pump limit governed by the single-laser linear self-injection locking theory, described by Eq.~\eqref{masterNL} with $\bar\zeta_\pm\to\zeta_\pm$ and $\bar\Gamma_\pm\to\Gamma_\pm$. 
In this limit, the two lasers do not affect each other. 
At large detuning from microresonator resonances, the lasers are in the free-running regime, so $\zeta_+ = \xi_+$ or $\zeta_- = \xi_-$. 
Close to resonance, self-injection locking ensues, characterized by the plateau $\partial\zeta_{\pm}/\partial\xi_{\pm}\ll 1$ \cite{Kondratiev:17}.
The colored curves in Fig.~\ref{fig:stationary_nonlinear}b correspond to pump amplitudes $f_+ = f_- = 0.5$, where nonlinear effects are significant. 
The locking regions (tuning curves' plateaus $\partial\zeta_+/\partial\xi_+\ll 1$) are shifted to lower frequency, resulting from the red shift of microresonator resonances caused by the nonlinearity. 
The overall shift is due to the nonlinearity of the ``$+$'' laser. 
The nonlinear effect of the ``$-$'' laser is manifested by additional surface shifts for different $\zeta_-$ values, and different hues show the latter. 
When this laser is closer to the resonance (lighter hues), the power inside the microresonator is higher, so the shift is more significant. 
The cross-influence of the lasers is also evident in the ($\zeta_-, \zeta_+$) view (Fig.~\ref{fig:stationary_nonlinear}c), manifesting as a red shift of the high color gradient region (corresponding to the locking regime of the second laser $\partial\zeta_-/\partial\xi_-\ll1$) in the central region of the plot.
The light-green and aqua colors in Fig.~\ref{fig:stationary_nonlinear}c show the locking regions of the two lasers, whose boundaries are defined by $\partial\zeta_\pm/\partial\xi_\pm = \infty$. 
Their intersection determines the dual-self-injection-locking region, displayed by the black and blue areas in Fig.~\ref{fig:stationary_nonlinear}d for the weak and strong pumps, respectively. 
The square shape of the black area is a manifestation of the lasers' mutual independence in the linear case. 
The blue area's overall red shift and the red-sided curvature of its boundaries represent the self-phase and cross-phase modulation of the fields inside the microresonator by the two lasers.

Recently several experiments were conducted to investigate the dual-self-injection-locking phenomenon. 
Reference~\cite{Jiang:21} shows the simultaneous locking of two vertical-cavity surface-emitting lasers (VCSELs) to a single WGM microresonator made by Hydex glass. 
Two-port configuration was used where 90$\%$ of the radiation from the drop-port is redirected back to the VCSELs output. 
Such feedback allowed significant compression of the linewidths of both lasers from 3.5 and 5~MHz to 20.9 and 24.1~kHz, respectively. 
It has shown frequency noise suppression of VCSELs in such a configuration by more than 60~dB for Fourier offset frequency from 100~kHz and higher. 

Another application of dual-self-injection-locking is to realize all-optical dissipative time crystals (DTC), as shown in Ref.~\cite{Taheri2022}. 
In this work, DTCs formed in Kerr-nonlinear microresonators is presented. 
Two independent CW DFB diode lasers were locked to different eigenmodes of a MgF$_2$ microresonator with FSR of 32.8 GHz and loaded resonance bandwidth of 200 kHz. 
It was demonstrated that the two locked lasers with normalized powers higher ($f > 1$) and lower ($f < 1$) than the nonlinear threshold for the one and the other laser, respectively, and frequency spacings in the locked regime of an arbitrary number $M$ of FSRs allow the generation of dissipative soliton time-crystals with periodicity $T = \frac{m}{M} T_R $, where the ratio $\frac{m}{M}$ is an integer and $T_R$ is a round-trip time. 
Instead, the fact that the temporal structure of the time crystals was not obtained experimentally from the measured spectrum, and its agreement with theoretical simulation leads to conclusion that the DTCs were achieved. 
These DTCs are stable over hundreds of lifetimes of cavity photons which makes their lifetime much longer than for DTCs in other physical systems. 
Such an approach opens new horizons to investigate all-optical DTCs at room temperature in dissipative Kerr systems. 

\begin{figure*}[t!]
\centering
\includegraphics[width=0.8\linewidth]{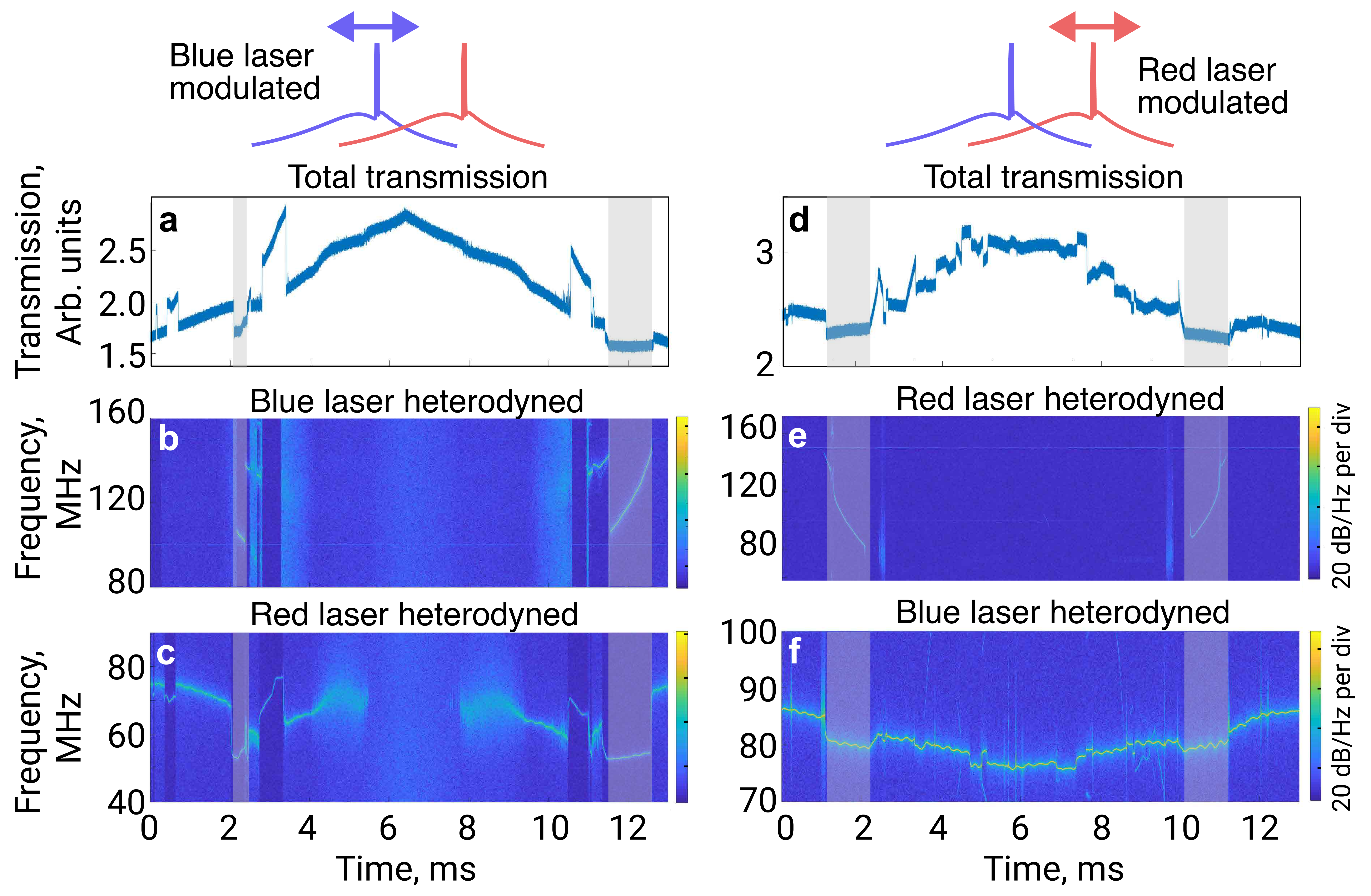}
\caption{
Dynamics of laser self-injection locking to two modes from different mode families. 
The transmission traces \textbf{a,d} and heterodyne spectrograms correspond to the frequency sweep of one laser (the blue laser in the left column and the red laser in the right column). 
In \textbf{b,f} the heterodyne measurement is performed on the same laser as the one being modulated; while in \textbf{c,e}, on the other laser. 
This figure is taken from Ref. \cite{Chermoshentsev2022}.
}
\label{fig:spectrogramm_differ}
\end{figure*}

The first experimental investigation of simultaneous locking of two multi-frequency FP lasers to a single integrated Si$_3$N$_4$ microresonator with 1~THz FSR and coupled to a single waveguide was demonstrated in Ref.~\cite{Chermoshentsev2022}. 
The authors experimentally investigate three possible configurations of dual-self-injection-locking: 
1. When both lasers are locked to the different modes of the same microresonator mode family with frequency spacing equal to 2 FSR; 
2. When lasers locked to the different modes of different mode families; 
3. When both lasers locked to the same eigenmode of the microresonator. 
In all cases, the spectral properties of dual-self-injection-locking, such as spectral collapse, phase-noise suppression, and linewidth compressing for both lasers, were observed by using heterodyne detection. 
The obtained laser linewidth calculated by Lorentz approximation gave the order of several kHz for both locked lasers. 
The phase-noise measurements showed that the phase noise level of reference had been achieved for the high offset frequency above $10^4$~Hz. 
The spectrogram showed that the lasers experienced nonlinear interaction inside the microresonator, and the dual-self-injection-locking's locking range was estimated. 
The spectrogram measurements for the case of locking to different modes of different mode families are presented in Fig.~\ref{fig:spectrogramm_differ}. 
The sharp dips in Fig.~\ref{fig:spectrogramm_differ}f in about $1\sim2$ ms and $10\sim11$ ms corresponded to nonlinear interaction between lasers when the dual-self-injection-locking regime was achieved. 
For the locking of two lasers to the same eigenmode, the authors also observed coherent addition of the laser output signal that can be a base for developing high-power narrow-linewidth compact laser sources.

\section{Special regimes and novel application} 

\begin{figure*}[t!]
\centering
\includegraphics[width=0.95\linewidth]{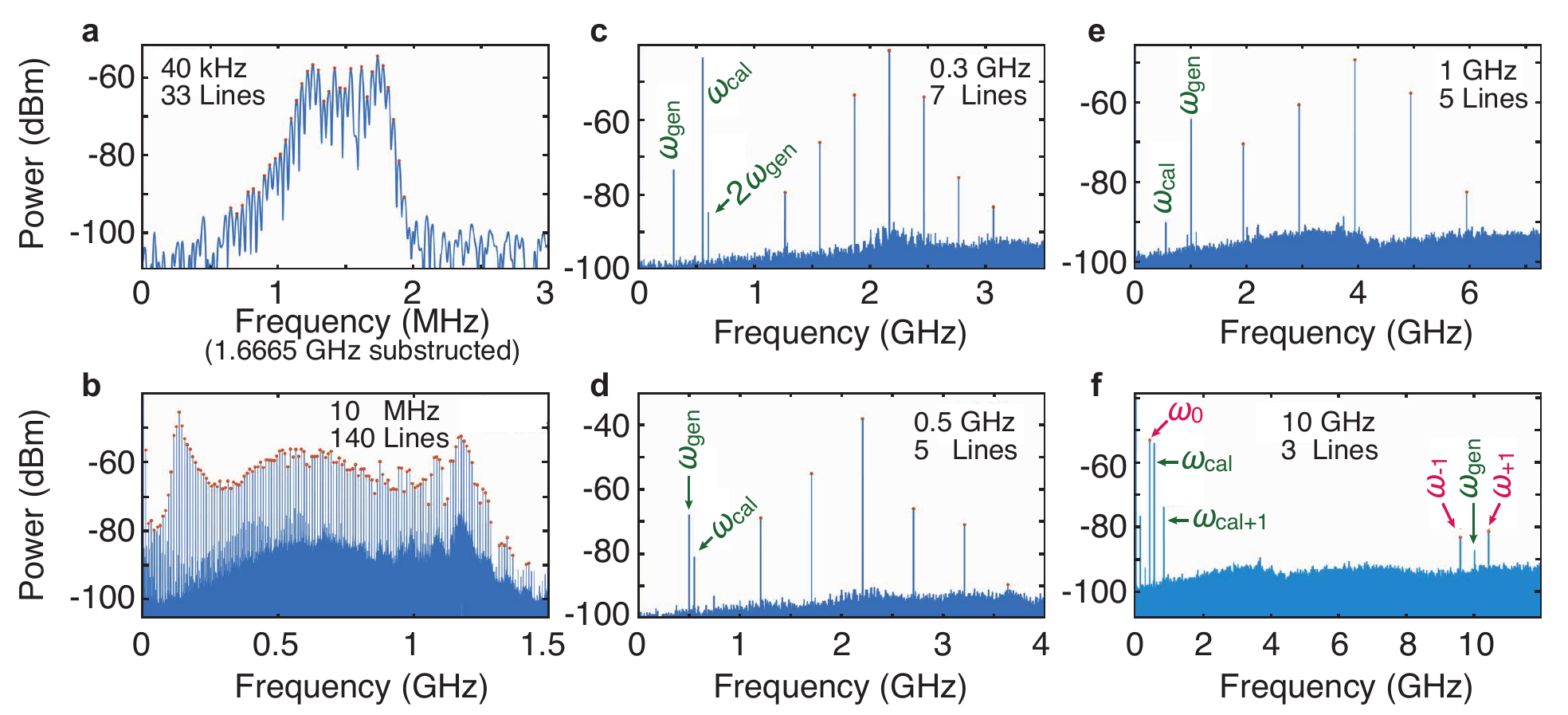}
\caption{
Frequency comb generation in the self-injection locking regime with a gain-switched laser. 
Spectra were obtained by means of the heterodyne method. 
Comb teeth are marked with red circles and red arrows, and green arrows correspond to the interference lines. 
\textbf{a}. The 40-kHz-spaced comb. 
The spacing between adjacent lines is presented on the inset. 
\textbf{b}. The 10-MHz-spaced comb with bandwidth wider than 1 GHz. 
\textbf{c}. The 0.3-GHz-spaced comb. 
\textbf{d}. The 0.5-GHz-spaced comb. 
\textbf{e}. The 1-GHz-spaced comb. 
\textbf{f}. The 10 GHz spaced comb. 
This figure is taken from \cite{shitikov2021self}.
}
\label{fig:gsCombSpacing}
\end{figure*}

In recent years, a number of interesting and promising applications of microresonator-stabilized lasers have been realized. 
First, an effective pump source for frequency comb generation \cite{Kippenberg2011, Kippenberg2018, PASQUAZI20181, Gaeta2019} has been demonstrated. 
The implementation of microresonators and compact laser diodes allows to decrease size and weight of comb generators in comparison with conventional mode-locked systems. 
Moreover, generation of coherent or solitonic frequency combs in the form of bright soliton \cite{Pavlov:18, Raja2019, Shen:20, Voloshin2021, Xiang:21} or platicon \cite{Lihachev:22a,Jin2021} trains have been shown. 
Such coherent optical frequency combs are actively used in different areas of science and technology such as high-precision metrology and optical clocks \cite{Papp:14,Newman:19}, high-resolution spectroscopy \cite{Suh600,Yang:19}, ultrafast optical ranging \cite{Suh2018,Trocha2018}, astrophysics \cite{Suh2019,Obrzud2019}, and high-volume telecommunication systems \cite{Marin-Palomo2017,fulop2018high,Helgason:19}. 
Besides, high-efficiency compact hybrid dual-comb system, important and necessary for spectroscopy, has been developed \cite{dmitriev2022hybrid}. 
Self-injection locking allows to compensate thermal effects inevitable in practical microresonator systems \cite{Ilchenko1992ThermalNE,Carmon:04,Fomin:05,Herr2014} and to facilitate access to soliton states. 
For the generation of dark pulses and platicons, self-injection locking can simplify the experimental setup and steps by avoiding complex multi-microresonator systems \cite{Xue2015,Kim:19} and pump modulation schemes \cite{Lobanov2019,Liu:22}.
Self-injection locking also provides possibilities for realization of another nonlinear processes, e.g., SHG \cite{Ling:22}.

Besides, it has been shown recently that self-injection locking can be applied to a laser in the gain-switching regime \cite{shitikov2021self}, where laser current is rapidly modulated above and below the lasing threshold. 
As a result, a frequency comb with a line spacing equal to modulation frequency is formed (see Fig. \ref{fig:gsCombSpacing}). 
In this case, self-injection locking allows for stabilizing and narrowing every comb line to a sub-100-Hz limit, as in a plain self-injection locking. 
Also it was demonstrated in an add-drop setup with Si$_3$N$_4$ on-chip microresonator with narrowing the linewidth to 4 kHz \cite{Shao:21, shao2022gain}. 
Interestingly, such stabilized combs can be tuned in terms of line spacing change with modulation frequency. 
Such high-contrast electrically tuned optical frequency combs with line spacing from 10 kHz to 10 GHz were investigated in Ref. \cite{shitikov2021self}. 
The adjustment of the modulation voltage can be used to control the width of the frequency comb in terms of the number of spectral lines.  
The unique combination of a gain-switched laser with self-injection locking enables a broad plain spectrum of the comb with sub-kilohertz linewidth.

Applications of self-injection-locked lasers for development of the quantum communication systems also have been shown in Ref. \cite{Lai:22,Wunderer:22}.
Such stabilized lasers allow to enhance characteristics of various laser-based devices, such as laser gyroscopes \cite{Geng:20,GENG2022127531,ZHANG2023129008}, LiDAR \cite{Dale:14,Lihachev:22}, sensors \cite{Lopez-Mercado:21,Farzia:22,Shitikov2022arx,Blumenthal:22}, and atomic clocks \cite{Lai:21}. 
Recently, it has been proposed a novel method to generate spectrally pure terahertz signals by beating two self-injection-locked lasers \cite{Amin:22,9895837}.

\section{Laser self-injection locking to photonic chips}

\subsection{Brief introduction to integrated photonics}

With advances of integrated photonics, particularly the development of low-loss photonic integrated circuit (PIC), high-$Q$ optical microresonators can now be realized on silicon chips.   
Integrated material platforms, which allow the fabrication of PIC-based microresonators using CMOS foundry processes, have been widely explored for linear and nonlinear photonics including frequency comb generation \cite{Kippenberg2018}, supercontinuum generation \cite{Gaeta2019}, wideband frequency translation \cite{Li:16}, and Brillouin lasers \cite{Gundavarapu:19}.
While silicon-on-insulator (SOI) wafers -- ubiquitously used for microelectronic circuits -- have also been the mainstream integrated platform for photonics, it is well known that silicon has intrinsic material limitations such as the two-photon absorption in the telecommunication bands that precludes high power handling and ultralow optical loss. 
In the past decade, a myriad of material platforms have emerged to complement or even to replace silicon, particularly for nonlinear photonic applications \cite{Gaeta2019,Kovach:20}.  
In addition to optical nonlinearity of the material itself, optical loss in the waveguide (i.e. inversely proportional to the microresonator $Q$ factor) is a critical figure of merit when comparing different platforms. 
The optical loss not only depends on material properties such as intrinsic optical absorption, but also on fabrication processes. 
Despite that currently the $Q$ factors of PIC-based microresonators remain orders of magnitude lower than those obtained in the best bulk fluoride crystals WGMs and suspended SiO$_2$ microdisks \cite{Wu:20}, main interest and focus have been put on integrated platforms with continuous effort to reduce optical loss.

The key advantages of PIC-based microresonators over bulk fluoride crystalline WGMs and SiO$_2$ microdisks are:

\begin{itemize}
\item Fabrication of microresonators can employ mature CMOS technology that have been developed for decades for microelectronic circuits.
The CMOS fabrication allow scalable manufacturing of integrated devices with high volume and low cost.

\item Both microresonators and bus waveguides can be directly fabricated together on the same chip, thus the coupling between them is much more robust as compared to the case where tapered fibers or prisms are used to couple light into fluoride crystalline WGMs and suspended SiO$_2$ microdisks. 
In addition, the coupling strength between the microresonator and the bus waveguide, as well as the back-reflection strength with loop mirrors in the drop port \cite{Siddharth:2022}, is lithographically controlled with high precision. 

\item Integrated microresonators do not have to be in perfect circular shapes, as long as the shapes are close.
For example, microresonators of microwave-rate FSR (e.g. less than 20 GHz) can be designed and fabricated in optimized racetracks microresonators \cite{JiX:21, Ye:21}, or even in spiral shapes to achieve extremely low FSR down to the RF domain, e.g. 135 MHz in Ref. \cite{Li2021}.
This design freedom allows significant reduction in device footprint on chip.

\item Heterogeneous integration enables co-fabrication of the laser and the external microresonator on the same monolithic substrate \cite{Xiang:21}, offering critical robustness and stability to the laser-microresonator coupled system. 
In addition, tuning elements can be simultaneously implemented, allowing fast (megahertz to gigahertz rate) resonance frequency actuation via e.g. piezoelectric MEMS \cite{Lihachev:22} or electro-optic lithium niobate \cite{Snigirev:21}. 
\end{itemize}

All these features highlight that the bridging with integrated photonics significantly broadens the technological scope and maturity of high-performance, self-injection-locked lasers and microcombs with large-volume and low-cost manufacturing. 

Among all material platforms used in integrated photonics so far, Si$_3$N$_4$ \cite{Gondarenko:09, Levy:10, Spencer:14, Xuan:16, Ji:17, Ye:19, WuK:20, Liu2021} has become the leading platform for applications that rely critically on ultralow loss \cite{Moss:13, Xiang:22a}.
Silicon nitride has a long history of being used as a CMOS material for diffusion barriers and etch masks in microelectronics. 
Its first use in integrated photonics can date back to 1987 \cite{Henry:87} or even earlier.
Silicon nitride has many properties that make it suitable for building ultralow-loss optical waveguides and high-$Q$ photonic microresonators.  
Its refractive index $n_0=2$ enables strip waveguides of tight optical confinement with SiO$_2$ cladding. 
Compared with silicon, the smaller difference in refractive indices between the Si$_3$N$_4$ waveguide core and SiO$_2$ cladding can reduce scattering losses induced by interface roughness and facilitate fiber-chip interface coupling with reduced mode mismatch. 
Amorphous Si$_3$N$_4$ has a wide transparency window from visible to mid-infrared and a large bandgap of 5 eV that makes Si$_3$N$_4$ immune to two-photon absorption in the telecommunication band around 1550 nm.  
In addition, Si$_3$N$_4$ has a dominant Kerr nonlinearity that is nearly an order of magnitude larger than that of SiO$_2$, but simultaneously negligible Raman and Brillouin nonlinearities that limit the maximum allowed optical power \cite{Gyger:20}. 
All these features highlight that Si$_3$N$_4$ is an excellent integrated platform to realize laser self-injection locking using chip devices. 

\subsection{Experimental progress}

\begin{figure*}[t!]
\centering
\includegraphics[width=1.0\linewidth]{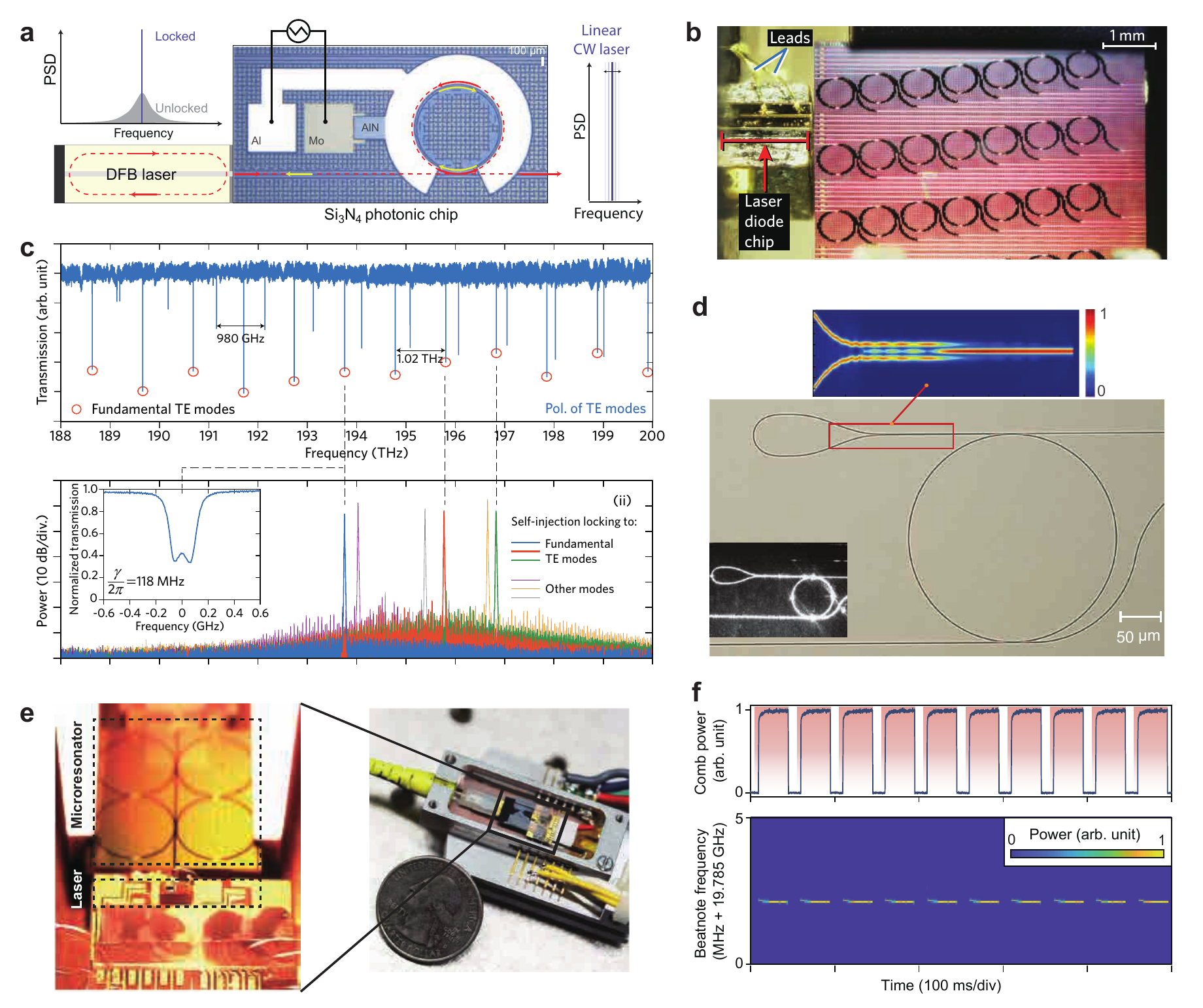}
\caption{Ultralow-noise lasers and turnkey soliton microcombs using laser self-injection locking to chip-based Si$_3$N$_4$ microresonator. 
\textbf{a}.  Schematic of a tunable, self-injection-locked laser \cite{Lihachev:22}. 
An DFB laser chip is butt-coupled to a chip-based Si$_3$N$_4$ microresonator. 
When the laser emission is coupled into the microresonator, backreflected light from the microresonator into the DFB laser can trigger laser self-injection locking that change the laser dynamics. 
As a result, the DFB laser frequency is locked to a resonance mode of the microresonator, and its linewidth is significantly reduced.
Piezoelectric actuators can be integrated directly on the Si$_3$N$_4$ microresonator, offering fast frequency tuning. 
\textbf{b}.  Close-range photo of self-injection-locked laser consisting a laser diode chip edge-coupled to a Si$_3$N$_4$ chip \cite{Raja2019}. 
\textbf{c}.   Experimental results of laser frequency locked to different microresonator resonances \cite{Raja2019}. 
Top panel shows the transmission spectrum of a Si$_3$N$_4$ microresonator of 1.02 THz FSR, where the fundamental TE mode family is marked with red circles. 
Bottom panel shows the laser emission spectrum, as well as a typical split resonance that triggers laser self-injection locking. 
\textbf{d}.   Increasing backreflection, thus to enhance laser self-injection locking, can be realized using a loop mirror in the microresonator drop-port \cite{Siddharth:2022}. 
\textbf{e}.   Images of a self-injection-locked soliton microcomb module in a compact butterfly package \cite{Shen:20}. 
\textbf{f}.   Demonstration of turnkey operation in the soliton module \cite{Shen:20}. 
Top panel shows measured comb power versus time upon power on and off. 
Bottom panel shows the spectrogram of the measured soliton repetition rate during power switching. 
Images are taken from Ref. \cite{Lihachev:22} (Panel \textbf{a}), Ref.\cite{Raja2019} (Panels \textbf{b, c}), Ref. \cite{Siddharth:2022} (Panel \textbf{d}), and Ref. \cite{Shen:20}(Panels \textbf{e, f}).
}
\vspace{10pt}
\label{fig:SIL_OnChip_1}
\end{figure*}

Self-injection locking of a semiconductor laser to a chip-based Si$_3$N$_4$ microresonator, without an optical isolator in between, can be realized via hybrid or heterogeneous integration \cite{Stern2018, Raja2019, Voloshin2021, Jin2021, Xiang:21}. 
Here we refer ``hybrid integration'' to the approach that a semiconductor laser diode or a gain chip is seamlessly edge-coupled to a Si$_3$N$_4$ chip, such that light is coupled from the laser into a Si$_3$N$_4$ microresonator, as shown in Figs.~\ref{fig:SIL_OnChip_1}(a, b). 
This approach has been widely used to build narrow-linewidth semiconductor chip lasers \cite{Stern:17, Fan:20} where the CW laser emission frequency is thermally controlled. 
Meanwhile,  once the circulating laser power inside the external Si$_3$N$_4$ microresonator exceeds a certain threshold, nonlinear parametric oscillation and soliton formation occur in the presence of laser self-injection locking \cite{Liang:15a, Pavlov:18}. 
Photodetection of the generated soliton stream produces an ultralow-noise microwave carrier at the soliton repetition rate \cite{Liang:15a}.

In addition, when using thin-core, ultrahigh-$Q$ Si$_3$N$_4$ microresonators, laser linewidth down to hertz level has be achieved \cite{Jin2021, Li2021, Guo2022}. 
In this case, the laser noise or linewidth is ultimately limited by the thermo-refractive noise (TRN) of the Si$_3$N$_4$ microresonators \cite{Huang:19} instead of the microresonator $Q$. 
Using ultrahigh-$Q$, spiral-shape Si$_3$N$_4$ microresonators with 135 MHz FSRs, 40 mHz Lorentzian laser linewidth has been achieved in Ref. \cite{Li2021}. 

To build low-noise chip-scale lasers, employing laser self-injection locking in a Si$_3$N$_4$ chip device enables a soliton microcomb module in a highly compact form, which consists only a laser chip or a laser diode with a high-$Q$ Si$_3$N$_4$ chip.
As shown in Fig.~\ref{fig:SIL_OnChip_1}b, a multi-mode laser diode chip with typical output power exceeding 100 mW is edge-coupled to a Si$_3$N$_4$ chip \cite{Raja2019}. 
By current tuning of the laser diode such that the laser emission frequency matches to a high-${Q}$ resonance of the Si$_3$N$_4$ microresonator, light is coupled from the diode into the microresonator. 
This triggers laser self-injection locking, and transforms the free-running, megahertz-linewidth, multi-longitudinal-mode laser diode into a single-mode laser with significantly reduced Lorentzian linewidth \cite{Kondratiev:17, Raja2019}. 
Meanwhile, as shown in Fig. \ref{fig:SIL_OnChip_1}c, it is observed that, resonances with prominent mode split -- indicating back-scattering \cite{Gorodetsky:00} -- can often be advantageous for laser self-injection locking. 
In addition, the intensity of backreflected light from the Si$_3$N$_4$ microresonator can be varied and controlled by adding a loop mirror in the drop port \cite{Siddharth:2022}, as shown in Fig.~\ref{fig:SIL_OnChip_1}d.

\begin{figure*}[t!]
\centering
\includegraphics[width=1.0\linewidth]{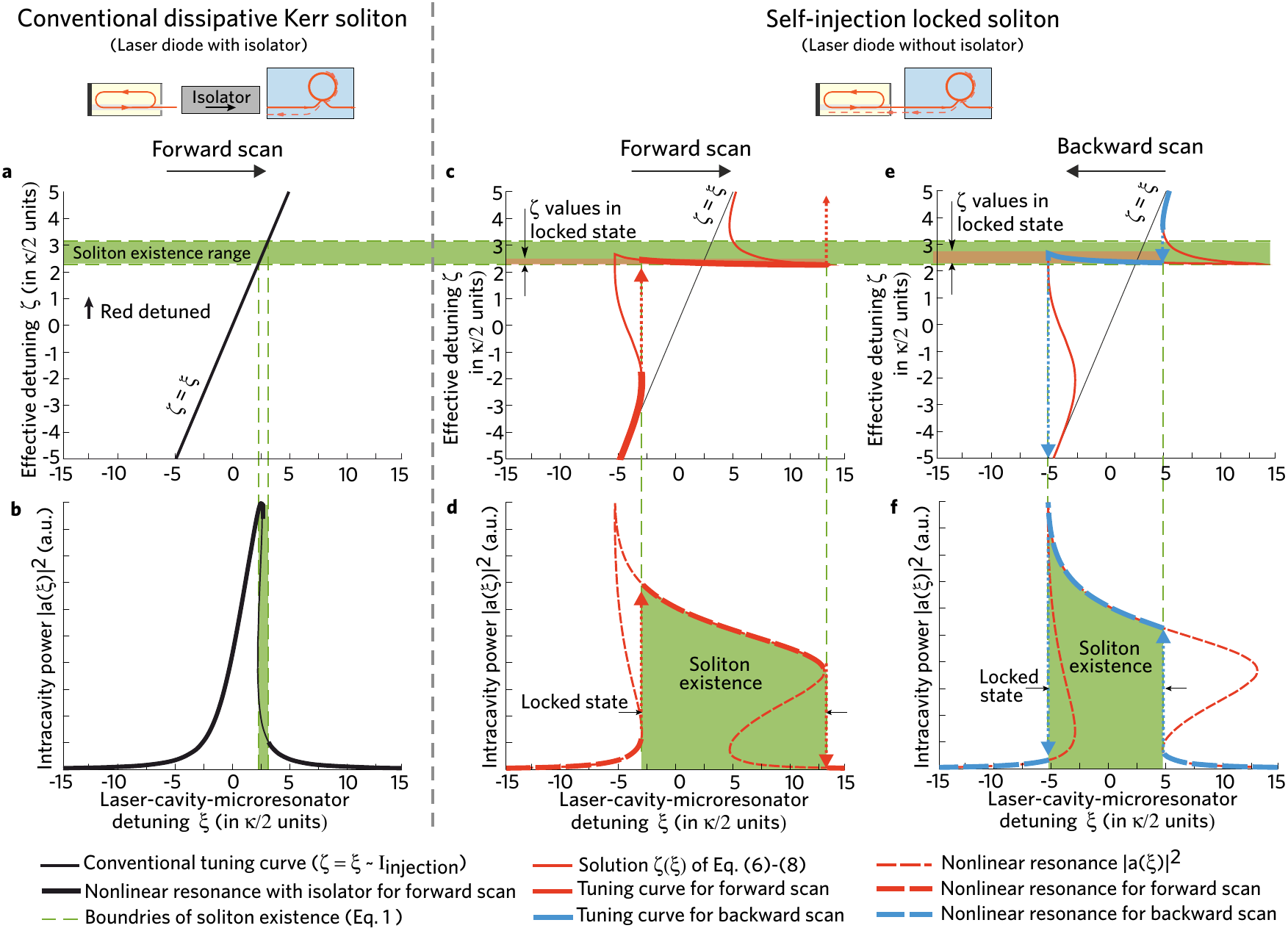}
\caption{Theoretical model of the nonlinear self-injection locking. 
Model parameters for \textbf{(a-f)}: the normalized pump amplitude $f = 1.6$, the normalized mode-coupling parameter $\Gamma = 0.11$, the self-injection locking coefficient $K_0 = 44$, the locking phase $\psi = 0.1\pi$.
\textbf{(a, b)} Model for the conventional case where the microresonator is pumped by the laser with an optical isolator, and $\zeta=\xi$ (dark green line). 
\textbf{(c-f)} Model for the self-injection locked regime. The solution of Eq. \eqref{master}-\eqref{NLbeta} (thin red curves in \textbf{(c, e)}) is compared with the linear tuning curve $\zeta = \xi$ (thin black lines in \textbf{(c, e)}).
While tuning the laser, the actual effective detuning $\zeta$ and the intracavity power $|a(\xi)|^2$ will follow red or blue lines with jumps due to the multistability of the tuning curve.
The triangular nonlinear resonance curve (thick black in \textbf{(b)}) is deformed when translated from $\zeta$ frame to the detuning $\xi$ frame \textbf{(d,f)} with corresponding tuning curve $\zeta(\xi)$ \textbf{(c,e)}.
The width of the locked state is larger for forward scan, but the backward scan can provide larger detuning $\zeta$, which is crucial for the soliton generation. 
This figure is taken from Ref. \cite{Voloshin2021}.
}
\label{fig:fig3}
\end{figure*}

With Si$_3$N$_4$ microresonators of $Q$ factors exceeding $10^7$ and anomalous GVD, the soliton generation threshold power of few tens of milliwatt can be easily satisfied in this scheme \cite{Liu:18a}. 
Via laser current tuning, a soliton microcomb can be electrically initiated, and its state can be controlled and switched from chaotic states to breathing soliton states, and finally to multi-soliton states and single soliton states \cite{Raja2019, Voloshin2021}.

In a conventional experimental setup, the light emitted from a laser passes through an isolator, and then is coupled into a Si$_3$N$_4$ microresonator. 
The soliton generation in Si$_3$N$_4$ requires complex initiation techniques where the laser quickly scans from the red-detuned side to the blue-detuned side across a resonance \cite{Herr2014, Guo2017}.
Meanwhile, accessing to the single soliton state requires delicate switching and feedback control.
This is due to the thermo-optic effect in the microresonator that the resonance experiences significant shift with drastic intracavity power variation during soliton initiation and switching \cite{Brasch:16, Li:17, Stone:18, Zhou:19}. 
It has been observed that, \textit{nonlinear} laser self-injection locking can overcome this issue, and enables soliton generation via ``turneky'' operation \cite{Shen:20}. 
As shown in Fig.~\ref{fig:SIL_OnChip_1}e, once the device is carefully assembled, packaged and stabilized, in the first trial a set of optimized parameters (e.g. laser current and laser-chip gap distance) is searched and found that allows the device to generate solitons. 
Later, as long as the setup is configured with this set of parameters, upon laser power-on, the same soliton state is immediately generated without any parameter tuning process \cite{Shen:20}, as shown in Fig.~\ref{fig:SIL_OnChip_1}f.
The turkey operation is resulted from the ultrafast (gigahertz-level bandwidth) feedback between the external photonic microresonator and the laser cavity, which is much faster than the thermo-optic speed (typically kilohertz to tens of kilohertz).
Therefore, this feature eliminates complex soliton initiation and electronic control, and offers a compact solution for field-deployable soliton microcomb modules. 

In Ref. \cite{Voloshin2021}, a set of practical parameters has been studied numerically and experimentally, leading to the following key conclusions. 
First, the effective detuning $\zeta$ predominantly locks into the red-detuned region, where the solitons can be initiated.
Second, in the self-injection locking regime, soliton generation can be observed for both directions of laser current sweep, as shown in Figs.~\ref{fig:fig3}(c, e). 
This is impossible in the conventional case where an isolator and an independent laser are used. 
Also, larger values of the detuning $\zeta$ can be obtained in the locking regime using backward tuning, as shown in Figs.~\ref{fig:fig3}(c, e).
Meanwhile, the span of detuning in the locked state can be shorter for the backward tuning than that for the forward tuning.
Third, while decreasing the diode current (i.e. backward tuning, increasing the free-running laser), as shown in Fig.~\ref{fig:fig3}e, the detuning $\zeta$ can grow that is counter-intuitive. 
Moreover, such non-monotonic behaviour of $\zeta$ can take place in the soliton existence domain and affect soliton dynamics. 
As it has been shown in Ref. \cite{Guo2017}, decreasing the detuning value in the soliton regime can trigger switching to different soliton states. 
In addition, besides bright dissipative solitons, it has been demonstrated that laser self-injection locking also allows dark pulse or platicon generation \cite{Jin2021, Yu2022, Lihachev:22a,Wang2022}.

\subsection{Heterogeneous integration advances laser self-injection locking}

\begin{figure*}[t!]
\centering
\includegraphics[width=1.0\linewidth]{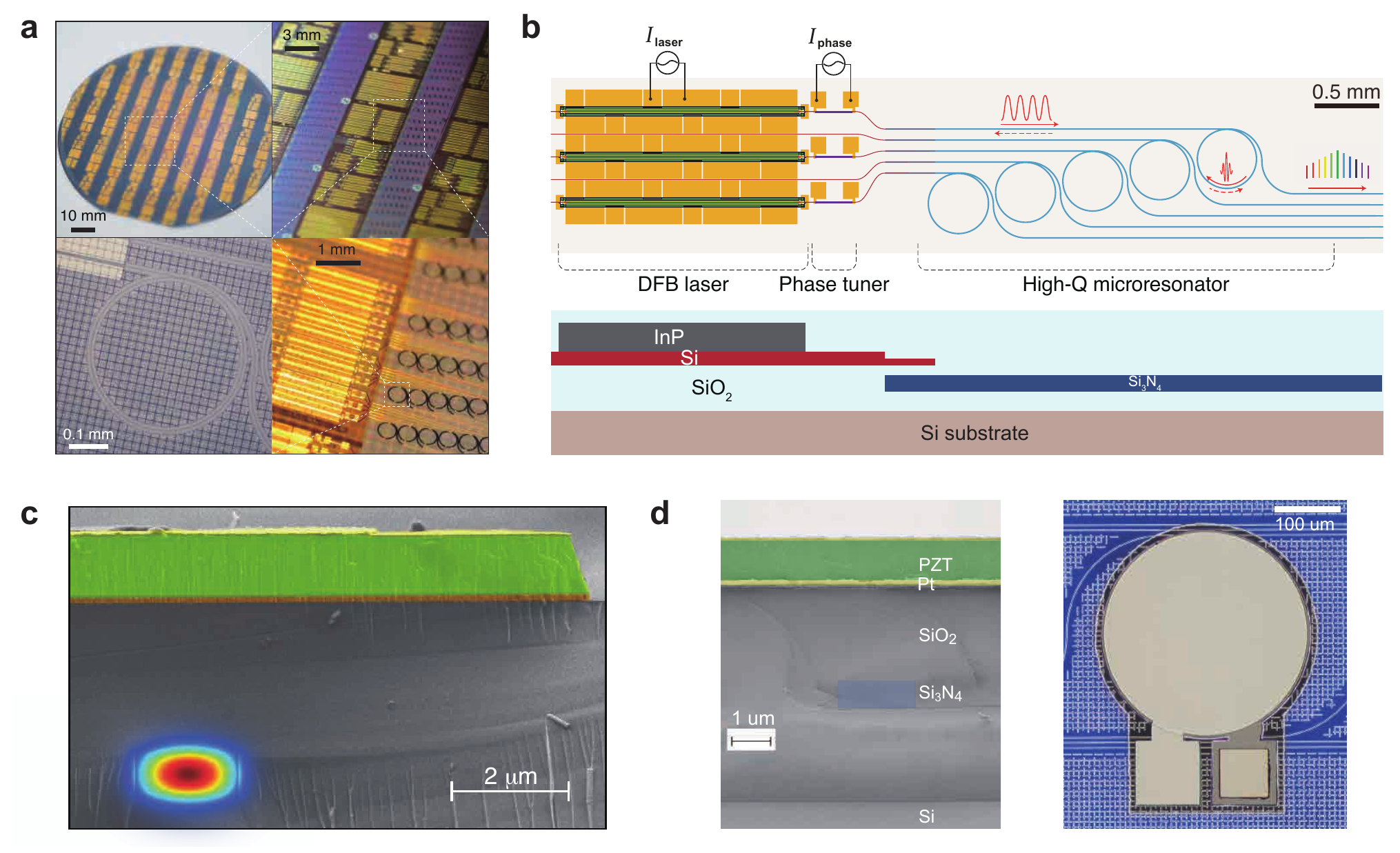}
\caption{Heterogeneous integration of lasers and modulators on Si$_3$N$_4$ photonic chips. 
\textbf{a}.  Photographs showing a completed 100-mm-diameter wafer and its zoom-in view of chips and elements \cite{Xiang:21}. 
A Si$_3$N$_4$ microring resonator and its interface with silicon is shown. 
\textbf{b}.  Schematic of laser soliton microcomb devices consisting of DFB lasers, phase tuners, and high-$Q$ microresonators on a monolithic substrate. 
Bottom panel shows the simplified device cross-section. 
The laser is based on InP/Si, and the microresonator is based on Si$_3$N$_4$. 
The intermediate silicon layer with two etch steps is used to deliver light from the InP/Si layer to the Si$_3$N$_4$ layer. 
\textbf{c}.  False-coloured scanning electron microscope (SEM) image of the Si$_3$N$_4$/AlN device cross-section, showing Al (yellow), AlN (green), Mo (red), Si$_3$N$_4$ (blue) and the optical mode (rainbow) \cite{Liu:20a}. 
\textbf{d}.  Left panel shows a false-coloured SEM image of the sample cross-section with a PZT actuator integrated on the Si$_3$N$_4$ photonic circuit. 
The piezoelectric actuator is composed of Pt (yellow), PZT (green) layers on top of Si$_3$N$_4$ (blue) buried in SiO$_2$ cladding \cite{Lihachev:22}. 
Right panel shows the optical micrograph of disk-shaped PZT actuator on top of Si$_3$N$_4$ microring with 100 GHz FSR \cite{Lihachev:22}. 
Images are taken from Ref. \cite{Xiang:21} (Panels \textbf{a, b}), Ref.\cite{Liu:20a} (Panel \textbf{c}), and Ref. \cite{Lihachev:22} (Panel \textbf{d}).
}
\label{fig:SIL_OnChip_2}
\end{figure*}

In parallel to the study of laser self-injection locking dynamics, there are equally important advances in photonic integration of laser self-injection locking for even more compact sizes and extra functions. 
The first example is heterogeneous integration of high-power, narrow-linewidth, InP/Si semiconductor lasers with ultralow-loss Si$_3$N$_4$ microresonators on a monolithic silicon substrate \cite{Xiang:21, Xiang:21a}, as shown in Fig.~\ref{fig:SIL_OnChip_2}a.
Heterogeneous integration \cite{Komljenovic:16, Park:20, Xiang:20} can further improve device stability and performance, and allows high-volume manufacturing.
Translating laser self-injection locking from hybrid to heterogeneous integration can enable thousands of narrow-linewidth lasers and soliton microcombs produced from a single wafer using CMOS-compatible techniques and foundry pilot lines.  

In Ref. \cite{Xiang:21}, a single soliton microcomb module, occupying a footprint less than 2 mm$^2$, has been demonstrated. 
This module consists of an InP/Si DFB laser, a thermo-optic resistive heater on silicon, and a high-$Q$ Si$_3$N$_4$ nonlinear microresonator. 
It is worth to mention that the thermo-optic resistive heater is used to tuned and stabilize the locking phase, while in the case of hybrid integration the gap distance between the laser chip and Si$_3$N$_4$ chip is mechanically controlled for phase tuning. 
As shown in Fig.~\ref{fig:SIL_OnChip_2}b, these elements are combined on a monolithic substrate by leveraging multilayer heterogeneous integration \cite{Xiang:20} through sequential wafer bonding of an SOI wafer and an InP multiple-quantum-well epitaxial wafer to a patterned and planarized Si$_3$N$_4$ substrate \cite{Liu2021}. 
The CW laser output from the DFB laser passes through the thermo-optic phase tuner and couples into the high-$Q$ Si$_3$N$_4$ microresonator where solitons are formed. 
Laser self-injection locking is optimized by electric control of the microresonator-laser relative locking phase using the thermo-optic phase tuner. 
The entire device outputs a CW laser with more than 1000 times linewidth reduction and a single soliton of 100 GHz repetition rate \cite{Xiang:21}. 
Furthermore, hertz-level instantaneous laser linewidth can be achieved using thin-core Si$_3$N$_4$ microresonators with higher $Q$ and lower FSR (down to 5 GHz) \cite{Xiang:21a}. 

The second example is monolithic integration of piezoelectric thin films for fast frequency modulation. 
For many metrology applications of lasers and optical frequency combs, frequency agility -- the ability to achieve megahertz to gigahertz bandwidth of frequency actuation -- is critical. 
In the case of laser self-injection locking, instead of direct modulation of the laser current, laser frequency actuation can be realized via actuating the external microresonator that drags the laser frequency to follow. 
In integrated photonics, metallic heaters deposited and patterned directly on microresonators are commonly used for frequency shift \cite{Joshi:16, Xue:16} and phase modulation \cite{Liang:21}, emplying the thermo-optic effect. 
However, heaters have several disadvantages, including kilohertz modulation bandwidth and strong cross-talk. 
Meanwhile, heaters have low tuning efficiency as the thermo-optic coefficient \cite{Arbabi:13} of Si$_3$N$_4$ is $\mathrm{d}n_\text{mat}/\mathrm{d}T=2.5\times10^{-5}$ K$^{-1}$, nearly one order of magnitude smaller than that of silicon \cite{Komma:12}. 

One approach to achieve high-speed on-chip actuators on Si$_3$N$_4$ microresonators is monolithic integration of piezoelectric actuators \cite{Tian:20, Liu:20a, Lihachev:22} on Si$_3$N$_4$ PIC. 
One suitable piezoelectric material is aluminium nitride (AlN) that is widely used in commercial micro-electro-mechanical-systems (MEMS) techniques for wireless communications. 
Figures \ref{fig:SIL_OnChip_1}a and \ref{fig:SIL_OnChip_2}c show the top-view optical microscope image and the scanning electron microscope (SEM) image of the cross-section. 
The piezoelectric actuators \cite{Tian:20, Liu:20a} are made from polycrystalline AlN as the main piezoelectric material, molybdenum (Mo) as the bottom electrode (ground) and the substrate to grow polycrystalline AlN, and aluminium (Al) as the top electrode.
The piezoelectric control employing the stress-optic effect \cite{Huang:03, vanderSlot:19} for actuation speed up to a megahertz \cite{Liu:20a, Lihachev:22}, and bulk-acoustic waves for megahertz to gigahertz actuation speed \cite{Tian:20}. 

Figure \ref{fig:SIL_OnChip_1}a shows the principle and structure of laser-self-injection locked, low-noise, frequency-agile lasers \cite{Lihachev:22}.
By fast piezoelectric actuation of the Si$_3$N$_4$ microresonator, the laser locked to the microresonator inherits the frequency actuation and can achieve a flat actuation response up to 10 MHz with optimized designs and mechanical damping. 
This low-noise, frequency-agile laser features a gigahertz frequency tuning range and megahertz tuning speed.
Similarly, electro-optic modulation of laser frequency can also be constructed on the heterogeneous Si$_3$N$_4$ -- LiNbO$_3$ platform with laser self-injection locking \cite{Snigirev:21}. 
The advantage of electro-optic modulation is the wider modulation bandwidth and smoother modulation response enabled by LiNbO$_3$, however at the cost of more complex fabrication process and lower device $Q$ factor.

Besides high speed, other key features of such piezoelectric AlN actuators are the high linearity, low hold-on electric power consumption, and maintained ultralow optical loss in the beneath Si$_3$N$_4$ PIC \cite{Liu:20a}. 
The main disadvantages are the uneven actuation response due to the presence of mechanical modes and the low stress-optic tuning efficiency (few tens of megahertz per volt)~\cite{Liu:20a}. 
The response flatness can be improved by dampening the mechanical modes \cite{Lihachev:22}, and the stress-optic tuning efficiency can be improved by using ferroelectric lead-zirconate-titanate \cite{Hosseini:15, Alexander:18, Lihachev:22} and employing geometry change \cite{Jin:18}. 
For example, cited from Ref. \cite{Lihachev:22}, Fig.~\ref{fig:SIL_OnChip_2}d shows the top-view optical microscope image and the SEM cross-section image of a PZT actuator integrated on Si$_3$N$_4$. 

\section{Outlook} 
Despite all the advances and milestones mentioned above, there are still many open questions and targets of laser self-injection locking. 
Below we outline few topics. 

\begin{itemize}
\item Exploring new physics of nonlinear laser self-injection locking dynamics. 
Currently for integrated photonics, Si$_3$N$_4$ is predominantly used as the material for external microresonators. 
However, as mentioned earlier, Si$_3$N$_4$ has dominant Kerr nonlinearity but simultaneous weak Raman and Brillouin nonlinearities. 
Therefore, if using materials other than Si$_3$N$_4$, novel dynamics can be observed in the presence of other optical nonlinearity such as Raman effect, Brillouin scattering and photo-refraction. 
How do these optical nonlinearities affect self-injection locking?

\item Laser self-injection locking to complex microresonator structures.
So far, nearly all reported works have used a single laser locked to a single microresonator. 
Can the nonlinear locking dynamics be very different if using coupled microresonators?
For example, it has been reported recently that dual-coupled-microresonator system can significantly boost the CW-to-soliton power conversion efficiency \cite{Xue2019,Boggio2022,Helgason:22}. 
Can this scheme be used with laser self-injection locking and simultaneously enable ``turnkey operation"\cite{Shen:20}? 

\item Improving long-term stability and overcoming frequency drift of self-injection-locked lasers.
Despite that there are promising progresses in achieving ultralow-noise lasers with hertz-level linewidth, the long-term frequency stability of self-injection-locked laser has not reached a level comparable to that of fiber lasers without active locking (e.g. using PDH lock). 
The laser frequency stability is ultimately impacted by the thermal stability of the external microresonator. 
For example, Si$_3$N$_4$ has a thermo-optic coefficient of $\mathrm{d}n_\text{mat}/\mathrm{d}T=2.5\times10^{-5}$ K$^{-1}$, thus 0.01$^\circ$C temperature change induces laser frequency drift of around 50 MHz at 1550 nm wavelength, much larger than the laser linewidth. 
Thermal stabilization and isolation of high-$Q$ external microresonators is becoming a central issue, particularly for chip-based devices where the laser chip and the external microresonator chip are closely packaged together. 
Therefore, an open question is: How to overcome this long-term frequency drift without significantly increasing the size, weight and power consumption of the chip module?

\end{itemize}

There are many open questions to be studied and answered on laser self-injection locking. 
It is encouraging to see that this field has today become an active field in optics (particularly in integrated photonics), and we are very sure that there will be more achievements in the future which will bring laser self-injection locking to next-gen chip-scale lasers and frequency combs.  

\section*{ADDITIONAL INFORMATION}
This review article is submitted to Frontiers of Physics, Special Topic on ``Embracing the Quantum Era: Celebrating the 5th Anniversary of Shenzhen Institute for Quantum Science and Engineering'' (Editors: Dapeng Yu, Dawei Lu and Zhimin Liao).

\begin{acknowledgments}
The authors are grateful for the fruitful discussion and collaboration with colleagues at EPFL, UCSB, Caltech, Purdue, and OEWaves, during and especially prior to the preparation of this review. 
The results presented in section 3.2 were obtained with the support of the Russian Science Foundation (project 22-22-00872).
The results presented in sections 2.3, 3.4 and 4 were obtained with the support of the Russian Science Foundation (project 20-12-00344).
Y.-H L. acknowledges support from the China Postdoctoral Science Foundation (Grant No. 2022M721482). 
W. L. acknowledges support from the National Natural Science Foundation of China (Grant No. 62075233) and the CAS Project for Young Scientists in Basic Research (Grant No. YSBR-69).
J. L. acknowledges support from Shenzhen-Hong Kong Cooperation Zone for Technology and Innovation (HZQB-KCZYB2020050) and from the Guangdong Provincial Key Laboratory (2019B121203002).
\end{acknowledgments}

\section*{AUTHOR CONTRIBUTIONS}
N.M.K., V.E.L., I.A.B., and J.L. contributed to conception and organization of the manuscript. 
N.M.K., V.E.L., W.L., J.L. and I.A.B. wrote the introduction. 
N.M.K., R.R.G., V.E.L., and D.A.C. organized and wrote the theoretical review part. 
A.E.S., D.A.C., N.Y.D., A.N.D., E.A.L., Y.-H.L., W.L., and J.L. organized and wrote the experimental review part. 
All authors contributed to manuscript revision, read, and approved the submitted version.

\bibliographystyle{apsrev4-2}

\end{document}